\documentclass[a4paper,11pt]{article}
\pdfoutput=1 

\usepackage{jheppub} 

\usepackage[T1]{fontenc} 
\usepackage[english]{babel}
\usepackage[utf8]{inputenc}

\usepackage{physics}
\usepackage{color}
\usepackage[dvipsnames]{xcolor}
\usepackage{graphicx}
\usepackage{wrapfig}
\usepackage{verbatim}
\usepackage{amsmath}
\usepackage{amssymb}
\usepackage{url}
\usepackage{bbold}
\usepackage{xspace}
\usepackage{slashed}
\usepackage{multirow}
\usepackage{threeparttable}
\usepackage{paralist}
\usepackage{subcaption}
\usepackage{floatrow}
\usepackage{multirow}
\usepackage{soul}
\usepackage[export]{adjustbox}
\usepackage[normalem]{ulem}
\usepackage{braket}
\usepackage{feynmp-auto}
\usepackage[version=4]{mhchem}
\usepackage{siunitx}
\usepackage{bm}

\newcommand{\orcid}[1]{\begingroup
  \hypersetup{hidelinks}\href{https://orcid.org/#1}{\includegraphics[width=10pt]{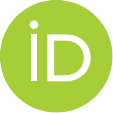}} \endgroup}

\newcommand{\ie}{\emph{i.e.}~}
\newcommand{\eg}{\emph{e.g.}~}

\newcommand{\MR}{($M-R$) }

\newcommand{\phiDwarf}{$\phi$-dwarf}
\newcommand{\phiDwarfs}{$\phi$-dwarfs}

\DeclareRobustCommand{\Sec}[1]{Sec.~\ref{#1}}
\DeclareRobustCommand{\Secs}[2]{Secs.~\ref{#1} and \ref{#2}}
\DeclareRobustCommand{\App}[1]{App.~\ref{#1}}

\DeclareRobustCommand{\Fig}[1]{Fig.~\ref{#1}}

\DeclareRobustCommand{\Eq}[1]{Eq.~(\ref{#1})}
\DeclareRobustCommand{\Eqs}[2]{Eqs.~(\ref{#1}), (\ref{#2})}

\newcommand{\dq}[1]{\ifmmode d_{#1}^{(2)} \else $d_{#1}^{(2)}$ \fi}
\newcommand{\dme}{\dq{m_e}}
\newcommand{\dmn}{\dq{m_N}}

\title{\boldmath $\phi$--Dwarfs: White Dwarfs probe Quadratically Coupled Scalars}

\author[a]{Kai Bartnick \orcid{0009-0001-8726-8485},}
\author[b]{Konstantin Springmann \orcid{0000-0002-9617-6136},}
\author[c]{Stefan Stelzl \orcid{0000-0001-5964-1054},}
\author[a]{Andreas Weiler \orcid{0000-0001-9253-0831}}

 \affiliation[a]{Technische Universität München, Physik-Department, James-Franck-Strasse 1, 85748 Garching, Germany}
 \affiliation[b]{Department of Particle Physics and Astrophysics,
Weizmann Institute of Science, Rehovot, Israel 7610001}
\affiliation[c]{Institute of Physics, Theoretical Particle Physics Laboratory, \'Ecole Polytechnique F\'ed\'erale de Lausanne, CH-1015 Lausanne, Switzerland}

\emailAdd{kai.bartnick@tum.de}
\emailAdd{konstantin.springmann@weizmann.ac.il}
\emailAdd{stefan.stelzl@epfl.ch}
\emailAdd{andreas.weiler@tum.de}

\preprint{TUM-HEP-1569/25}

\abstract{
We study ultralight scalar fields with quadratic couplings to Standard-Model fermions and derive strong constraints from white-dwarf mass–radius data. Such couplings source scalar profiles inside compact stars, shift fermion masses, and can produce a new ground state of matter. We analyze couplings to electrons and to nucleons, incorporating composition and finite-temperature effects in white dwarf structure and equations of state. We identify two robust observables: (i) \emph{forbidden gaps}—ranges of radii with no stable configurations—and (ii) characteristic \emph{shape} distortions that drive white dwarf masses toward the Chandrasekhar limit (electron couplings) or shift the maximum mass (nucleon couplings). Confronting these predictions with precise measurements for  Sirius B and Procyon B, together with the global white dwarf population, excludes large regions of unexplored parameter space and extends earlier QCD-axion–specific bounds to a broader class of scalar theories. Our stellar constraints rely only on sourcing and do not assume the scalar constitutes dark matter; where mass reductions are small, precision laboratory searches remain competitive. White-dwarf astrophysics thus provides a powerful, largely assumption-minimal probe of ultralight, quadratically coupled scalars.
}

\begin{document} 
\maketitle
\flushbottom

\section{Introduction}

Light spin-0 particles arise in many solutions to open problems of the Standard Model (SM), such as the strong-CP problem \cite{Kim:1979if,Shifman:1979if,Dine:1981rt,Zhitnitsky:1980tq}, the Higgs hierarchy problem \cite{Graham:2015cka,Dvali:2003br,Dvali:2004tma}, the flavor-puzzle~\cite{Froggatt:1978nt}, the dark matter (DM) problem \cite{Preskill:1982cy,Abbott:1982af,Dine:1982ah,Arvanitaki:2014faa} or in string-theory \cite{Damour:1994zq,Arvanitaki:2009fg}.

Arguably, the simplest way in which an ultralight spin-0 particle can couple to the SM is linearly,
$$\frac{\phi}{\Lambda} \mathcal{O}_{\mathrm{SM}}.
$$
Scalars that are coupled in this way generate new long-range forces (5th force), which are strongly constrained by tests of EP \cite{Schlamminger:2007ht,Lee:2020zjt,Tan:2020vpf}.
However, such couplings do not arise in theories where the scalar is odd under a $\mathbb{Z}_2$-symmetry, such as \eg the QCD axion \cite{Peccei:1977hh, Peccei:1977ur, Wilczek:1977pj, Weinberg:1977ma}. 
Instead, the dominant (non-derivative) couplings to the SM are quadratic \cite{GrillidiCortona:2015jxo,Hook:2017psm,Balkin:2020dsr,Kim:2022ype},
$$\frac{\phi^2}{\Lambda^2} \mathcal{O}_{\mathrm{SM}}~.
$$
Scalars with a quadratic coupling are typically much less constrained, as a long-range force only comes at one-loop \cite{Brax:2017xho}.
Such couplings have recently received increasing attention in the context of ultralight dark matter models, including QCD axion-like theories \cite{Hees:2018fpg,Balkin:2021wea,Balkin:2021zfd,Zhang:2021mks,Balkin:2022qer,Bouley:2022eer,Banerjee:2022sqg,Balkin:2023xtr,Kim:2023pvt,Beadle:2023flm,Gomez-Banon:2024oux,Kumamoto:2024wjd,VanTilburg:2024xib,Bauer:2024hfv,Burrage:2024mxn,Barbosa:2024pkl,Banerjee:2025dlo,delCastillo:2025rbr,Arakawa:2025hcn,Delaunay:2025pho,Burrage:2025grx,Grossman:2025cov,Gan:2025nlu} but also as a solution to the Hubble tension \cite{Kamionkowski:2024axz} or the hierarchy problem \cite{Chattopadhyay:2024rha}. Note that theories in which a scalar couples to fermion masses can be mapped to specific modified gravity theories (see \cite{Damour:1993hw,Banerjee:2017uwz,Saltas:2018mxc,Abac:2020drc,Wojnar:2020ckw,Kalita:2022trq,Astashenok:2022uhw,Garain:2023ptj,Vidal:2024wto,Wojnar:2024kzy,Sextl:2021lvo} and \eg \cite{Olmo:2019flu} for a recent review) by a Weyl transformation \cite{Balkin:2023xtr}. 

Assuming a quadratically coupled scalar makes up a significant fraction of dark matter, there are promising new avenues to probe its couplings, such as searches for varying fundamental constants with quantum clocks \cite{Arvanitaki:2014faa,Brzeminski:2020uhm} and interferometry \cite{Stadnik:2014tta,DeRocco:2018jwe}. 
Variations in SM parameters can be probed by comparing the frequencies of two clocks with different parameter dependencies~\cite{Flambaum_2009,Ludlow:2015gnc,Brewer:2019gyz,Stadnik:2015kia}, most of which rely on electronic transitions. 
The exciting prospect of a nuclear clock based on the ultra-low energy nuclear transition of Thorium-229 
~\cite{Peik:2003,Campbell:2012,Kazakov:2012,vonderWense:2020bbs} has the potential to extend current best bounds on scalars coupled to the electronic and nuclear sector by many orders of magnitude \cite{Fuchs:2024edo,Delaunay:2025lgk}, depending on the cancellation of electromagnetic and nuclear contributions to the transition energy~\cite{Flambaum:2006ak,Peik:2020cwm,Caputo:2024doz}.
This is particularly exciting as, for the first time, it can offer direct sensitivity on the nuclear sector.
We note, however, that it has recently been shown that the dark matter profile around Earth can be different from the naive DM distribution expected at spatial infinity \cite{Banerjee:2025dlo,delCastillo:2025rbr}, leading to a reduced scalar amplitude at Earth's surface.

It is worth noting that the region of parameter space currently accessible to laboratory experiments is generically fine-tuned, since the scalar mass receives one-loop corrections. 
For the theory to remain natural, there needs to be a symmetry-based mechanism that parametrically suppresses them. 
Naturalness can be restored with a discrete $\mathbb{Z}_\mathcal{N}$ symmetry with $\mathcal{N}>2$ copies of the SM \cite{Hook:2018jle,Banerjee:2025kov}. 
Furthermore, in \cite{Delaunay:2025pho}, a discrete twin-symmetry is introduced that cancels the divergence at linear order in the coupling.

While laboratory experiments advance, the strongest constraints on quadratically coupled scalars come from astrophysical sources and do not assume the scalar to be dark matter, such as \eg from the cooling of supernovae or neutron stars, see \eg \cite{Raffelt:1996wa,Buschmann:2021juv,Springmann:2024mjp,Springmann:2024ret} or superradiance \cite{zel1971generation,penrose1971extraction,Damour:1976kh,Detweiler:1977gy,Cardoso:2004nk,Brito:2015oca,Arvanitaki:2009fg,Arvanitaki:2010sy,Witte:2024drg}.
Recent studies have revealed an alternative mechanism inside compact astrophysical objects that now imposes the most stringent limits across large regions of parameter space.
For the QCD axion and its lighter cousins, it has been realized that a finite background of nucleons can destabilize the minimum at $\theta=0$, which leads to a phase transition within stars \cite{Hook:2017psm,Balkin:2020dsr,Balkin:2021zfd,Balkin:2021wea,Balkin:2022qer,Balkin:2023xtr}, \ie $\phi$-stars.
The non-trivial axion field profile resulting from such a phase transition allows to probe QCD axion parameter space since it would predict a strong CP angle of order one within the Earth, a different neutrino spectrum from the Sun \cite{Hook:2017psm}, and modified gravitational wave inspiral dynamics \cite{Hook:2017psm,Zhang:2021mks}.

On the other hand, it has been realized that the same coupling leads to a reduced nucleon mass once the axion is sourced, which can have a dramatic backreaction on the star \cite{Balkin:2022qer,Balkin:2023xtr,Gomez-Banon:2024oux,Kumamoto:2024wjd}.
In \cite{Balkin:2022qer}, it is shown that the reduction of the nucleon mass allows for a new ground state (NGS) of matter which is energetically preferred to ordinary matter. 
The onset of such a phase of matter within white dwarfs leads to mass-radius relationships exhibiting a gap in radii where no stable configuration can be found, hence implying that white dwarf stars of certain radii should not exist (\emph{forbidden gaps}).
Since this contradicts mass and radius observations of white dwarfs, strong constraints have been put forward \cite{Balkin:2022qer}.
Furthermore, in the NGS phase, the reduction of the nucleon mass can stiffen the equation of state (EOS) of neutron stars \cite{Balkin:2023xtr}, rendering them heavier than in the SM, potentially addressing the hyperon puzzle \cite{Glendenning:1982nc,Chatterjee:2015pua,Vidana:2024ngv}.
Additionally, for axion models that allow a new ground state in neutron stars, it has recently been shown that heat-blanketing envelopes, an insulating phase in the outermost regions of a neutron star, are drastically thinner or even completely absent \cite{Gomez-Banon:2024oux}. 
The absence of such a phase dramatically accelerates the cooling behavior of neutron stars, which is incompatible with observations.
Additionally, cooling and other crust effects, such as thermal relaxation of the crust as well as pulsar glitches, have been used in \cite{Kumamoto:2024wjd} to further constrain the axion parameter space.
None of these effects relies on the axion to constitute any fraction of the observed DM abundance. 

In this work, we go beyond QCD axion models and probe general quadratically coupled scalar fields using white dwarf stars.
We consider quadratic couplings to electrons and nucleons that lead to new ground states and thus \emph{forbidden gaps} in white dwarfs and use mass and radius data to put stringent constraints on previously unexplored parameter space.
The previous analysis for axions \cite{Balkin:2022qer} is improved in several ways:
We 
\begin{itemize}
    \item include temperature effects in the white dwarf equation of state,
    \item allow for either electron or nucleon couplings and keep the fermion mass reduction general,
    \item consider their combined couplings, and map out their phase diagram.
\end{itemize}
Temperature effects are most relevant for low-mass white dwarfs.
As expected, we show that the analysis of \cite{Balkin:2022qer} for QCD axions is not affected by temperature effects.
In scenarios where the scalar field renders the electrons lighter, they become ultra-relativistic at low densities, which leads to maximum mass white dwarfs (\ie at the Chandrasekhar mass limit) at large radii.
We find that even in the absence of a \emph{forbidden gap}, due to the changed \emph{shape} in the \MR curve, such models are not compatible with current mass-radius observations.
In the case where nucleons become much lighter, the maximum mass of white dwarfs can be shifted well beyond the Chandrasekhar mass limit.
Going beyond the limit of a new ground state to usual first- or second-order phase transitions requires additional precision in the astrophysical modeling and will be discussed in a companion work \cite{Bartnick:2025lbv}.

The white dwarf mass-radius \MR relation is now a sharp probe of physics beyond the SM.
The discovery of Sirius B in 1844 as a faint companion to Sirius A \cite{Bessel1841, Adams1915} revealed an object of extraordinary density, whose small radius and high mass led to the concept of degenerate matter and the white dwarf equation of state \cite{Chandrasekhar:1931ftj}. Sirius B thus became the archetype for the mass–radius relation, a cornerstone of stellar astrophysics, now playing a central role in setting novel constraints on scenarios of physics beyond the SM. 

This work is organized as follows. 
In \Sec{sec:EFT}, we discuss the effective field theory (EFT) of quadratically coupled scalars. 
In particular, we discuss the naturalness of the scalar mass and induced quadratic couplings to other SM particles such as photons.
We set the stage in \Sec{sec:SMWDs} by discussing ordinary white dwarfs, including temperature envelopes and composition.
In \Sec{sec:ScalarWDs} we study how this picture is altered by the presence of quadratically coupled scalars and derive bounds on previously unexplored parameter space. 
First, we give a brief review on the sourcing of quadratically coupled scalars.
Then we proceed to discuss couplings to electrons.
Next, we review the bounds on light QCD axion, including temperature effects in the equation of state.
We end the section by investigating the quadratic nucleon coupling.
In \Sec{sec:SunEarthBounds} we discuss complementary probes of scalar sourcing beyond the white dwarf \MR relation, such as constraints arising from sourcing in Sun and Earth and neutron stars.
We conclude in \Sec{sec:Conclusion}.

\section{EFT of quadratically coupled scalars}\label{sec:EFT}
We investigate models of a new scalar $\phi$ coupled to SM fermions, which is charged under a $\mathbb{Z}_2$-symmetry that forbids linear terms of $\phi$.
At energies below the QCD scale, the relevant Lagrangian is
\begin{equation}
	\mathcal{L} = \bar{\psi_i}i\slashed{\partial}\psi_i - m_{\psi_i}\left(1 + d_{m_i}^{(2)} \frac{\phi^2}{2M_p^2} +\dots\right) \bar{\psi}_i\psi_i + \frac{1}{2} \partial_\mu \phi \partial^\mu \phi- V(\phi), 
    \label{eq:LScalarAndFermion}
\end{equation}
where $M_p=1/\sqrt{8\pi G_N}$ is the reduced Planck mass, and the fermions $\psi_i$ we consider are $i=e,n,p$.
$V(\phi)$ is the potential which is minimized at $\phi=0$.
A priori $d_{m_i}^{(2)}$ can have either sign. In this work, we are interested in the case where this interaction leads to an \emph{attractive} force, \ie 
\begin{equation}
    d_{m_i}^{(2)}<0,
\end{equation} 
since this will turn out to be crucial for the existence of \phiDwarfs.
Note, that without a $\mathbb{Z}_2$-symmetry, linear couplings of the form $d_{m_i}^{(1)}m_{\psi_i}\phi/M_p$ would also be present. Their magnitude is strongly constrained from fifth force experiments and equivalence principle tests \cite{Schlamminger:2007ht,Lee:2020zjt,Tan:2020vpf}, and in the allowed parameter space, they are too small to lead to the investigated phenomenology \cite{Balkin:2023xtr}. 

A notable example of a $\mathbb{Z}_2$-symmetric scalar with $d_{m_N}^{(2)} \sim - \sigma_{\pi N}/m_N M_p^2/f^2_a<0$ is the standard QCD axion \cite{Peccei:1977hh, Peccei:1977ur, Wilczek:1977pj, Weinberg:1977ma}, where the corresponding operator gets generated non-perturbatively. Here,
$N$ collectively denotes neutrons and protons, $\sigma_{\pi N}\simeq 50\text{MeV}$ is the nucleon sigma term, and $f_a$ is the axion decay constant. The associated $\mathbb{Z}_2$-symmetry is identified with the $\mathcal{CP}$ symmetry.
The effect of this operator on the QCD axion potential at finite baryon densities has been studied in \cite{Balkin:2020dsr}.
For lighter axions, this interaction can source axion fields within compact objects, which has been extensively studied in \cite{Hook:2017psm,Zhang:2021mks,Balkin:2021zfd,Balkin:2023xtr,Balkin:2022qer,Gomez-Banon:2024oux,Kumamoto:2024wjd}.

In this work, we broaden our focus beyond scalar-nucleon couplings and also investigate couplings to other SM fermions, notably electrons.
A possible, though tuned, UV completion, where a quadratic coupling to all massive Standard Model fermions arises, is achievable through a Higgs portal coupling.
Alternatively, such couplings directly emerge in certain scalar-tensor formulations of modified gravity, see \cite{Balkin:2023xtr}.
For our analysis, the potential of the scalar field is of particular importance, which we parametrize as,
\begin{equation}
	V(\phi) = \frac{1}{2} m_\phi^2 \phi^2  + \frac{\lambda}{4} \phi^4 +\dots= \frac{1}{2} m_\phi^2 \phi^2 \left(1 + \kappa \frac{\phi^2}{f_\phi^2} + \dots
    \right) \label{eq:V0Scalar},
\end{equation}
such that its minimum is at $\phi=0$ where $V(0)=0$ and with $f^2_\phi\equiv 2 m^2_\phi/\abs{\lambda}$ as well as $\kappa = \pm 1$. 
Higher order terms in the potential are generically expected to be present and of the form $c_{2n} (\phi/f_\phi)^{2n}$ with $c_{2n} \sim \mathcal{O}(1)$.

For our analysis, we focus on the parameter space for which it is sufficient to truncate the potential at quadratic order; however, to evaluate the region of validity of our expansion, it is crucial to keep the quartic term. 
In that case, sourcing leads to large field excursion for $\phi$, which approaches the maximum field value $\phi^\ast \equiv \sqrt{2} M_p/\sqrt{\dq{m_\psi}}$.
At this field value, the fermion becomes massless, which inevitably prevents a runaway to arbitrarily large field values
(see \Sec{sec:ScalarWDs} and \cite{Balkin:2023xtr}).
If it is a star that triggers sourcing, due to its finite size, gradient effects can prevent the field from reaching $\phi^\ast$, see also \Eq{eq:phiMaxGrad}.
If $\phi \gtrsim f_\phi$, which is the case for the region of the parameter space where
\begin{equation}
    \frac{2 M_p^2}{\dq{m_\psi}} \gtrsim f_\phi^2, \label{eq:condHigherOrderImportant}
\end{equation}
the quartic term and, in fact, all higher-order terms become relevant.\footnote{Depending on the UV completion, higher order terms might be suppressed by an additional factorial, \ie $c_{n} \sim 1/n!$ \cite{Falkowski:2019tft}. While this amounts to a large suppression at larger $n$, the lower terms, like $\phi^6$, still become relevant at field values close to $f_\phi$.}
\begin{figure}[h]
    \centering
    \includegraphics{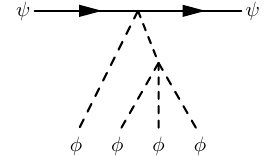}
    \caption{Generation of higher order scalar fermion couplings due to a quartic scalar self-interaction}
    \label{fig:higherOrderTermGen}
\end{figure}

Similarly, in this case, the potential generates higher order fermion interactions (see \Fig{fig:higherOrderTermGen}), which become important for $\phi\simeq f_\phi$ and not as naively guessed at $\phi \simeq \phi^\ast$.\footnote{This aligns with the generic EFT prediction, where one would expect the coupling to fermions to be given by a prefactor times an even function $F(\phi/f_\phi)$ with $\mathcal{O}(1)$ Taylor coefficients. If the prefactor is large (small), this leads exactly to the situation described above, where just the quadratic scalar fermion coupling is suppressed by a scale lower (higher) than $f_\phi$.}
Consequently, in the limit \Eq{eq:condHigherOrderImportant}, truncating the EFT is not reliable for sourcing observables.
Since we are interested in generic potentials and couplings, we thus first limit our study to $\phi \lesssim f_\phi$. 
However, there are models where the potential and its non-derivative couplings to fermions are known to all orders in $\phi/f_\phi$.
This is the case in pseudo Nambu Goldstone constructions, assuming all spurions are known, as for \eg (lighter) QCD axions.
In this case, the energetically favorable field configuration can be reliably calculated, up to the largest field excursions. For instance, for light QCD axions, the new finite density minimum is at $\phi=\pi f_\phi$ around which excitations have to be considered.
Motivated by examples with just one spurion, we subsequently study specific benchmarks and focus on cases where the full potential and coupling function coincide.

\subsection{Quantum corrections}
\begin{figure}[h]
    \centering
    \includegraphics{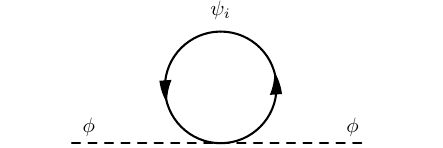}
    \caption{Fermion Loop contribution to the scalar mass}
    \label{fig:FermionLoop}
\end{figure}

 We now estimate the ultraviolet cutoff and demonstrate that the resulting Lagrangian remains natural for the characteristic electron energies encountered in white dwarfs. 
 Quantum corrections can otherwise drive the scalar mass large, where the dominant correction arises only from the fermion loop displayed in Fig.~\ref{fig:FermionLoop}.
Demanding this new contribution to be smaller than the bare scalar mass, we find the naturalness constraint
\begin{equation}
    \dq{m_i} \lesssim \num{1e16} \left(\frac{m_\phi}{\SI{e-10}{eV}}\right)^2 \left(\frac{m_e}{m_i}\right)^2  \left(\frac{\SI{0.3}{MeV}}{\Lambda_\mathrm{UV}}\right)^2. 
    \label{eq:naturalness_constraints}
\end{equation}
Let us briefly justify our choice of the ultraviolet cutoff $\Lambda_\text{UV}$. 
We are interested in sourcing the scalar field within white dwarfs.
Most observed white dwarfs that will be used to place constraints on the scalar field have masses $M\sim 0.6 M_\odot$. 
Typical electron kinetic energies within these white dwarfs are $\sim \SI{0.3}{MeV}$.
We now demand that this is the minimal UV-cutoff, $\Lambda_\text{UV}$, as suggested in \Eq{eq:naturalness_constraints}, and shown throughout the paper. If the scalar mass was instead stabilised by an additional discrete symmetry, $\Lambda_\text{UV}$ could be raised without re‑introducing fine‑tuning; see~\eg \cite{Hook:2018jle,Banerjee:2025kov,Delaunay:2025pho}.

\begin{figure}[h]
    \centering
    \begin{subfigure}{0.4\linewidth}
        \centering
        \includegraphics{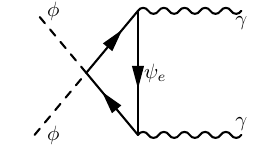}
    \end{subfigure}
    \begin{subfigure}{0.4\linewidth}
        \centering
        \includegraphics{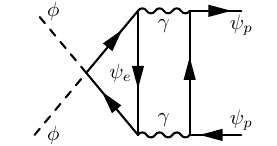}
    \end{subfigure}
    \caption{Radiative generation of additional interactions.
\textbf{Left:} A single electron loop converts the primary $\phi^2\bar\psi_e\psi_e$ vertex into a one‑loop coupling to photons.
\textbf{Right:} Inserting an extra photon rung yields a two‑loop diagram that induces a coupling to protons.}
    \label{fig:loopInducedCouplings}
\end{figure}

At one loop, the interaction in Eq.(\ref{eq:LScalarAndFermion}) radiatively generates a photon coupling,
\begin{equation}
    \mathcal{L} \supset \frac{\phi^2}{2}\frac{\dq{e}}{4 M_P^2} F_{\mu\nu}F^{\mu\nu}.
\end{equation}
via an electron loop (left panel of \Fig{fig:loopInducedCouplings}).
Matching at scales below the electron mass yields
\begin{equation}
    \dq{e} = \frac{2}{3\pi}\alpha_\mathrm{QED} \cdot \dme \approx \num{1.5e-3} \cdot \dme, \label{eq:inducedScalarPhoton}
\end{equation}
confirming the NDA expectation.
As naively expected, even in a background of maximal magnetic fields in white dwarfs of $B\simeq 10^{12}\, \text{G}$ \cite{Drewes:2021fjx}, which contributes positively to the effective scalar mass, effects due to such couplings are always subdominant.

At two‑loop order, the electron interaction radiatively generates couplings to every other charged fermion suppressed by two extra powers of $\alpha_\mathrm{QED}$.  For the proton example (right panel of Fig.~\ref{fig:loopInducedCouplings}), in the limit of small momentum exchange, one obtains
\begin{equation}
    \dq{m_p} \sim \frac{\alpha_\mathrm{QED}^2}{\left(4\pi\right)^2} \dme \sim \num{e-7} \dme,
\end{equation}
and thus this loop-induced coupling is always negligible.

\section{White Dwarfs in the SM}\label{sec:SMWDs}

White dwarfs balance the force of gravity via electron degeneracy pressure, while their mass is primarily composed of light but non-relativistic nuclei. 
Due to charge neutrality, the number density of electrons is linked to the energy density of nucleons by the relation $\varepsilon_{\psi}\simeq Y_e m_N n_e$, where $Y_e=A/Z$ represents the ratio of nucleons per electron and $n_e$ is the electron number density. 
Since white dwarfs consist of light nuclei ranging from helium (\ce{{}^4He}) to magnesium (\ce{{}^24Mg}), the nucleon-to-electron ratio is well-approximated by $Y_e\simeq 2$.
Temperatures in the core of white dwarfs can range from roughly $10^5\,\text{K}$ to $10^8\,\text{K}$ \cite{Saumon:2022gtu}.

The equation of state for a degenerate white dwarf can be modeled as a free Fermi gas of non-interacting electrons and an ideal gas of nuclei. 
We consider positively charged non-relativistic nuclei, denoted by $\psi$, with a mass of $m_{\psi}=A m_N$. 
The total pressure is dominated by the electron contribution, $p=p_e+p_{\psi}\simeq p_e$, while the nuclei provide the bulk of the energy or matter density, $\varepsilon=\varepsilon_e+\varepsilon_{\psi}\simeq \varepsilon_{\psi}$. 
Due to charge neutrality, $Z n_{\psi}=n_e \equiv n$, the electron Fermi momentum is related to the energy density as $k_F(\varepsilon)=(3\pi^2\varepsilon/Y_e m_N)^{1/3}$, and the EOS can be derived, see \eg \cite{1990sse..book.....K},

\begin{equation}
     p(\varepsilon)=\frac{2}{3}\int_0^{k_F(\varepsilon)}\frac{\text{d}^3k}{\left(2\pi\right)^3}\frac{k^2}{\sqrt{k^2+m_e^2}} .
    \label{eq:WDEOS}
\end{equation}

Because $T/\mu_e\ll1$ for ordinary white dwarfs ($\mu_e$ is the electron chemical potential), the zero‑temperature approximation is usually sufficient.  Finite‑$T$ effects mainly displace the \MR curve toward larger radii at fixed mass, a shift most pronounced for the lightest, most dilute white dwarfs as detailed in Refs.~\cite{10.1093/mnras/stx1522,Nunes:2021jdt,PhysRevC.89.015801,1972A&A....16...72S,1970A&A.....8..398K,cite-key,1975ApJ...200..306L} and reviewed in Ref.~\cite{1990RPPh...53..837K}.

\begin{figure}[h]
    \centering
    \includegraphics[width=0.7\linewidth]{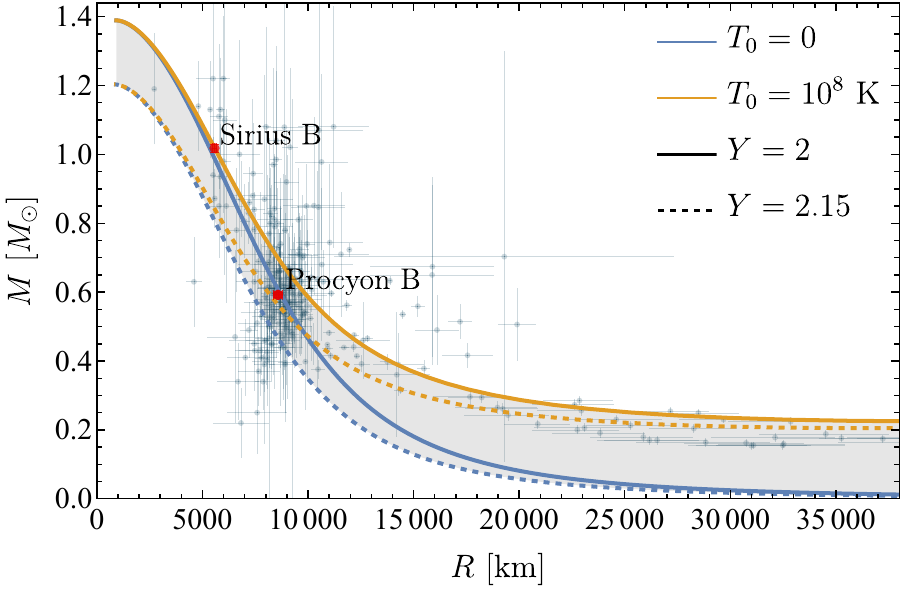}
    \caption{Mass–radius relation for white dwarfs in the free Fermi‑gas model with a radiative, He‑dominated envelope.
    The gray band shows the expected spread of the mass-radius curve, obtained by varying the composition from Helium/Carbon ($Y=2$), to Iron ($Y=2.15$), as well as the central temperature from $T_0 = 0$ up to $T_0=\SI{e8}{K}$.
    Colored curves illustrate specific choices to isolate these effects: warmer cores (blue~$\to$~orange) push the high‑radius portion of the curve to larger masses, while a higher neutron excess (dashed vs. solid) lowers the maximum mass at large radii.  
    Observed white dwarf mass-radius data is shown in blue \cite{B_dard_2017,tremblay_gaia_2016,2018MNRAS.479.1612J,Bond_2015,Bond_2017,10.1093/mnras/stx1522,Brown_2020}. The two closest white dwarfs, Sirius B and Procyon B, are highlighted in red (\cite{Bond_2015,Bond_2017}).
    Note that the errors on mass and radius measurements of those nearby stars are too small to be visible in the plot.
    }
    \label{fig:MR_SM}
\end{figure}

The EOS completes the set of equations that describe the balance between pressure and gravity, known as the Tolman-Oppenheimer-Volkoff (TOV) equations~\cite{Oppenheimer:1939ne,Tolman:1939jz}:

\begin{subequations}
\label{eq:TOV}
\begin{align}
&p'= -\frac{G_N M\varepsilon}{r^2}\bigg[1+\frac{p}{\varepsilon}\bigg]\left[1-\frac{2G_N M}{r}\right]^{-1}\left[1+\frac{4\pi r^3 p}{M}\right], \label{eq:TOV_p}\\
&M'=   \, 4\pi r^2 \varepsilon, \label{eq:TOV_M}
\end{align}
\end{subequations}
where $M(r)$ is the enclosed white dwarf mass, and all derivatives are with respect to the radial coordinate. Post‑Newtonian corrections matter only for ultra‑compact white dwarfs close to the Chandrasekhar limit and are numerically negligible for the stars we compare with observations.
Solutions with varying central pressures lead to a ($M\,$-$\,R$) relationship that aligns well with current data, see Fig.~\ref{fig:MR_SM}.

We now include temperature effects in the EOS. 
Temperature effects only play a significant role for the white dwarf structure in the outermost, least dense layers of the star, where the temperature drops from the approximately isothermal core to the surface temperature, the so-called envelope.
We are using an envelope model, which assumes the envelope to be thin compared to the bulk of the star, described in \cite{1990sse..book.....K}, which we summarize in \App{App:finiteT}.
We provide more details on precision white dwarf modeling in a companion paper \cite{Bartnick:2025lbv}.

As derived in \App{App:finiteT}, we use the temperature-dependent EOS in the envelope
\begin{equation}
 p(\varepsilon,T_0)=  \frac{\varepsilon}{ Y_e m_N} \left( \frac{3 \pi^2}{ 5^{3/2} }\left(\frac{Z + 1}{Z}\right)^{7/4} \frac{ \varepsilon}{ Y_e m_N }  \frac{T_0^{7/4}}{{m_e}^{3/2}}\right)^{4/13}, \label{eq:EOSfiniteT}
\end{equation}
 where $T_0$ is the central temperature and $Z$ is the average number of protons per ion in the envelope.
 In this work, we will fix $Z=2$, which corresponds to a helium envelope.
It is used at low energy densities below the boundary value $\varepsilon_b$, where
 \begin{equation}
    \varepsilon_b = \frac{5^{3/2}}{3 \pi^2} \left(\frac{Z+1}{Z}\right)^{3/2} Y_e m_N \left( m_e T_0\right)^{3/2} \approx \SI{0.38}{MeV^4} \left(\frac{T_0}{\SI{e8}{K}}\right)^{3/2},
    \label{eq:EOSfinteTMatchingEnergy}
\end{equation}
while above this energy density, the zero temperature EOS \Eq{eq:WDEOS} is used.
As shown in Fig. \ref{fig:MR_SM}, this simple finite $T$ EOS, in addition to varying compositions, leads to ($M\,$-$\,R$) curves that fit the white dwarf data to a very good approximation and are in good agreement with measured surface temperature where available.
Finite-temperature effects matter most for low-mass white dwarfs and are necessary to explain these observed white dwarfs.

However, note that the thin-envelope approximation fails at large radii, as the envelope holds most of the mass. For a central temperature $T_0 = \SI{e8}{K}$, this transition sets in at roughly  $R\sim \SI{40000}{km}$. We therefore limit our analysis to configurations with $R<\SI{40000}{km}$.

\section[\texorpdfstring{$\phi$--Dwarfs}{phi--Dwarfs}]{\boldmath $\phi$--Dwarfs} \label{sec:ScalarWDs}

In the presence of light scalar fields, the stellar composition of white dwarfs can change drastically. We consider a scalar field coupling to a fermion species $\psi$, representing either electrons or nucleons. 
At sufficiently large densities, the scalar can be displaced from its vacuum expectation value, which in turn changes the fermion mass.
This can lead to the formation of a new ground state of matter, \ie, a minimum energy configuration in which the scalar is sourced and has less energy per particle than matter at zero density. This phenomenon, which was analyzed in Refs.~\cite{Balkin:2022qer,Balkin:2023xtr}, is briefly reviewed below.

When gravitational effects are included, the TOV equations for standard-model matter (\Eq{eq:TOV}) couple non-trivially to the scalar field equation of motion (EOM). 
Solving this coupled system numerically provides a complete description of a \phiDwarf, a white dwarf modified by a scalar field.
However, numerical solutions can become cumbersome; fortunately, there is a simplifying and instructive limit:
If the typical length scale $\lambda_\phi\simeq m_\phi^{-1}(n)$ (see \Eq{eq:mPhiDens} below) over which the scalar field changes its value is much smaller than the radius $R$ of the white dwarf
\begin{equation}
    \lambda_\phi \sim \sqrt{\frac{M_p^2}{\dq{m_\psi}m_\psi n }} \ll R, \label{eq:GradientScaleEstimate}
\end{equation}
all scalar gradient terms in the coupled equations can be neglected. 
Here, $n$ is given by typical white dwarf number densities, of the order of $n_{\text{WD}}\sim m_e^3$.
In particular, in this so-called  \emph{negligible-gradient limit}, the scalar and matter equations of motion decouple, and the scalar field tracks the minimum of the total energy density, or equivalently, the effective in-density potential.
Consequently, in this limit, the effect of the scalar can be understood as a modification of the EOS \cite{Balkin:2022qer,Balkin:2023xtr}. 
Gravity continues to be governed by the standard TOV equations \Eq{eq:TOV}, while the scalar field equation of motion at fixed number density $n$ is
\begin{equation}
    \left.\pdv{\varepsilon(\phi,n)}{\phi}\right|_{n} =\left.\pdv{V(\phi)}{\phi}+n_s\pdv{m_\psi(\phi)}{\phi}\right|_{n}= 0, \label{eq:ScalarEOM}
\end{equation}
where $n_s\equiv\braket{\bar{\psi}\psi}=\partial \varepsilon/\partial m_\psi$ is the scalar density.
In the non relativistic limit $n_s \approx n$, while in the ultrarelativistic limit, $n_s \sim m_\psi n^{2/3} \ll n$.
The total energy density of the system 
\begin{equation}
    \varepsilon(n, \phi)=\varepsilon_\psi(n,m_\psi(\phi))+V(\phi)  
\end{equation}
includes the contribution of the fluid $\varepsilon_\psi$ and the scalar potential. 
Correspondingly, the total pressure now also includes the scalar potential with the opposite sign \begin{equation}
    p(n, \phi)=p_\psi(n,m_\psi(\phi))-V(\phi).
\end{equation}

From the scalar-fermion coupling introduced in Eq.~\eqref{eq:LScalarAndFermion}, the scalar-dependent fermion mass takes the form
\begin{equation}\label{eq:EffectiveFermionMass}
	m_\psi(\phi)  = m_\psi \left(1 - \frac{|\dq{m_\psi}|}{2 M_{p}^2}\phi^2\right) = m_\psi \left(1 - \theta(\phi)^2\right),\quad \theta(\phi) \equiv \frac{\sqrt{|\dq{m_\psi}|}}{\sqrt{2} M_p}\phi,
\end{equation}
where $\theta$ is dimensionless.
Solving \Eq{eq:ScalarEOM} yields the field value as a function of fermion-number density $\phi(n)$.
Solutions $\phi(n)$ give the energy density $\varepsilon(n)$ and pressure $p(n)$.

It is convenient to rewrite the scalar potential, originally given in Eq.~\eqref{eq:V0Scalar}, in terms of the dimensionless field $\theta(x)$
\begin{equation}
    V(\theta) = m_\psi^4 c_{m_\psi} \theta^2
\end{equation}
with coefficient
\begin{equation}
    c_{m_\psi} = \frac{m_\phi^2 M_p^2}{\dq{m_\psi} m_\psi^4}.
\end{equation}
This field redefinition makes it explicit that in the negligible-gradient limit, the effective number of parameters is reduced to one.
At constant fermion density  $n$ the scalar acquires an effective mass around $\phi=0$,
\begin{equation}
    m^2_\phi(n) = m_\phi^2 + n_s \left.\frac{\partial^2 m_\psi}{\partial\phi^2} \right|_{\phi = 0}, \label{eq:mPhiDens}
\end{equation}
For $n_s < c_{m_\psi} m_\psi^3 \equiv n_s^\text{tach}$, the tachyonic density, the effective mass-squared is positive and the scalar field remains at zero. 
For $n_s> n_s^\text{tach}$, a tachyonic instability occurs, and the scalar develops a non-zero vacuum expectation value $\braket{\phi}\neq0$.\footnote{In the sourced case, typically (see \Sec{sec:QuadCouplingElectrons}), linear scalar-fermion-couplings $\frac{m_{\psi_i}} {M_p} \sqrt{{\dq{m_{\psi_i}}}/{2}} \,\phi \bar{\psi_i} \psi_i$ are induced.
They provide an additional energy loss channel in stellar cooling and are thus tightly constrained, see \cite{Bottaro:2023gep}.}
If potential and coupling to electrons are purely quadratic, this tachyonic density corresponds to the critical density $n_c$ at which the EOS becomes unstable, $p<0$. 
This also holds if higher-order terms are relevant, but potential and coupling have the same functional form; however, for generic higher-order terms, the instability of the EOS can be delayed to $n_c > n^\text{tach}$.

Let us now discuss the maximum field excursion induced by sourcing in the regime where it suffices to keep only the quadratic terms in the potential and the coupling.
This will, in particular, allow us to derive the validity condition (see also \Eq{eq:condHigherOrderImportant})
\begin{equation}
    \frac{2 M_p^2}{\dq{m_\psi}} \le f_\phi^2 \label{eq:condHigherOrderImportant2}
\end{equation} 
for this truncation.
Although a tachyonic mass might suggest otherwise, sourcing does \textit{not} drive the field to arbitrarily large values, even in the purely quadratic limit.
This can be seen by studying the scalar field equation of motion \Eq{eq:ScalarEOM} at fixed number density $n > n^\mathrm{tach}$. For small $\phi$, by definition, the term containing the derivative of the mass dominates over the potential, leading to $\partial \varepsilon/\partial\phi<0$. On the other hand, for large $\phi$, the fermion becomes lighter and thus more relativistic, and as $m_\psi(\phi)$ approaches $0$ the ultrarelativistic approximation $n_s \sim m_\psi n^{2/3}$ becomes valid, which shows that at fixed $n$, $n_s \to 0$, and thus $\partial \varepsilon/\partial\phi>0$. Consequently, there is always a solution to \Eq{eq:ScalarEOM} at finite $\phi$, where $m_\psi(\phi)>0$, see also \cite{Balkin:2023xtr}.
Alternatively, the maximum field excursion due to sourcing can be understood considering the total energy density $\varepsilon$:
The vacuum potential is always minimized at $\theta=0$, while the finite density contribution is minimized at $\theta=1$, where the fermions are massless, see \Eq{eq:EffectiveFermionMass}.
Consequently, $\theta < 1$ always, approaching $1$ from below only at asymptotically high densities.
Away from the nearly massless limit, both contributions are dominated by their quadratic terms, so $\theta$ is close to $1$ for general $n>n_c$.
Translating back to canonically normalized kinetic terms, we find that the field excursions due to sourcing are limited to
\begin{equation}
    \phi \le\phi^\ast= \frac{\sqrt{2} M_p}{\sqrt{\dq{m_\psi}}}.
\end{equation}
Imposing $\phi < f_\phi$ for the validity of the quadratic truncation and combining with the bound above yields \Eq{eq:condHigherOrderImportant2}, which specifies when higher-order terms can be neglected.

Comparing the energy gain due to sourcing, $\Delta \varepsilon R^3 \sim \Delta m_\psi n_s R^3$, with the energy contribution of scalar gradients, which scales like $\Delta \phi^2 R$ in a compact of object of size $R$, limits the maximum field excursion in the ultrarelativistic limit to\footnote{This would not be the case if the scalar were sourced in the early universe due to the medium present in all space, at least for $m_\phi > H$.}
\begin{equation}
    \phi \lesssim \phi_\mathrm{max}^\mathrm{grad} = \phi^\ast \sqrt{1- \frac{\phi^{\ast 2}}{m_\psi^2n^{2/3} R^2}}, \label{eq:phiMaxGrad1}
\end{equation}
Maximizing over $\phi^\ast$, in the ultrarelativistic limit, this gives an upper bound on $\phi$,\footnote{Note that there is a more stringent upper limit for $\phi$ if the fermions under consideration are nucleons, as the ultrarelativstic limit breaks down before.}
\begin{equation}
    \phi \lesssim R m_\psi n^{1/3} \label{eq:phiMaxGrad}
\end{equation}
where $R$ and $n\sim m_e^3$ are typical white dwarf radii and densities, respectively.
In particular this can be understood as the maximum field value obtained from scalar sourcing in the whole $m_\phi$-$\dq{m_\psi}$ parameter plane.
If $f_\phi > R m_\psi n^{1/3}$, truncating the EFT expansion at quadratic order in $\phi$ is always valid for white dwarf sourcing.\footnote{In figures \ref{fig:exclusionElectron}, we indicate the region where higher order terms are important by shading areas where $\phi^\ast > f_\phi$ when $f_\phi < R m_\psi n^{1/3}$. Note that depending on the size of the object, gradient effects limit the maximal $\phi$ also for small $\dq{m_\psi}$, see \Eq{eq:phiMaxGrad1}, so the higher order terms only become important in a region of intermediate $\dq{m_\psi}$ where large field excursion are reached. At small $\dq{m_\psi}$, the field excursion are again small, and there are no bounds from the modification of the stellar structure.}

\begin{figure}[h]
    \centering
    \includegraphics[width=0.7\linewidth]{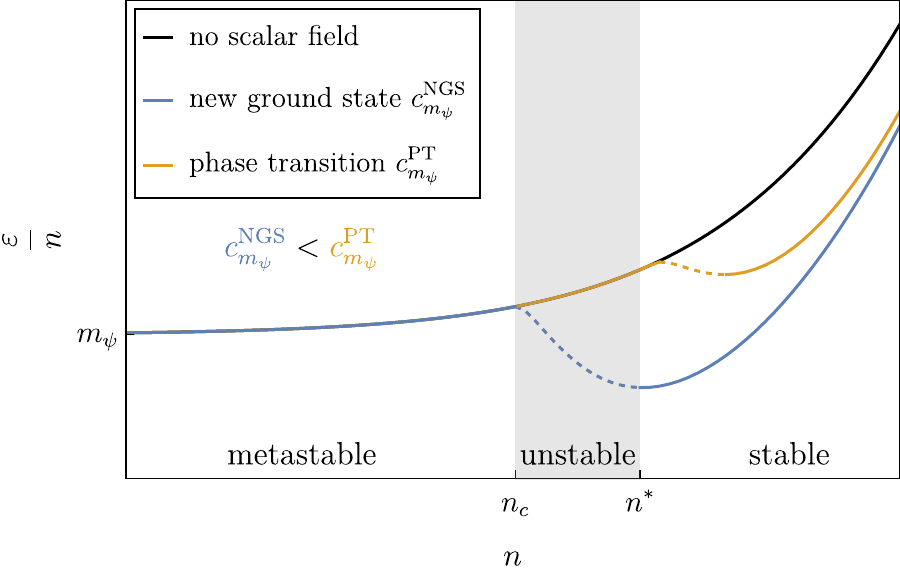}
    \caption{Energy per particle with (blue and orange) and without (black) a light, sourced scalar field.
    The lower value of $n_c$ for the new ground state (blue) compared to that of the phase transition (orange) originates from a different choice of the scalar potential, in particular, a smaller $c_{m_\psi}$ for the new ground state.
    Dashed segments mark unstable regions ($p<0$). On the new ground state curve (blue), we also distinguish the metastable and stable parts.}
    \label{fig:explainNGS}
\end{figure}

Investigating the energy per particle $\varepsilon/n$, one finds that the system can settle to a new ground state of matter, see \cite{Balkin:2023xtr} and \Fig{fig:explainNGS}:
Without a scalar field, the energy per particle rises monotonically with density because of the Fermi degeneracy energy. This can equally well be seen as the Fermi pressure is proportional to the slope of the energy per particle
\begin{equation}
    p=n^2\frac{\partial(\varepsilon/n)}{\partial n}.
\end{equation}
For our choice of the coupling, non-zero scalar field values lead to lighter fermions, which reduces the energy of the system.
This can lead to the formation of a bound state, a global minimum of the energy per particle at finite density $n^\ast>n_c$, which corresponds to a new, absolutely stable, ground state (see blue curve in \Fig{fig:explainNGS}).
For larger critical densities, \emph{i.e.} higher $c_{m_\psi}$, the scalar potential and the degeneracy energy contributions can exceed the effects of the fermion mass reduction.
In this case, the energy per particle has its global minimum at $n=0$ with $\varepsilon/n=m_\psi$ and the scalar induces a first (or second)-order phase transition (see orange curve in \Fig{fig:explainNGS}).

Although this work focuses on the case of an NGS, we investigate the case of first- (and second-) order phase transitions in a separate publication \cite{Bartnick:2025lbv}, since it requires a more detailed study of the properties of white dwarfs. 

To sum up, in the NGS scenario, the equation of state splits into three regimes
\[
\begin{aligned}
n<n_c &:\quad \text{ordinary matter (metastable)},\\[4pt]
n_c<n<n^\ast &:\quad p(n)<0 \quad (\text{unstable}),\\[4pt]
n>n^\ast &:\quad \text{absolutely stable NGS.}
\end{aligned}
\]
This rich structure leads to drastic changes in the stellar structure of white dwarfs, which we explore in subsequent sections.

\begin{figure}[t]
    \centering
    \includegraphics[width=0.6\linewidth]{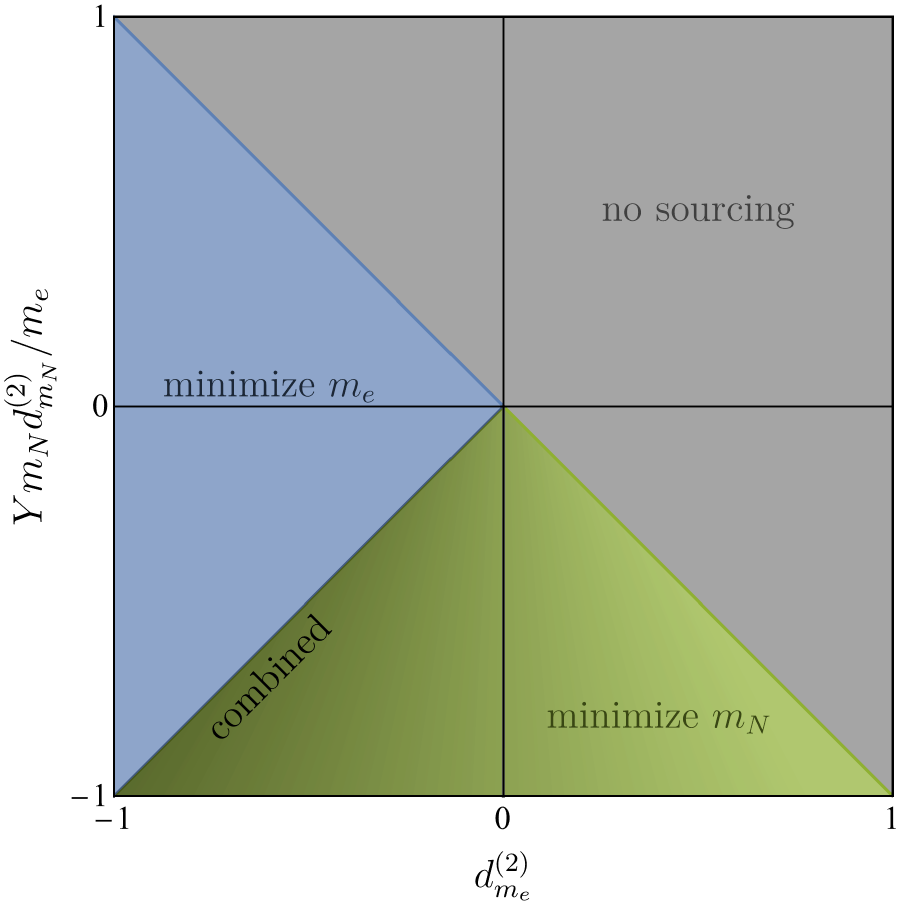}
    \caption{Sourcing in the combined parameter space of nucleon and electron coupling. 
    In most of the parameter space, one of the two couplings dominates, while in a small region, both contribute equally, and the phenomenology can be more complex, see \App{app:combinedCouplingsMR}.
    }
    \label{fig:combinedCouplingParamSpace}
\end{figure}

\subsection{Quadratic couplings to matter} \label{sec:MixedQuadraticCouplings}
A generic $\mathbb{Z}_2$-symmetric scalar can couple to all standard model fermions.
To leading order we write
\begin{equation}
    \begin{aligned}
        \mathcal{L} =&\, \bar{\psi}_N \left(i\slashed{\partial}-m_N\right) \psi_N + \bar{\psi}_e \left(i\slashed{\partial}-m_e\right) \psi_e + \frac{1}{2}\partial_\mu \phi \partial^\mu \phi - \frac{1}{2} m_\phi^2 \phi^2 \\
    & - \frac{\dmn}{2 M_p^2} m_N \phi^2 \bar{\psi}_N \psi_N - \frac{\dme}{2 M_p^2} m_e \phi^2 \bar{\psi}_e \psi_e,
    \end{aligned} \label{eq:LScalarAllFermions}
\end{equation}
where $\psi_e$ is the electron field, while $\psi_N$ with $N = (p, n)$ collectively denotes the proton and neutron, whose couplings we have taken to be isospin-symmetric. 
In a background of non-relativistic electrons and nucleons, the effective scalar mass $m_\phi^\mathrm{eff}$ is thus given as
\begin{equation}
    (m_\phi^\mathrm{eff})^2 = m_\phi^2 + \frac{\dmn}{M_p^2} m_N n_N + \frac{\dme}{M_p^2} m_e n_e = m_\phi^2 + \frac{n}{M_p^2} \left(\dmn Y m_N + \dme m_e\right),
\end{equation}
using charge neutrality ($n_p=n_e\equiv n$ and $n_N\simeq Y n$) and the nucleon-to-electron ratio $Y$ for white dwarf matter in the second equality.
Sourcing will now only occur if the effective mass correction is negative, \ie if both couplings are negative, or
\begin{equation}
  \dmn <0 \;\; \text{and} \;\;   \dme < |\dmn| \frac{Y m_n}{m_e}, \qquad \text{or} \qquad \dme < 0 \;\; \text{and} \;\; \dmn < |\dme| \frac{m_e}{Y m_N}.
\end{equation}
Since $Y m_N/m_e \sim 4000$, if the absolute values of $\dme$ and $\dmn$ are similar, sourcing does not happen for positive $\dmn$, see \Fig{fig:combinedCouplingParamSpace}.

To summarize, if both couplings to electrons and nucleons are active, the nucleon mass is minimized in most of the parameter space.
While in principle, the electron mass might also change (in both directions), the details are model-dependent.
Consequently, in the following, we will focus on the two simple cases where just one of the two couplings is active (\Secs{sec:QuadCouplingElectrons}{sec:NucleonCoupling}). 
In \App{app:combinedCouplingsMR}, we consider two benchmark points where both couplings are present.

\subsection{Coupling to electrons} \label{sec:QuadCouplingElectrons}

\begin{figure}[h]
    \centering
    \includegraphics[width=0.7\linewidth]{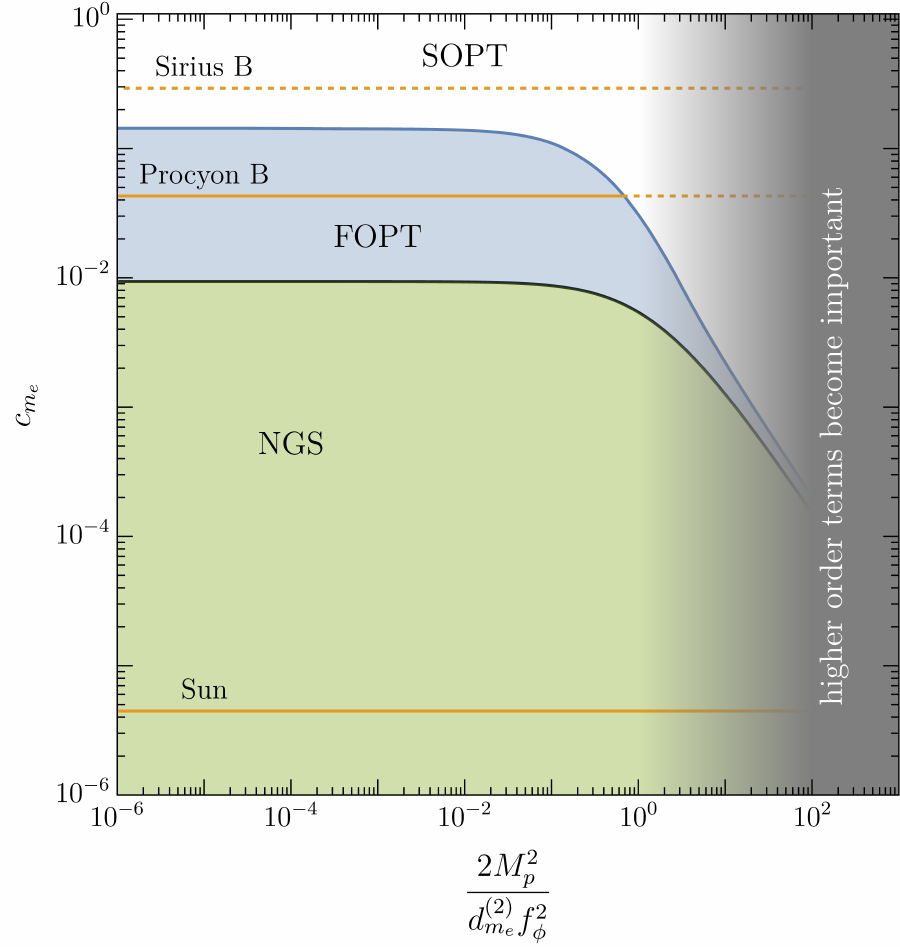}
    \caption{
    Phase diagram as a function of scalar-field parameter $c_{m_e}$ and field excursion in units of the scale of higher order terms $2 M_p^2/(\dme f_\phi^2)$.
    In the green-shaded region, one finds an NGS, in the blue-shaded region, there is a scalar-induced first-order phase transition (FOPT, see \cite{Bartnick:2025lbv}), while it becomes second-order (SOPT) in the white area.
    If the field excursions due to sourcing become too large, all higher-order terms in $\phi$ have to be included, as indicated by the gray area.
    Yellow lines mark the points where critical densities correspond to central densities of relevant exemplary objects, our Sun, as well as the two closest white dwarfs, Sirius B and Procyon B.
    Below the solid lines, a thermodynamic instability occurs inside the object, \ie at critical density $n_c < n_0$, where $n_0$ is the central density, meaning the object follows the absolutely stable branch of the EOS. 
    This only happens in the NGS or FOPT case.
    Below the dashed lines, the effective mass at $\phi=0$ becomes tachyonic inside the object, the scalar is displaced from its vacuum value, triggering the SOPT.
    }
    \label{fig:NGSBoundaryCCl_electrons}
\end{figure}

We now consider the case $d^{(2)}_{m_e}<0$, with $\dmn=0$.
At zero temperature, if higher order terms are not relevant (\ie $M_p^2/\dme \ll f_\phi^2$), we find an NGS for\footnote{This result agrees with Sec. 3.2.2 of reference \cite{Balkin:2023xtr} for neutron stars and a coupling to neutrons, since in their case the pressure is given by neutron degeneracy.
In white dwarfs, only electrons are relevant for the phase structure (\ie they play the role of degenerate neutrons), while nucleons simply track the electrons.
}
\begin{equation}
    c_{m_e}  \leq c_{m_e}^\mathrm{NGS} = \num{0.0093}.
    \label{eq:cMaxAndclMaxNGS}
\end{equation} 
We show the full phase diagram in \Fig{fig:NGSBoundaryCCl_electrons}, where we include in the potential both quadratic and quartic terms. Once the quartic becomes important, one approaches the region where all higher order terms become relevant, \ie, which can not be described with the quadratic or quartic expansion.

The NGS region is defined by the presence of a global minimum in the energy per particle, $\varepsilon/n$, at a finite number density.
A first-order phase transition (FOPT) arises when that global minimum becomes a local one, so a Maxwell construction is required to determine the stable thermodynamic state (see \Fig{fig:explainNGS}). 
A second-order transition appears once the pressure $p(n) = n^2 \partial(\varepsilon/n)/\partial n$ becomes strictly monotonically increasing with density, eliminating the need for a Maxwell construction.
In that case, $\partial(\varepsilon/n)/\partial n$ still flattens for densities above $n^{\text{tach}}$, but it never decreases.

The shape of the phase–space boundary in \Fig{fig:NGSBoundaryCCl_electrons} reflects a few competing effects.
First, larger \(c_{m_e}\) pushes scalar sourcing to higher densities.
When, at the sourcing density, the Fermi contribution to the energy per particle exceeds the gain from the electron–mass reduction, the system undergoes a phase transition rather than forming an NGS.
As the critical density increases further, electrons become increasingly relativistic and the scalar responds more weakly with \(n\); the would-be first-order transition then turns second order.
Additionally, as \(|\dme|\) decreases (\ie the field excursion approaches \(f_\phi\)), the quartic term in the potential begins to matter: it raises the potential and counteracts the gain from reducing \(m_e\).
Consequently, the NGS region (and the FOPT line) terminate at lower \(c_{m_e}\).
For \(2 M_p^2/(|\dme| f_\phi^2)>1\), higher-order terms are no longer suppressed; both the potential and the electron coupling must be treated to all orders in \(\phi\), and the naive quadratic truncation breaks down.

\begin{figure}[h]
    \centering
    \begin{subfigure}{0.49\textwidth}
        \centering
        \includegraphics[width=\linewidth]{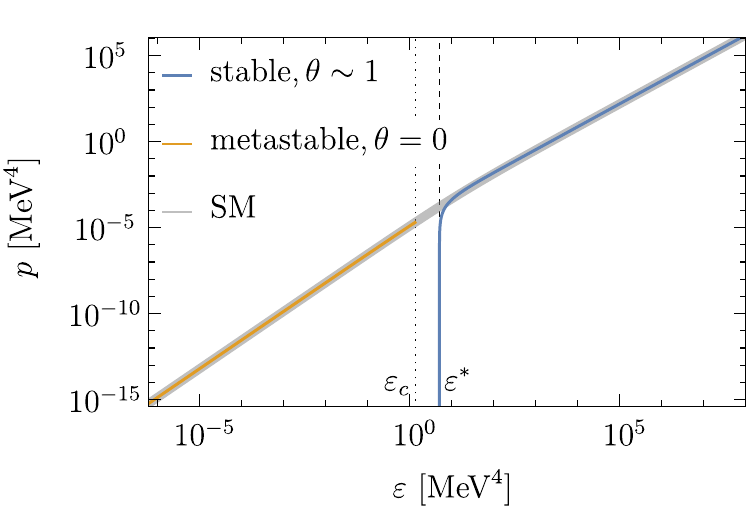}
    \end{subfigure}
    \hfill
    \begin{subfigure}{0.49\textwidth}
        \centering
        \includegraphics[width=\linewidth]{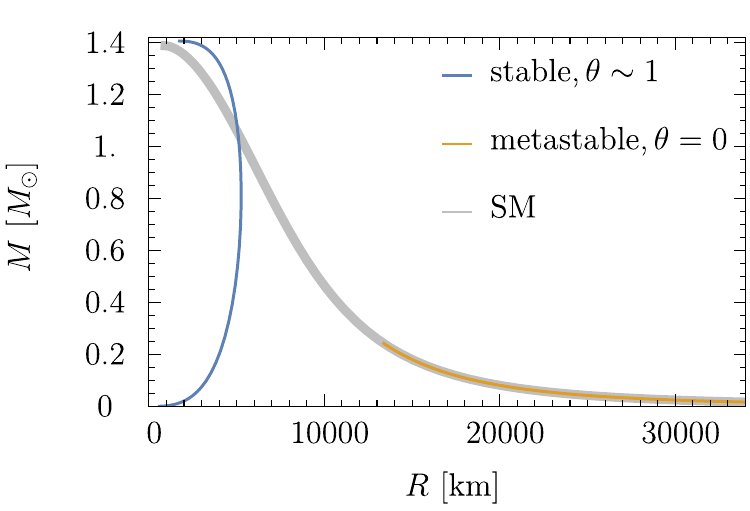}
    \end{subfigure}
    \caption{\textbf{Left:} White dwarf EOS including the effects of a scalar coupling quadratically to electrons. The EOS splits into two branches. The metastable branch (orange) below the critical density $\varepsilon_c$ (dotted line) and the stable branch (blue), which starts at the NGS density $\varepsilon^*$ (dashed line). SM EOS without a scalar field shown in gray for comparison. We consider the benchmark point $c_{m_e} = 0.005$ and $Y=2$.
    \textbf{Right:} Mass-radius relation obtained from this EOS. The two branches of the EOS correspond to two branches in the ($M\,$-$\,R$) relation. At intermediate radii, none of those predict white dwarfs. We show the SM prediction in gray.}
    \label{fig:EOSMRScalarElectron}
\end{figure}

We concentrate exclusively on the NGS scenario when comparing with data, because it produces the most pronounced changes in the EOS, a point already demonstrated for nucleon couplings in~\cite{Balkin:2022qer,Balkin:2023xtr}. 
The same qualitative effect arises when coupling a scalar to electrons, as shown in the left panel of~\Fig{fig:EOSMRScalarElectron}.
Solving the TOV equations with this scalar-modified EOS yields two distinct branches in the mass–radius relation, each corresponding to one of the positive-pressure phases; these branches are displayed in the right panel of Fig.~\ref{fig:EOSMRScalarElectron}.

As demonstrated in the previous section, the EOS without a scalar field admits white-dwarf configurations across the full interval \(R\sim\SI{40000}{km} \,\text{to}\,R\sim \SI{1000}{km}\). Introducing the scalar field carves out a \emph{forbidden gap}: no equilibrium solutions exist at radii between the stable and metastable branch. Therefore, any values of \(c_{m_e}\) that would place even a handful of observed white dwarfs inside that radius gap are already ruled out by current mass-radius data. In~\cite{Balkin:2022qer}, we have performed a full statistical analysis and compared it to this simple exclusion criterion and found no quantitative difference in the excluded parameter spaces.

\begin{figure}[h]
    \centering
    \includegraphics[width=0.7\linewidth]{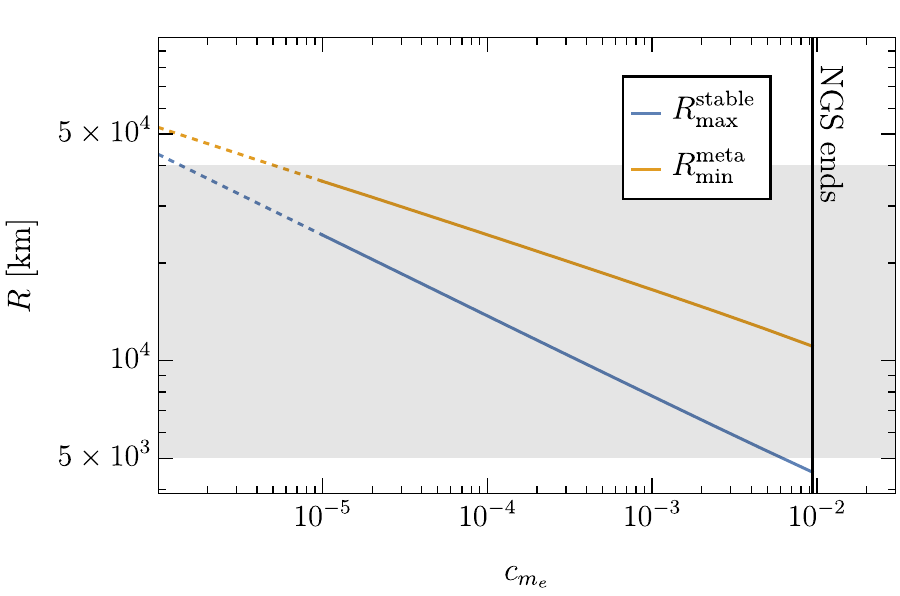}
    \caption{Numeric results and analytic estimates for the position of the radius gap as a function of scalar mass to coupling ratio $c_{m_e}$.
    For $c_{m_e}\geq 10^{-5}$ an explicit numerical calculation of the ($M\,$-$\,R$) curves was performed (solid), at smaller $c_{m_e}$, the gap-position is extrapolated using the analytic estimates Eqs.~\eqref{eq:RminMetaAna}, \eqref{eq:RmaxStableAna} (dashed).
    The gray shared area marks typical white dwarf radii.}
    \label{fig:RGap}
\end{figure}

In the negligible gradient limit, the location of the gap only depends on the scalar field parameter, $c_{m_e}$.
Before discussing our numerical results, it is instructive to derive analytic estimates for the position of the gap, following the procedure developed in~\cite{Balkin:2022qer}, see~\App{App:RadiusGap} for more details.
Using a non-relativistic (ultrarelativistic) approximation for the EOS, the minimum radius on the metastable branch (maximum radius on the NGS branch) is given by
\begin{subequations}
\begin{align}
    R_\mathrm{min}^\mathrm{meta} &\simeq 1.124 \cdot \frac{\sqrt{8\pi}M_p}{m_N m_e} (c_{m_e})^{-1/6}\label{eq:RminMetaAna},\\
	R_\mathrm{max}^\mathrm{stable} &\simeq 0.291 \frac{\sqrt{8\pi}M_p}{m_N m_e} \left(\frac{m_e^3}{n^\ast}\right)^{1/3}, \label{eq:RmaxStableAna}
\end{align}
\label{eq:RgapAnalyticEstimates}
\end{subequations}
where a parametric estimate for $n^\ast$ is given in \Eq{eq:rhoNGSAnalytic}.
The $\mathcal{O}(1)$ prefactor is set by comparing with numerical results.
As shown in \Fig{fig:RGap}, we combine the numerical results (where available) with these analytical estimates to place constraints.

\begin{figure}[h]
    \centering
    \begin{subfigure}{.49\textwidth}
        \centering
        \includegraphics[width=\textwidth]{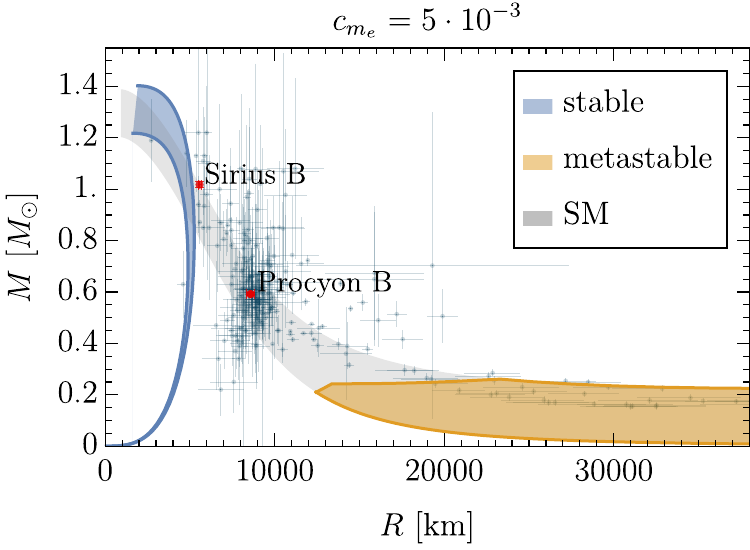}
    \end{subfigure}
    \hfill
     \begin{subfigure}{.49\textwidth}
        \centering
        \includegraphics[width=\textwidth]{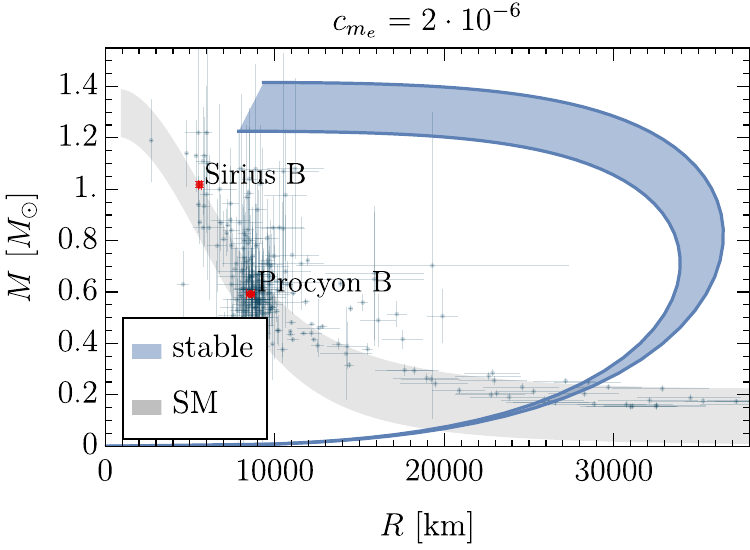}
    \end{subfigure}
    \caption{
    Scalar modified ($M\,$-$\,R$) curves.
    Bands are obtained by varying central temperature $T_0$ from $0$ to \SI{e8}{K} and composition $Y_e$ from $2$ (Helium or Carbon) to $2.15$ (Iron).
    We show the stable (blue) and metastable branch (orange), and for comparison, the ($M\,$-$\,R$) curves without scalar (gray).
    Data in blue with Sirius B and Procyon B in red as in \Fig{fig:MR_SM}.
    \textbf{Left:} \MR curve for the same mass to coupling ratio as in \Fig{fig:EOSMRScalarElectron}, $c_{m_e}=\num{5e-3}$.
    \textbf{Right:} Same as left, but for stronger coupling or lighter scalar mass, $c_{m_e}=\num{2e-6}$.
    The metastable branch is at radii beyond the range of the plot, where the simple envelope approximation would break down.
    While there is, in principle, still a radius gap at large $R$, the prediction with the scalar field is also in direct tension with the observation of Sirius B and Procyon B.
    }
    \label{fig:MRScalarElectron}
\end{figure}

Even when the gap's location is not ruled out by existing white-dwarf measurements, the \textit{shape} of the mass-radius (($M\,$-$\,R$)) relation remains a powerful discriminator.  
In the NGS limit, the electron mass is negligible, so electrons are ultrarelativistic already at low densities; this flattens the ($M\,$-$\,R$) curve near the Chandrasekhar mass, \(M\simeq1.44\,M_\odot\). When the critical density is small, the resulting flattened branches are in contradiction with observations, excluding the associated $c_{m_e}$ values (see the right panel of Fig.~\ref{fig:MRScalarElectron}). In particular, any parameter set that remains incompatible with the accurately measured masses and radii of Sirius B and Procyon B is ruled out.

\begin{figure}[h!]
    \centering
    \includegraphics[width=0.49\linewidth]{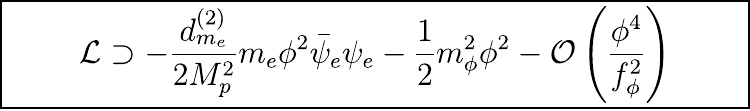} \vfill         
    \centering
    \begin{subfigure}{0.49\linewidth}
        \centering
        \includegraphics[width=\linewidth]{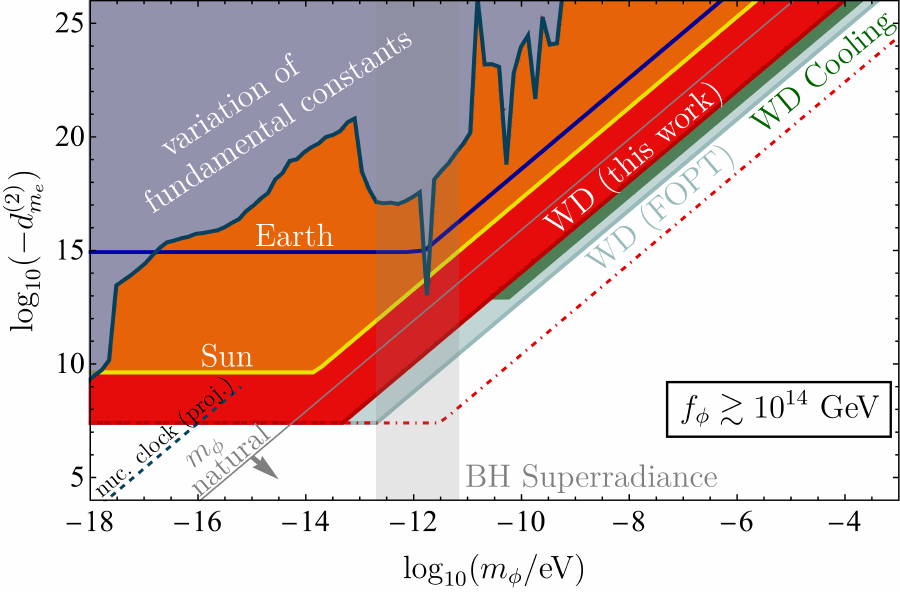}
    \end{subfigure}
    \begin{subfigure}{0.49\linewidth}
        \centering
        \includegraphics[width=\linewidth]{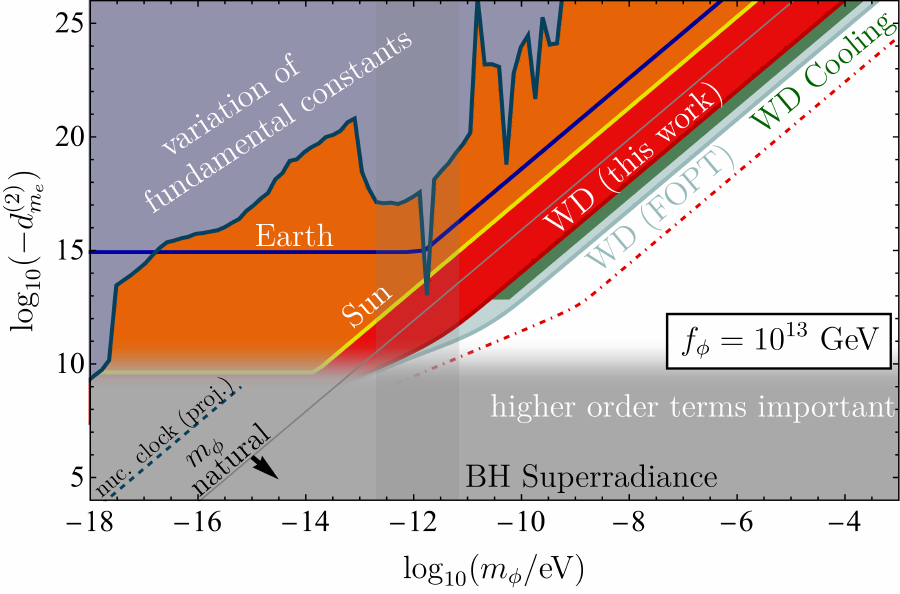}
    \end{subfigure}
    
    \begin{subfigure}{0.49\linewidth}
        \centering
        \includegraphics[width=\linewidth]{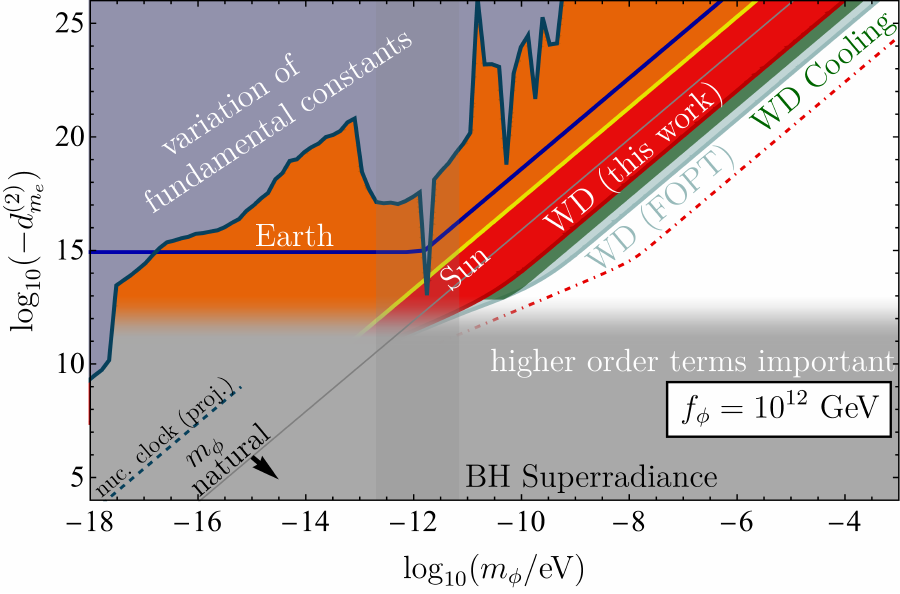}
    \end{subfigure}
    \begin{subfigure}{0.49\linewidth}
        \centering
        \includegraphics[width=\linewidth]{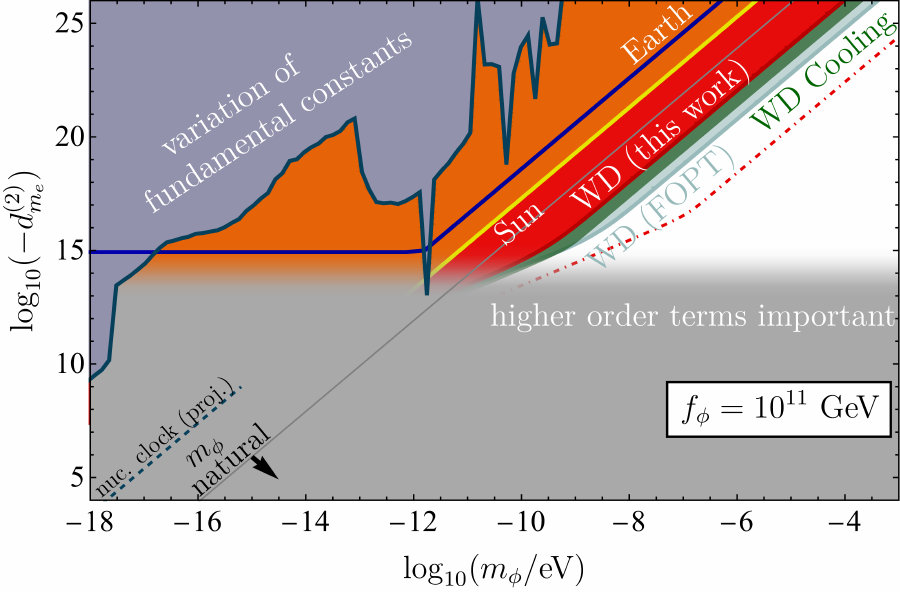}
    \end{subfigure}
    \caption{Allowed parameter space showing the limits from \phiDwarfs~(red). Different panels show different values of $f_\phi$ where the self-interaction becomes important.
    If sourcing predicts field excursions of order $f_\phi$ and larger, the quadratic expansion breaks down and the full shape for potential and coupling is needed, as marked by the gray areas.
    The red dash-dotted line marks the maximum density where the scalar field would still be sourced in heavy white dwarfs.
    In the area marked in cyan, scalar sourcing leads to a FOPT, which is also constrained from changes in the white dwarf \MR curve \cite{Bartnick:2025lbv}.
    The light green shaded area translates the cooling bounds from \cite{Bottaro:2023gep} to the case of a sourced scalar. 
    If the scalar field is sourced in the Sun or Earth, it is excluded by solar spectra or falling objects, respectively, as discussed in \Sec{sec:SunEarthBounds} below, which is shown in yellow (blue).
    The gray-shaded area marks where black hole superradiance is relevant \cite{Baryakhtar:2020gao,Mehta:2020kwu,Unal:2020jiy,Hoof:2024quk,Witte:2024drg}.
    If the scalar field is a significant portion of dark matter, bounds from the variation of fundamental constants \cite{Kennedy:2020bac,Oswald:2021vtc,Vermeulen:2021epa,Savalle:2020vgz,Aharony:2019iad,Antypas:2019qji,Aiello:2021wlp,Branca:2016rez,Sherrill:2023zah,Kobayashi:2022vsf,Campbell:2020fvq,Zhang:2022ewz,Oswald:2025bih}, shown in dark blue, apply. The projected sensitivity of a nuclear clock is shown with the dashed blue line \cite{Caputo:2024doz,Kim:2022ype}.
    To the right of the gray line, the scalar mass can be natural without additional model building.
    }
    \label{fig:exclusionElectron}
\end{figure}

We have included finite temperature effects in much the same way as in the SM EOS.
In particular, the metastable branch exhibits the same \MR changes as in the SM, for low-density white dwarfs, at a given mass, the radius becomes larger.
In the case of an NGS, the envelope is drastically modified: 
If $n^*>n_b$, the star ends before the envelope, and finite temperature effects on the ($M\,$-$\,R$) curve are not present in our approximation. Here $n_b$ is the density where the radiative envelope starts as defined in \Eq{eq:EOSfinteTMatchingEnergy}. 
This also dramatically changes the cooling of the star.\footnote{Analogous effects for the case of neutron stars are discussed in \cite{Gomez-Banon:2024oux}. 
Similarly, pulsar emissions \cite{Draft2025} are also strongly affected.}
If $n^*<n_b$, an envelope exists but ends abruptly once $p = p_T - V(\theta) = 0$.
As in the SM case, thermal effects lead to larger radii at a given mass.
These finite temperature effects are also included in \Fig{fig:MRScalarElectron}.
Combining the constraints from the \textit{gap} and the \textit{shape} of the ($M\,$-$\,R$) curve leads to the exclusion shown in \Fig{fig:exclusionElectron}.

Let us briefly discuss the shape of our exclusion. 
If the mass at a given coupling is too large, there is no NGS but an FOPT, which we do not constrain in this paper; see, however, \cite{Bartnick:2025lbv}. 
If an NGS is present, so for small enough scalar masses, there are always at least a few white dwarfs in the \emph{forbidden gap} (\Fig{fig:MRScalarElectron}, left) or the NGS ($M\,$-$\,R$) curve \emph{shape} is such that it can no longer explain the bulk of the observed white dwarf population (\Fig{fig:MRScalarElectron}, right), therefore the exclusion always extends to arbitrary low masses.
For small couplings, gradient effects prevent sourcing, and the exclusion ends.
If higher order terms in the potential (and interaction) become relevant at a scale  $f_\phi \gtrsim \SI{e14}{GeV} \sim \phi_\mathrm{max}^\mathrm{grad}$, then in the negligible–gradient limit the field excursion remains below \(f_\phi\) and the quadratic expansion is reliable, see \Eq{eq:phiMaxGrad}.
Consequently (up to small thermal corrections), $c_{m_e} < c_{m_e}^\mathrm{NGS}$ (see \Eq{eq:cMaxAndclMaxNGS}) fixes the high mass boundary of our exclusion (see top left panel of \Fig{fig:exclusionElectron}).
This translates to 
\begin{equation}
    \dme < \frac{m_\phi^2 M_p^2}{c_{m_e}^\mathrm{NGS} m_e^4}
\end{equation}
which yields the characteristic quadratic scaling of the excluded coupling with \(m_\phi\).
For smaller $f_\phi$, the quadratic expansion breaks down at larger $\dme$ as marked by the gray areas in \Fig{fig:exclusionElectron}.
In this region, the potential and coupling need to be included to all orders in $\phi$; see below for examples, including the QCD axion.

As discussed above, the sourcing of the scalar field in a white dwarf also leads to linear couplings to electrons, which are constrained by white dwarf cooling \cite{Bottaro:2023gep} (green area in \Fig{fig:exclusionElectron}).
In deriving those limits, the white dwarf luminosity function at $M=0.6~M_\odot$ is used. 
Therefore, we apply these limits only if the critical density is below the central density of a $0.6~M_\odot$ white dwarf.
If the scalar field leads to an NGS of matter, much stronger effects on white dwarf cooling are expected, similarly to what has been explored for neutron stars in \cite{Gomez-Banon:2024oux}.

An additional important class of limits is due to black-hole (BH) superradiance 
\cite{zel1971generation,penrose1971extraction,Damour:1976kh,Detweiler:1977gy,Cardoso:2004nk,Brito:2015oca,Arvanitaki:2009fg, Baryakhtar:2020gao,Mehta:2020kwu,Unal:2020jiy,Hoof:2024quk,Witte:2024drg}.
These bounds are independent of the coupling to standard model matter and are relevant for ultra-light scalars, see gray shaded area in \Fig{fig:exclusionElectron}.
The energy-extraction rate from the BH spin due to the scalar field depends on the scalar self-interaction.
Only sufficiently weak self-interactions, $f_\phi \gtrsim \SI{8.7e11}{GeV}$, are constrained by the observation of spinning black holes \cite{Witte:2024drg}. 

If one additionally assumes that the scalar field is responsible for an $\mathcal{O}(1)$ fraction of the dark matter abundance, further limits apply.
For ultra-light dark matter (ULDM) limits from bounds on oscillating fundamental constants (shown in dark blue in \Fig{fig:exclusionElectron}), we assume the background dark matter density. 
For our plots, we recast the linear coupling bounds to the case of quadratic coupling by doubling the frequency, assuming a local dark matter density $\rho_\mathrm{DM} = \SI{0.4}{\GeV/\cm^3}$ \cite{McMillan:2011wd}.
Via the one-loop induced scalar photon coupling (\Eq{eq:inducedScalarPhoton}), the scalar electron coupling will also be probed with a nuclear clock.
We show the projected sensitivity for a $^{229}\mathrm{Th}$ clock with a dashed dark blue line in \Fig{fig:exclusionElectron} \cite{Caputo:2024doz,Kim:2022ype}.
Note that these limits might be affected by nontrivial dark matter profiles around the Earth or the Sun, see \cite{Hees:2018fpg,Banerjee:2025dlo,delCastillo:2025rbr,Budker:2023sex}.

In addition, the early universe also provides an excellent probe for ULDM since the dark matter abundance and thus the oscillation amplitudes of the fermion masses are considerably higher.
In particular, the abundance of light elements produced during Big Bang nucleosynthesis (BBN) depends on fundamental constants like the electron mass, which can be used to constrain light scalar fields \cite{Bouley:2022eer}.
To obtain the correct scalar field amplitude during BBN, thermal corrections have to be included, which depend on the sign of the coupling. In particular, in \cite{Bouley:2022eer}, the case of positive scalar fermion couplings has been worked out.
If the coupling is negative, the discussion is more involved, and the thermal correction might render the scalar mass tachyonic.

If the scalar field accounts for most of the dark matter and is extremely light ($m_\phi\lesssim \SI{e-19}{eV}$), there are additional probes.
In pulsar timing arrays, the dark matter can induce oscillations in the reference clocks on Earth as well as in the pulsar masses, leading to a shift in the pulse frequency.
In the case of a linear coupling between the scalar field and the standard model, this yields competitive constraints for very low masses $m_\phi \lesssim \SI{e-20}{eV}$ \cite{Kaplan:2022lmz,NANOGrav:2023hvm}.
While in principle also applying to quadratically coupled dark matter, the translation of the limits is nontrivial, since the scalar field can be sourced in the pulsar.
Moreover, structure formation as probed by the Lyman-alpha forest \cite{Rogers:2020ltq} can be affected, and extremely light masses are in tension with ultra-faint dwarf galaxies \cite{Dalal:2022rmp}.
For even lighter fields ($m_\phi \lesssim \SI{e-28}{eV}$), cosmological bounds on time-varying fundamental constants can become relevant even when $\phi$ constitutes only a subdominant fraction of the dark matter \cite{Baryakhtar:2024rky,Baryakhtar:2025uxs}.

\subsubsection[\texorpdfstring{Bounded $m_e(\theta)$ case}{Bounded me(theta) case}]{Bounded  \boldmath $m_e(\theta)$ case} \label{sec:BoundedMe}
So far, we have restricted ourselves to quadratic couplings between the scalar and fermion bilinears, which is only valid for $\phi \lesssim f_\phi$.
We now extend the interaction beyond quadratic order in $\phi$ while preserving the internal $\mathbb{Z}_2$-symmetry.
Higher order terms are particularly relevant if they prevent the scalar field from making the electrons massless at finite density.
We thus consider $m_e(\theta)$ bounded from below and parameterize the maximal electron mass reduction by $0<g\le1$, such that $m_e(\theta_\ast)=m_e(1-g)$.
We consider\footnote{The canonical example is the QCD axion, where the shift-symmetry is broken by non-perturbative effects. 
This leads to a periodic potential and couplings to the nucleon mass which, due to the common origin, exhibit the same functional form to leading order in the dominant shift-symmetry-breaking spurion; see \Eqs{eq:AxionPotential}{eq:AxionNucleonCoupling}.}
\begin{equation}
    \label{eq:boundElectronCoupling}
\begin{aligned}
    \mathcal{L} &\supset - g \frac{m_\phi^2 M_p^2}{|\dme|} F\left(\theta\right) - m_e \left(1 - g F(\theta)\right) \bar{\psi}_e \psi_e \\
    &= -\frac{1}{2}m_\phi^2\phi^2-m_e(1-\frac{\dme}{2M_p^2}\phi^2+\dots)\bar{\psi}_e \psi_e+\dots, \qquad \qquad \theta = \sqrt{\frac{|\dme|}{g}} \frac{\phi}{M_p}.
\end{aligned}
\end{equation}
We assume that $F(\theta) \leq 1$, with $F(\theta_*)=1$ for some $\theta_* \sim \mathcal{O}(1)$.\footnote{Unless the quadratic term is tuned small compared to the higher order terms, the last statement is expected to hold in general.} 
Additionally, we require that $F(\theta)$ is invariant under the $\mathbb{Z}_2$, leading \eg to  $F'(0)=0$, \ie no linear coupling of the scalar to electrons. 
Fixing the normalization to $F(0)=0$, and $F''(0)=1$ ensures that the scalar mass is $m_\phi$ and that the leading interaction with electrons is quadratic.
For low-field values, this exactly reproduces the quadratic coupling \Eq{eq:LScalarAndFermion} used above.

We note that in the negligible gradient limit, the explicit form of $F$ does not play a role. One can see this as follows.
By definition $F(\theta)$ monotonically increases until $\theta_*$, which allows to perform a field redefinition $\phi \to \chi = F(\theta(\phi))$.
Both potential and coupling to electrons are now linear in the new field $\chi$.
Since $\chi$ is constrained to be $\chi\geq0$, sourcing still proceeds as before.
The gradient term gets changed by the field redefinition and again introduces a dependence in $\mathcal{L}$ on the functional form of the coupling; however, in the negligible gradient limit, it can be dropped.
In particular, we note that the $g = 1$ limit exactly reproduces the case of quadratic couplings ($\lambda=0$) discussed above.

\begin{figure}[h]
    \centering
    \begin{subfigure}[c]{0.49\linewidth}
        \centering
        \vspace{.7em}
        \includegraphics[width=\linewidth]{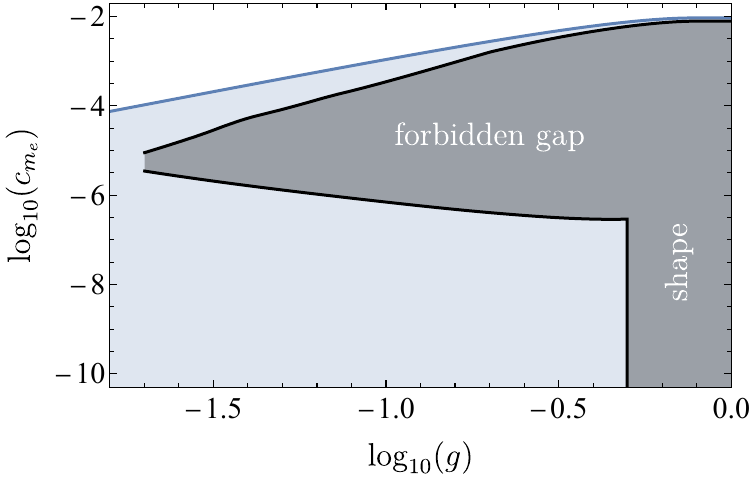}
    \end{subfigure}
    \begin{subfigure}[c]{0.49\linewidth}
        \centering
        \includegraphics[width=\linewidth]{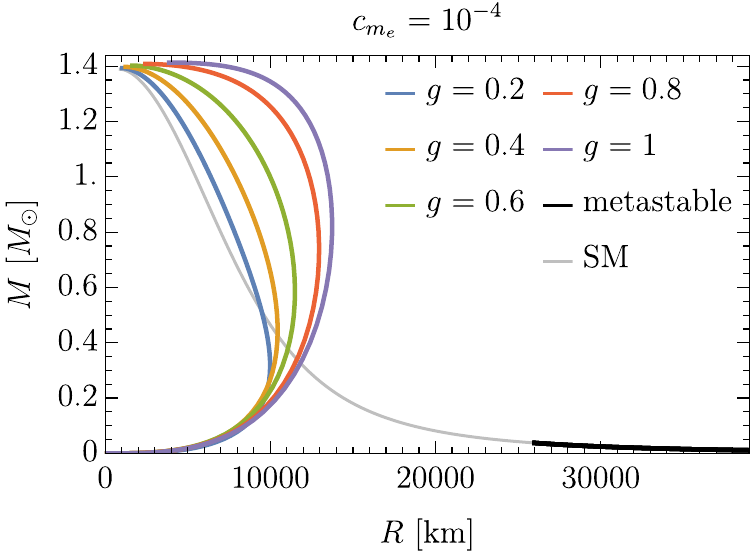}
    \end{subfigure}
    \caption{\textbf{Left:} Scalar parameter space for periodic electron coupling. Below the blue line, an NGS is found in zero-temperature White-Dwarf matter. Including temperature corrections, the NGS boundary is given by the upper black line.
    The area shaded in black is incompatible with the white dwarf \MR relation, either due to the presence of a \emph{forbidden gap} or due to the incompatible \emph{shape}, as discussed in the text.
    \textbf{Right:} \MR curves for different mass-reduction $g$, $c_{m_e}=10^{-4}$ is kept fixed. Stable branches for varying $g$ are shown in different colors; the metastable branch, which is $g$ independent, is shown in black.}
    \label{fig:NGSBoundaryAndMRBoundElectrons}
\end{figure}

\begin{figure}[h]
    \centering
    \begin{subfigure}[c]{0.49\linewidth}
        \centering
        \vspace{.3em}
        \includegraphics[width=\linewidth]{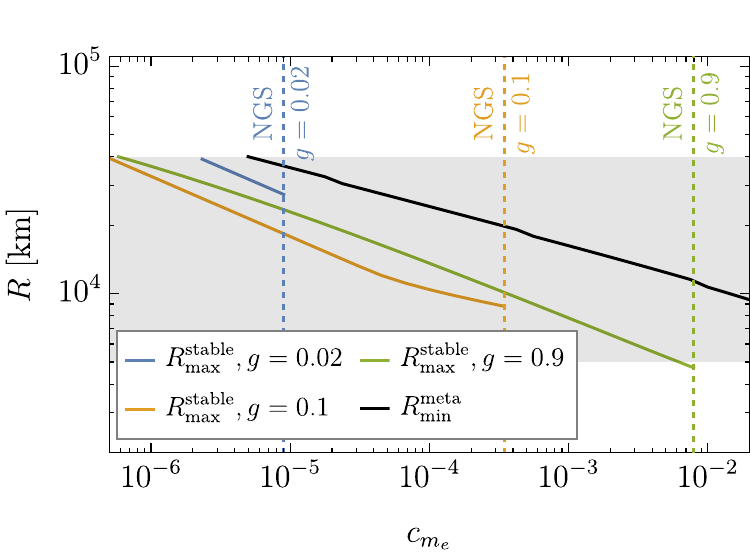}
    \end{subfigure}
    \begin{subfigure}[c]{0.49\linewidth}
        \centering
        \includegraphics[width=\linewidth]{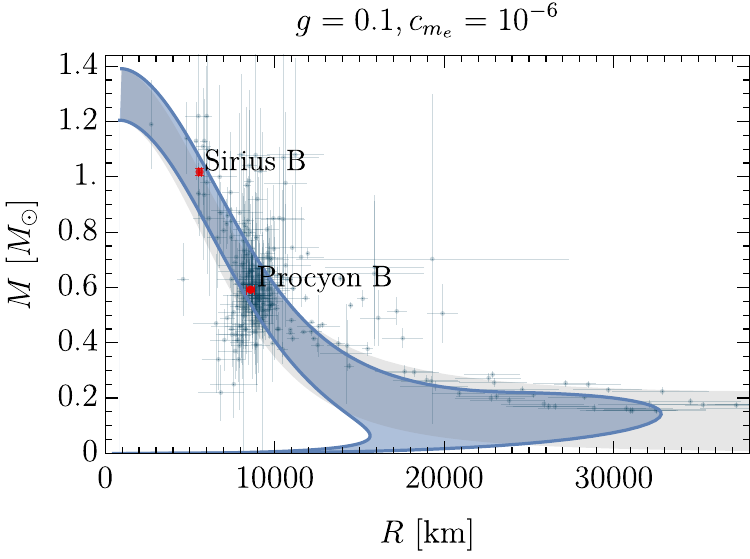}
    \end{subfigure}
    \caption{\textbf{Left:} Radius gap as a function of $c_{m_e}$. The minimal radius of the metastable branch (black) is independent of the mass reduction $g$, since $n_c$ is $ g$-independent. 
    At fixed $c_{m_e}$, as $g$ is decreased, the maximal radii of the stable branch (colors) first get smaller (due to the electrons becoming ultra-relativistic later) before getting larger again (due to the NGS-density decreasing).
    Additionally, at larger $g$, the NGS prevails until higher values of $c_{m_e}$ as marked by the vertical line in the corresponding color.
    Lastly, we only show the curves for $R\lesssim \SI{40000}{km}$, where our envelope approximation holds.
    The gray band shows typical white dwarf radii.
    \textbf{Right:} Mass-radius curves for a small electron mass reduction $g=0.1$ and coupling $c_{m_e} = 10^{-6}$. The blue band shows the stable branch (varying the composition between helium and iron and the central temperature between $T_0=0$ and $T_0=\SI{e8}{K}$). The gray band shows the standard model prediction. The metastable branch ends at radii much larger than shown. White dwarf data as in previous plots. }
    \label{fig:RGapAndLowCMRBoundEl}
\end{figure}

Defining $c_{m_e}= m_\phi^2 M_p^2/(|\dme| m_e^4)$, the theory just has two free parameters.
Note that with this definition, all quadratic effects are captured in $c_{m_e}$ (see \Eq{eq:boundElectronCoupling}), in particular the mass becomes tachyonic at $n_c = m_e^3 c_{m_e}$, independent of $g$.
Meanwhile, $g$, which parameterizes the fermion mass reduction at finite density, also captures the relative importance of higher-order effects, in particular $V(\theta_\ast) = m_e^4 \cdot g \cdot c_{m_e}$.
A new ground state of matter is found for small enough $c_{m_e}$, as shown in the left panel of \Fig{fig:NGSBoundaryAndMRBoundElectrons}.
As expected, the NGS ends at lower $c_{m_e}$ for smaller mass reductions $g$.
If an NGS is present, the EOS shows the same gapped behavior as in \Fig{fig:EOSMRScalarElectron}. 
This again leads to a \emph{forbidden gap} in the ($M\,$-$\,R$) curves, see \Fig{fig:NGSBoundaryAndMRBoundElectrons} (right).
Considering the gap position as a function of $c_{m_e}$ (\Fig{fig:RGapAndLowCMRBoundEl}, left), we exclude all points in the parameter space where a few white dwarfs are found within the gap.
At fixed $c_{m_e}$, lower $g$ corresponds to a lower NGS-density.
This leads to a smaller gap and means the excluded area ends earlier (see left panel of \Fig{fig:NGSBoundaryAndMRBoundElectrons}). 

Furthermore, if the mass reduction $g$ is large enough, the electrons become ultrarelativistic at low densities.
Requiring compatibility with our two benchmark stars, Sirius and Procyon B, any mass reduction of $g\gtrsim0.5$ is excluded for arbitrary low $c_{m_e}$ (see right part of \Fig{fig:NGSBoundaryAndMRBoundElectrons}, left).

We also include finite-temperature effects as discussed above, which are particularly relevant for low $c_{m_e}$ (\Fig{fig:RGapAndLowCMRBoundEl}, right).
Additionally, temperature effects become relevant for the existence of an NGS:
In particular, given the scalar minimum as a function of density $\phi(n)$, we can calculate the temperature distribution in a white dwarf as a function of density $T(n)$ in our envelope approximation.
At the NGS density $n^\ast$, the white dwarf matter has a non-zero temperature, which increases the energy per particle by $\sim T(n^\ast)$.
Consequently, thermal effects can be enough to turn the global minimum in the energy per particle (of the NGS at $T=0$) into a local one. 
This lowers the $c_{m_e}$ value of the NGS boundary compared to $T=0$  (blue line in \Fig{fig:NGSBoundaryAndMRBoundElectrons}, left).
At large $g \gtrsim 0.2$, the phase transition would occur within the isothermal core, where our approximation remains reliable.
At lower $g$, the transition extends into the envelope region, which generically backreacts onto the $T(n)$. This feedback may alter the location of the new ground state–phase transition boundary.\footnote{A fully consistent treatment would require incorporating temperature self-consistently into the stellar structure solution. As the white dwarf cools, it may evolve from the PT regime into the NGS phase. Capturing this behavior would demand a full time-dependent treatment, along with temperature-based selection of relevant observational data. 
Both effects exceed the scope of this work. 
Our approach is conservative in that these effects are expected to extend the exclusion boundaries.
}

\begin{figure}[h]
    \centering 
    \includegraphics[width=0.49\linewidth]{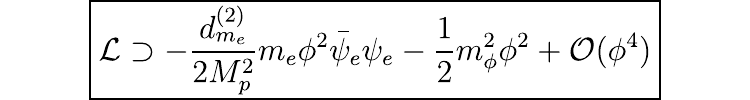} \vfill         
    \centering
    \begin{subfigure}{0.49\linewidth}
        \centering
        \includegraphics[width=\linewidth]{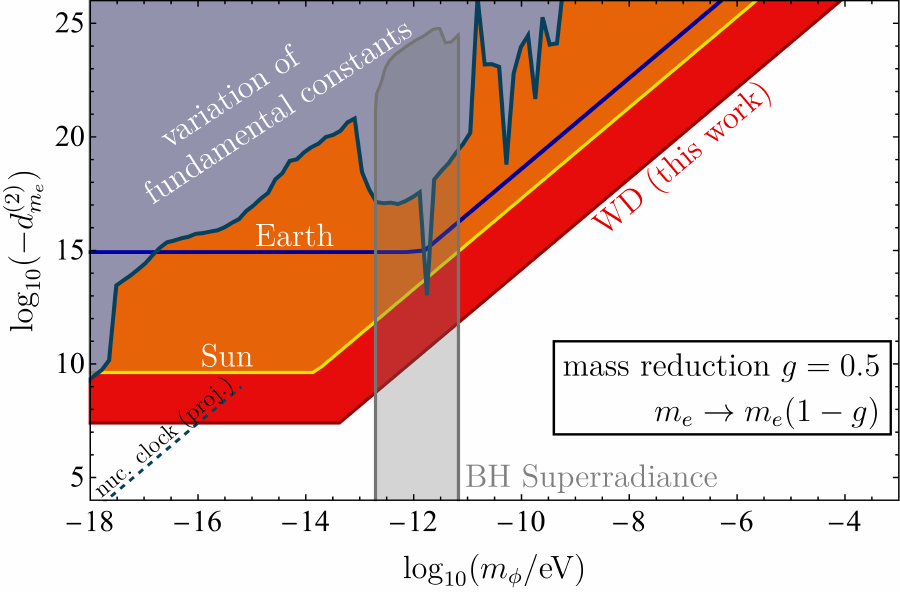}
    \end{subfigure}
    \begin{subfigure}{0.49\linewidth}
        \centering
        \includegraphics[width=\linewidth]{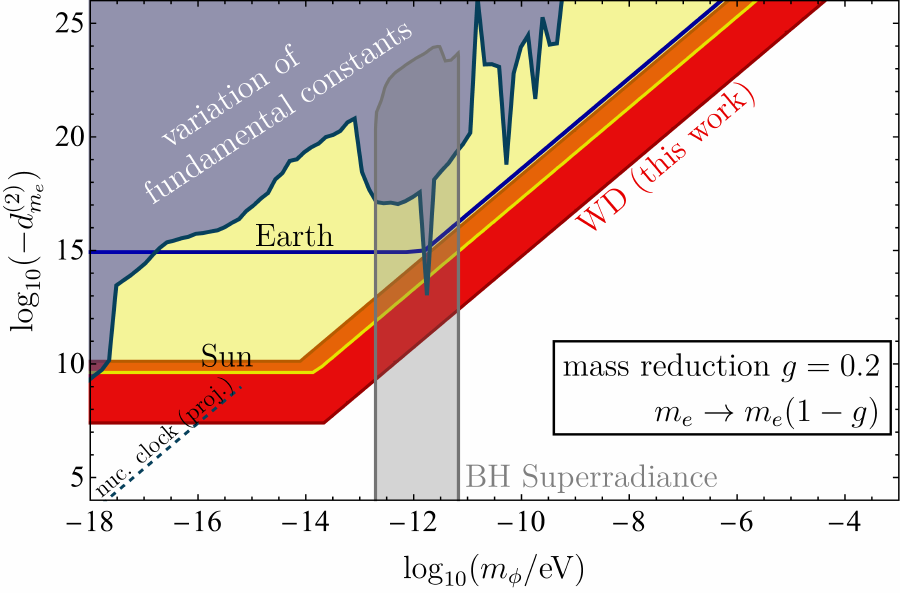}
    \end{subfigure}
    
    \begin{subfigure}{0.49\linewidth}
        \centering
        \includegraphics[width=\linewidth]{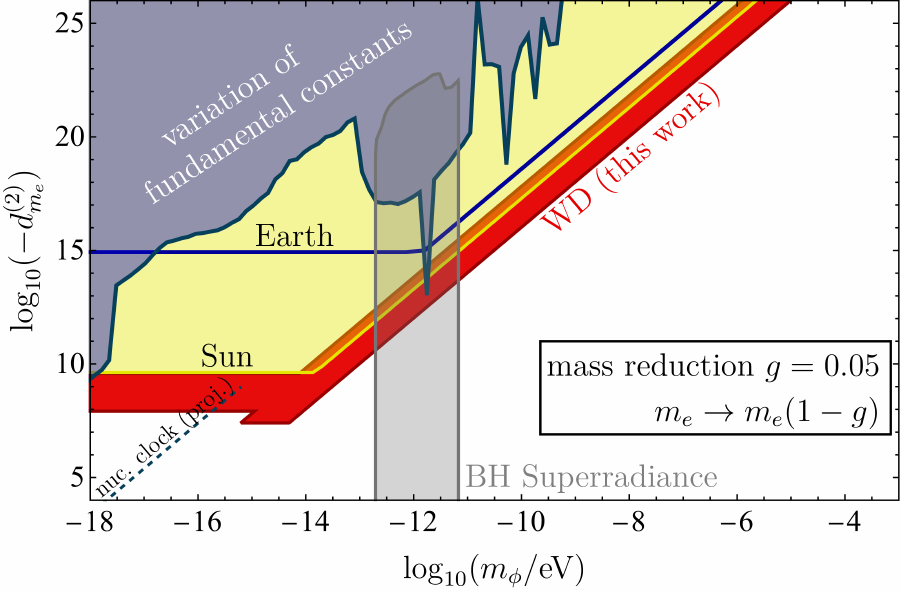}
    \end{subfigure}
    \begin{subfigure}{0.49\linewidth}
        \centering
        \includegraphics[width=\linewidth]{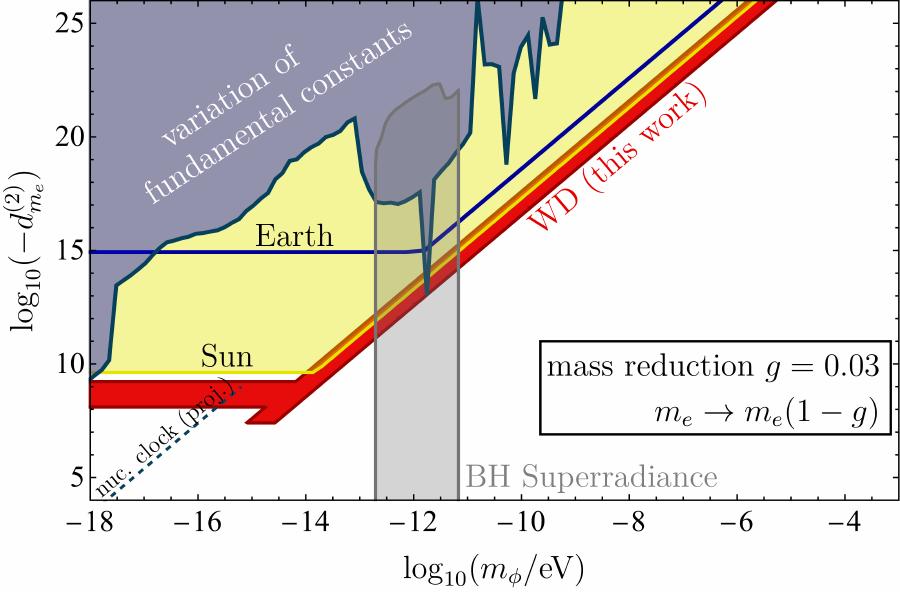}
    \end{subfigure}

    \caption{
    Limits on bounded $m_e(\theta)$ models from sourcing in white dwarfs (red). Different panels show different values of the mass reduction $g$.
    In the top left panel, the full exclusion is obtained from explicit solutions in the negligible gradient limit as discussed above.
    In the other three panels, gradient effects lead to an additional exclusion at low masses and small couplings (see \App{app:finiteGradient}), this corresponds to the horizontal component of the region excluded by white dwarfs.
    Other limits as in figure \ref{fig:exclusionElectron}, see also \ref{sec:SunEarthBounds} for a discussion of Sun and Earth bounds.
    For the superradiance limits, $\mathrm{d}^4F/\mathrm{d}\theta^4 \sim 1$ is assumed.
   }
    \label{fig:exclusionBoundElectron}
\end{figure}

Mapping these results onto the $\dme-m_\phi$ parameter space yields the excluded region shown in \Fig{fig:exclusionBoundElectron}.
We show four different panels corresponding to four values of the electron mass reduction $g$.
Again, the white dwarf constraints provide the best sensitivity as long as $\Delta m_e/m_e = g\gtrsim0.02$.
At smaller $g$, the scalar field is still sourced, but the changes in the \MR are too small to lead to constraints, which is the case for the model proposed in \cite{Delaunay:2025pho}.
Given the sourcing, other observables might be sensitive to the existence of a \phiDwarf, \eg changes in the spectra, the cooling, or gravitational waves from \phiDwarf~mergers.
This is left for future work.

Note that at small scalar field masses, finite gradient effects are known to stabilize the gap position even if the potential becomes negligibly small \cite{Balkin:2022qer}. 
This leads to an additional exclusion at small couplings and arbitrarily small scalar field masses in \Fig{fig:exclusionBoundElectron}. To quantify gradient effects, we use the analytic estimates presented in \App{app:finiteGradient}. We expect the general shape and position of these bounds to be correct, while the exact $\mathcal{O}(1)$ factors require a precise numerical study, which is beyond the scope of this work.
Also note that this gradient-dominated exclusion depends on the model-dependent value of $\theta_\ast$. In \Fig{fig:exclusionBoundElectron}, we fix $\theta_\ast = \pi$, see \App{app:finiteGradient} for the full $\theta_\ast$ dependence.

Lastly, we note that some other bounds shown in \Fig{fig:exclusionBoundElectron} also depend on the mass reduction $g$.
This is most evident for the superradiance bound \cite{Witte:2024drg}.
This is because, without fine-tuning, quartic scalar self-interactions of order $\dme m_\phi^2/g^2 M_p^2$ are expected, which can prevent the growth of the superradiant scalar cloud.
This can be seen from the upper boundary of the exclusion limit, which moves by two orders of magnitude from $g=0.5$ to $g=0.05$.
At smaller $g$ than shown, the Sun exclusion would also pick up a $g$-dependence; see the discussion in section \ref{sec:SunEarthBounds}.

\subsection{Light QCD axions} \label{sec:LightAxions}

We now focus on light QCD axions \cite{Hook:2018jle,Banerjee:2025kov}, see also \cite{DiLuzio:2021pxd}. Axions couple nonderivatively and quadratically to nucleons, see \eg \cite{Hook:2017psm,Springmann:2024mjp}. 
Recently, it has been shown in \cite{Balkin:2022qer} that within white dwarfs an NGS can appear, which leads to a \emph{forbidden gap}, similarly to the effects discussed in the preceding sections.
In the previous analysis, temperature effects have been neglected. 
We demonstrate here that the bounds are robust even when temperature corrections are included.

We consider the axion potential
\begin{equation}
    V(a)=-\epsilon m_\pi^2 f_\pi^2\left[\sqrt{1-\frac{4z}{(1+z)^2}\sin^2\left(\frac{a}{2f_a}\right)}-1\right], \label{eq:AxionPotential}
\end{equation}
where $f_a$ is the axion decay constant, $m_\pi$ and $f_\pi$ are the pion mass and the decay constant, respectively, and $z=m_u/m_d$ is the quark mass ratio.
The dimensionless parameter ${\epsilon\le 1}$ specifies how much lighter the axion is than the QCD expectation for a given $f_{a}$.
A technically natural UV-completion of light QCD axions invokes a $\mathbb{Z}_N$-symmetry to protect the QCD axion mass, see \cite{Hook:2018jle,DiLuzio:2021pxd,Banerjee:2025kov}.\footnote{The explicit form of the potential in \Eq{eq:AxionPotential} emerges when the $\mathbb{Z}_N$ is softly broken by early universe dynamics \cite{Banerjee:2025kov}, while the original proposal in \cite{Hook:2018jle} predicts a slightly different form, see also Appendix C in \cite{Balkin:2022qer}.
We further study sourcing for the $\mathbb{Z}_N$-potential in \cite{Bartnick:2025lbv}.}
Moreover, axions also couple quadratically to nucleons, as derived in \cite{Hook:2017psm,Balkin:2020dsr,Springmann:2024mjp}. This coupling is given by
\begin{equation}
\mathcal{L}\simeq-\sigma_{\pi N}\bar{N}N\left[\sqrt{1-\frac{4z}{(1+z)^2}\sin^2\left(\frac{a}{2f_a}\right)}-1\right], \label{eq:AxionNucleonCoupling}
\end{equation}
where $\sigma_{\pi N} \simeq \SI{50}{MeV}$ is the nuclear sigma term.
Expanding to second order in the axion field, it can be written in the notation of \Eq{eq:LScalarAndFermion} as $d_{m_N}^{(2)}=-z\sigma_NM_\text{Pl}^2/[(1+z)^2m_Nf_a^2]$.

For $\epsilon\lesssim 10^{-7}$, white dwarfs are dense enough to source the axion, which can result in \phiDwarfs.
For the potential considered, \Eq{eq:AxionPotential}, all $\epsilon$ that lead to sourcing in white dwarfs will also lead to an NGS.

\begin{figure}
    \centering
    \begin{subfigure}{.49\textwidth}
        \centering
        \includegraphics[width=\textwidth]{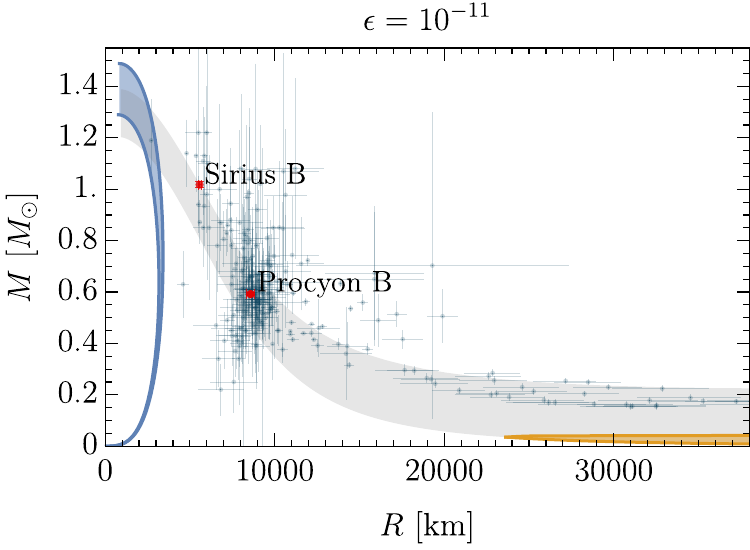}
    \end{subfigure}
     \begin{subfigure}{.49\textwidth}
        \centering
        \includegraphics[width=\textwidth]{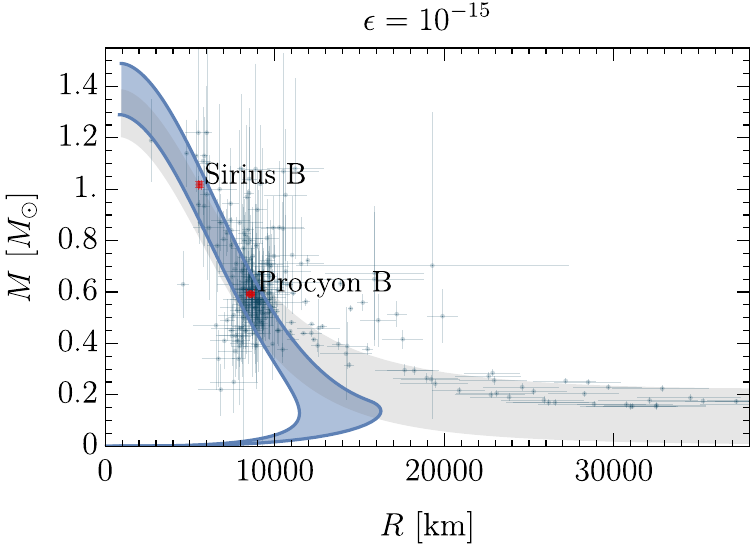}
    \end{subfigure}
    \caption{
    ($M\,$-$\,R$) curves for light QCD axions at finite temperature using the simplified envelope model described above.
    The stable branch is shown in blue, the metastable in orange, and ($M\,$-$\,R$) curves without axion as a gray band.
    The bands are each obtained by varying the central temperature from $T_0 = 0$ to $\SI{e8}{K}$ as well as the central composition from Helium $Y=2$ to iron $Y=2.15$. 
    Data as in \Fig{fig:MR_SM}.
    \textbf{Left:} If the star is in the NGS, temperature effects are not visible since the star ends at densities higher than typical envelope densities. 
    In the metastable phase, temperature effects are important. 
    As temperature generally pushes white dwarfs of the same mass to larger radii, within the range of the plot, only relatively low temperatures can be seen.
    \textbf{Right:} Same as left, but for smaller $\epsilon$. 
    In this case, within the NGS, a thin thermal envelope at the edge of the stars remains. 
    This leads to slightly larger radii in the \MR curve around $M_\text{WD}\simeq 0.2M_\odot$.
    Meanwhile, the metastable branch is beyond the range of the plot, even in the $T=0$ limit. Clearly, the very light white dwarfs still fall into the \emph{forbidden gap}.
    }
    \label{fig:MRAxionFiniteT}
\end{figure}

The inclusion of temperature effects proceeds exactly as described in previous sections for couplings to electrons.
The results for $\epsilon=10^{-11}$ and $\epsilon=10^{-15}$ are shown in \Fig{fig:MRAxionFiniteT}.
We find that including temperature, the stable \MR curves are only significantly affected for very small $\epsilon\lesssim 2 \times 10^{-14}$.
This is because, for larger values of $\epsilon$, the NGS density $n^\ast$ is so large that for any $n>n^\ast$, electron degeneracy dominates over thermal corrections to the EOS.
The bound can thus only be affected by thermal corrections for small values of $\epsilon$, at which the axion would already be sourced within the Sun, or possibly Earth, which is ruled out, see \cite{Hook:2017psm}.
We therefore conclude that the bound from white dwarf \MR changes put forward in \cite{Balkin:2022qer} is robust against temperature corrections.

\subsection{Coupling to nucleons} \label{sec:NucleonCoupling}
We now generalize the QCD axion case and consider generic light scalar fields that couple quadratically to nucleons.
Energies and densities in white dwarfs are well below the QCD scale, and we can focus on an effective coupling to nucleons without specifying the couplings to quarks or gluons.\footnote{The translation to an UV gluon coupling is discussed in \Sec{sec:DilatonicCharges}.}
We consider the simpler case where the nucleon does not become massless for any value of $\phi$. 
We comment on the unbound case, where the nucleon mass can become zero at finite density, at the end of this section.

Similarly to the case of a bound electron mass (\Eq{eq:boundElectronCoupling}), we parameterize the theory as
\begin{equation}
    \mathcal{L} \supset - g \frac{m_\phi^2 M_p^2}{|\dmn|} F\left(\theta\right) - m_N \left(1 - g F(\theta)\right) \bar{\psi}_N \psi_N, \qquad \theta = \sqrt{\frac{|\dmn|}{g}} \frac{\phi}{M_p}, \label{eq:boundNucleonCoupling}
\end{equation}
with coupling $\dmn$, maximal nucleon mass reduction $g$ and where $F$ satisfies the same properties as in \Eq{eq:boundElectronCoupling}.
We focus on the isospin symmetric case, and combine neutrons and protons into a single field $\psi_N$ with mass $m_N = \SI{939}{MeV}$.
Isospin breaking corrections are beyond the scope of this work, as changing the neutron to proton mass ratio might change the stable nuclei inside white dwarfs.\footnote{This could lead to drastic effects: If the neutron becomes lighter than the proton inside a white dwarf, the protons could decay, potentially triggering additional instabilities.}

With a suitable field redefinition, as in the case for bounded electron masses (\Sec{sec:BoundedMe}), we find that the sourcing of the scalar field is independent of the function $F$, and defining, 
\begin{equation}
    c_{m_N} = \frac{m_\phi^2M_p^2}{|\dmn| m_e^4},
\end{equation}
there are only two independent parameters, $c_{m_N}$ and $g$ in the negligible gradient limit.
We note that for the mass reduction $g$ we consider, nuclei always stay non-relativistic to a good approximation in white dwarfs.
Consequently, the critical density is given by
\begin{equation}
    n_c = m_e^3 \frac{c_{m_N}}{Y} \frac{m_e}{m_N} \label{eq:rhoCBoundNuc}
\end{equation}
at which the scalar field directly jumps to its minimum and the nucleon mass is effectively reduced to $m_N(1-g)$.
Similarly, for non-relativistic electrons, we estimate the NGS density as
\begin{equation}
    n^\ast = m_e^3 \frac{5^{3/5}}{(3\pi^2)^{2/5}} \cdot (c_{m_N} g)^{3/5}. \label{eq:nNGSBoundNuc}
\end{equation}

\begin{figure}[h]
    \centering
    \begin{subfigure}[c]{0.49\linewidth}
        \centering
        \vspace{.7em}
        \includegraphics[width=\linewidth]{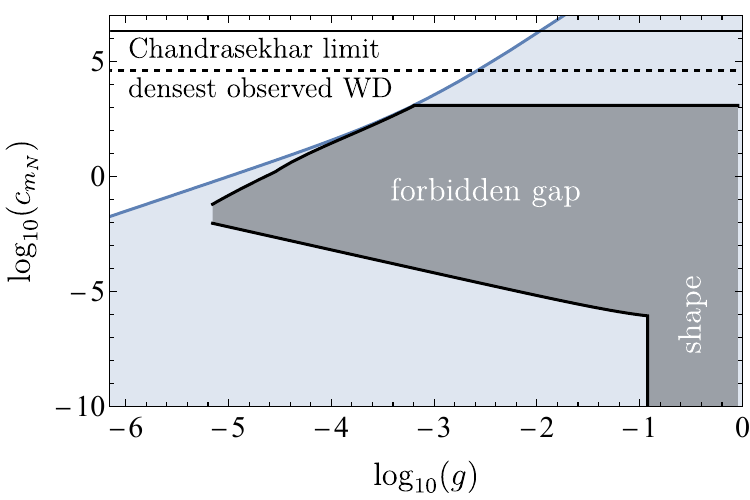}
    \end{subfigure}
    \begin{subfigure}[c]{0.49\linewidth}
        \centering
        \includegraphics[width=\linewidth]{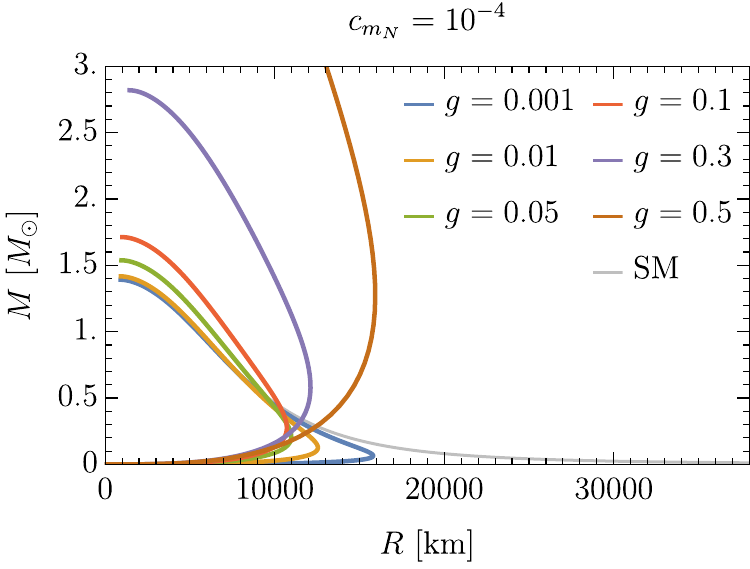}
    \end{subfigure}
    \caption{\textbf{Left:} Scalar parameter space for couplings to nucleons with finite in-medium nucleon masses. 
    Below the blue line, an NGS is found in White-Dwarf matter (we fix $Y=2$ and $T=0$). The black solid (dashed) line marks where the critical density equals the central density of a Chandrasekhar limit white dwarf (WD) (densest white dwarf in the dataset used).
    The area shaded in black is incompatible with white dwarf observation, either due to the presence of a \emph{forbidden gap} or due to the incompatible \emph{shape} as outlined below. 
    \textbf{Right:} \MR curves for different mass-reduction $g$, $c_{m_N}=10^{-4}$ is kept fixed, as are $Y=2$ and $T=0$.
    Stable branches for varying $g$ are shown in different colors. The metastable branch, which is $g$ independent, ends at radii outside of the plot range.}
    \label{fig:NGSBoundaryAndMRBoundNucleons}
\end{figure}

We show the parameter space for $c_{m_N}$ and $g$ in the left panel of \Fig{fig:NGSBoundaryAndMRBoundNucleons}.
For large values of $g\gtrsim m_e/m_N$, if the scalar is sourced within the white dwarf, it always leads to an NGS.
This is indicated by the blue shaded area in the left panel of \Fig{fig:NGSBoundaryAndMRBoundNucleons}.
Compared to typical electron degeneracy energies, reducing the nucleon mass with $g\gtrsim m_e/m_N$, provides a large energetic advantage.

In the right panel of \Fig{fig:NGSBoundaryAndMRBoundNucleons}, we show \MR curves for $c_{m_N}=10^{-4}$ and various values of $g$.
First, note that the meta-stable branch, which is independent of $g$, lies at larger radii than shown in the plot.
For the stable (NGS) branch, there are two relevant effects.
The larger $g$, the larger the potential and consequently, the larger the NGS-density $n^\ast$ (see \Eq{eq:nNGSBoundNuc}).
This, in principle, leads to a smaller maximal radius of the NGS as $g$ is increased (see \eg blue and orange curves in \Fig{fig:NGSBoundaryAndMRBoundNucleons} (right)).
Additionally, the maximum mass of a white dwarf is given by the Chandrasekhar limit \cite{Chandrasekhar:1931ftj}, 
\begin{equation}\label{eq:Chandrasekhar}
    M_\mathrm{max} \simeq 1.4 M_\odot \left(\frac{2 }{Y(1-g)}\right)^2,
\end{equation}
which implies that lighter nucleons lead to heavier white dwarfs, which is analogous to the behavior in neutron stars \cite{Balkin:2023xtr}.
In this case, the NGS-solutions deviate from the $R^3$-behavior of roughly constant density self-bound objects (see \cite{Balkin:2023xtr}) at larger radii and therefore lead to larger maximal radii overall.
This can be seen for the larger values of $g$ shown in \Fig{fig:NGSBoundaryAndMRBoundNucleons}.

\begin{figure}[h]
    \centering
    \begin{subfigure}[c]{0.49\linewidth}
        \centering
        \vspace{.7em}
        \includegraphics[width=\linewidth]{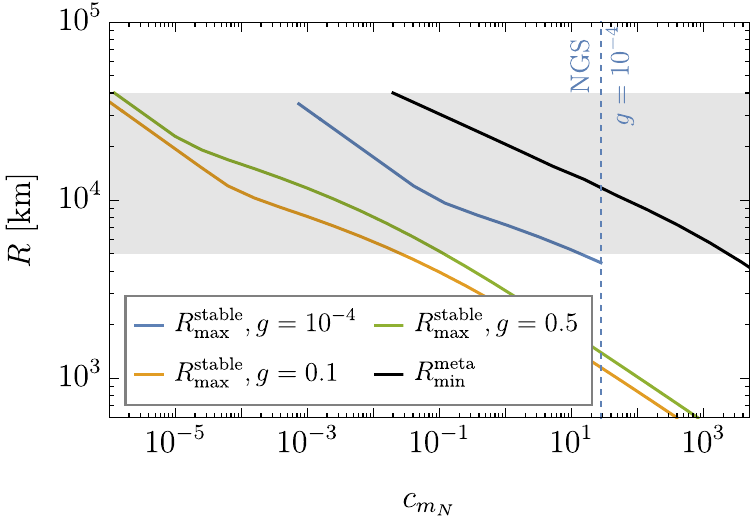}
    \end{subfigure}
    \begin{subfigure}[c]{0.49\linewidth}
        \centering
        \includegraphics[width=\linewidth]{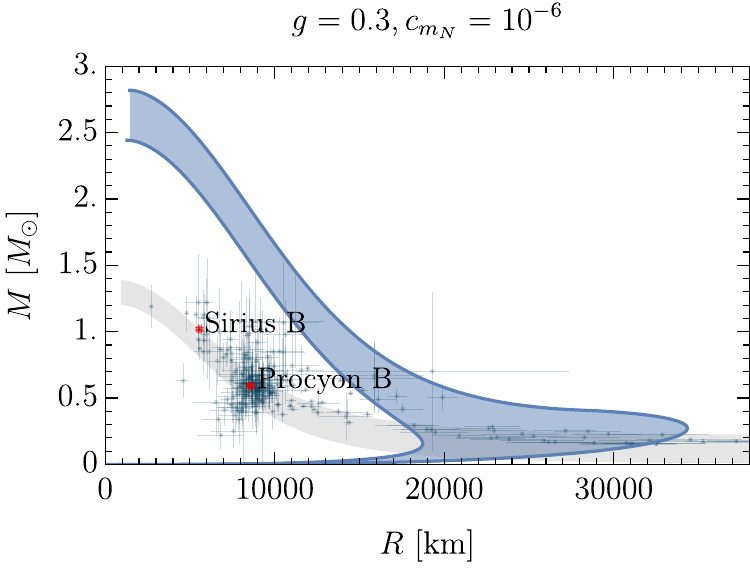}
    \end{subfigure}
    \caption{\textbf{Left:} Maximum radius of the stable branch $R_\mathrm{max}^\mathrm{stable}$ as function of $c$ for different values of $g$ and minimum radius of the metastable branch $R_\mathrm{min}^\mathrm{meta}$ ($g$ independent, shown in black). The $g=10^{-4}$ curve for $R_\mathrm{max}^\mathrm{stable}$ ends at $c_{m_N} \approx 28$, as marked by the blue dashed line, since there is no NGS for larger $c_{m_N}$ in this case. 
    At small $c_{m_N}$, we stop all curves when they reach $R\approx \SI{40 000}{km}$ since then the envelope approximation becomes problematic.
    For $g=10^{-4}$, the breakdown occurs slightly earlier at $R\approx\SI{35 000}{km}$.
    For comparison, the gray band shows typical white dwarf radii.
    \textbf{Right:} Mass-radius curve for low $c_{m_N}$ but large mass reductions $g$ (stable branch, blue, the metastable branch is not visible, bands obtained by varying temperature and composition as described above). For comparison, the SM prediction is shown in gray, and observed white dwarfs are shown as above.
    While the gap is already at rather larger radii, due to the increased maximum mass, most of the observed white dwarfs are no longer explained by the prediction with the scalar field.}
    \label{fig:RGapAndLargeGMRBoundNuc}
\end{figure}

Including the quadratic scalar also leads here to \emph{forbidden gaps}. 
We show the gap as a function of $c_{m_N}$ for various values of $g$ in \Fig{fig:RGapAndLargeGMRBoundNuc} (left).
As above, we exclude values of $c_{m_N}$ and $g$ where a few white dwarfs have been observed within the gap.
This exclusion ends for small $c_{m_N}$, since the gap falls outside of the range of observed white dwarf radii, see left panel of \Fig{fig:NGSBoundaryAndMRBoundNucleons}.
Meanwhile, if the nucleon mass reduction is large enough, the stable branch lies at considerably larger masses, as can be understood from the Chandrasekhar formula \Eq{eq:Chandrasekhar}.
Nearly all observed white dwarfs - including Sirius B and Procyon B - would then occupy the disallowed region, effectively ruling out such a coupling. 
We show an illustrative example in the right panel of \Fig{fig:RGapAndLargeGMRBoundNuc}.
This change in \emph{shape} allows us to place an exclusion for arbitrarily small values of $c_{m_N}$ assuming that $g\gtrsim 0.12$, compare \Fig{fig:NGSBoundaryAndMRBoundNucleons} (left).

We included temperature effects in the same way as above. 
In particular, we also included thermal corrections to the NGS-FOPT boundary as described in detail for the case of bounded electrons in \Sec{sec:BoundedMe}.
This is the reason why the upper boundary in $c_{m_N}$ deviates from the $T=0$ NGS line in \Fig{fig:NGSBoundaryAndMRBoundNucleons}.

\subsubsection{Comparison with precision experiments} \label{sec:DilatonicCharges}
In the following, we discuss how the white dwarf sensitivity compares to laboratory-based searches for time-dependent fundamental constants.
These experiments, however, require the additional assumption of $\phi$ constituting a large fraction of the dark matter abundance.
Quadratic scalar–nucleon couplings are typically not measured directly in these precision tests because these experiments work with large nuclei whose binding energies also shift under scalar interactions. Consequently, bounds are expressed in terms of the underlying constituent couplings.

For simplicity, we focus on the scenario where the scalar nucleon coupling is induced by a scalar-gluon coupling in the UV\footnote{The analysis proceeds analogously, if the nucleon coupling comes from a UV scalar quark coupling.}
\begin{equation}
\begin{aligned}
     \mathcal{L}_{\phi\mathrm{SM}} &= g\xi F(\theta)\left[\frac{ \beta_{g_s}}{2  g_s}G_{\mu\nu}G^{\mu\nu} + \gamma_{\hat{m}}\hat{m}\bar qq\right]\\
     &= -\dq{g}\frac{\phi^2}{2M_p^2}\left[  \frac{ \beta_{g_s}}{2  g_s}G_{\mu\nu}G^{\mu\nu} + \gamma_{\hat{m}}\hat{m}\bar qq\right]
        +\dots,
\end{aligned}
\label{eq:ScalarGluonCoupling}
\end{equation}
where $g_s$ and $\beta_{g_s}$ are strong coupling and the beta-function of the strong coupling respectively, $2\hat m=m_u+m_d$ and $\xi \simeq 1.1$, as discussed below.
The inclusion of $\beta_{g_s}$ and the quark mass anomalous dimension $\gamma_{\hat{m}}$ ensures that the scalar field coupling to the quark masses vanishes at the QCD scale \cite{Damour:2010rp}.  

For small values of $g\ll1$, $F(\theta)$ has the same functional form as in the IR Lagrangian \Eq{eq:boundNucleonCoupling}.
Expanding to quadratic order in the fields, we then obtain the quadratic Lagrangian given in \Eq{eq:ScalarGluonCoupling}.
The scalar coupling to nucleons inside a white dwarf can be written as
\begin{equation}\label{eq:Couplings}
    \dmn = \frac{Q_g^\mathrm{WD}}{A^\mathrm{WD}} \dq{g},
\end{equation}
with dilatonic charge for white dwarf matter $Q_g^\mathrm{WD}$ and average nucleon number $A^\mathrm{WD}$ depending on the interior composition of the white dwarf.
One finds \cite{Damour:2010rp,Banerjee:2022sqg}
\begin{equation}
\xi \equiv A^\mathrm{WD}/ Q_g^\mathrm{WD} \simeq 1.1,
\end{equation}
independent of the precise nucleus present.

\begin{figure}[h]
    \centering
    \includegraphics[width=0.588\linewidth]{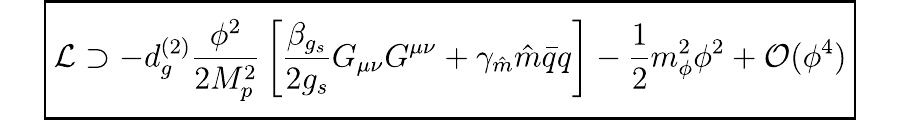} \vfill         
    \centering
    \begin{subfigure}{0.49\linewidth}
        \centering
        \includegraphics[width=\linewidth]{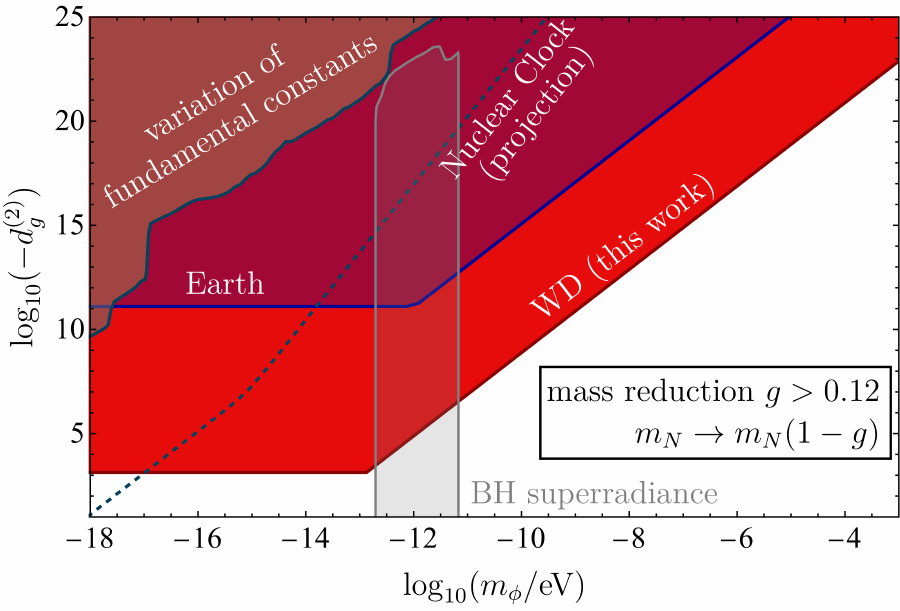}
    \end{subfigure}
    \begin{subfigure}{0.49\linewidth}
        \centering
        \includegraphics[width=\linewidth]{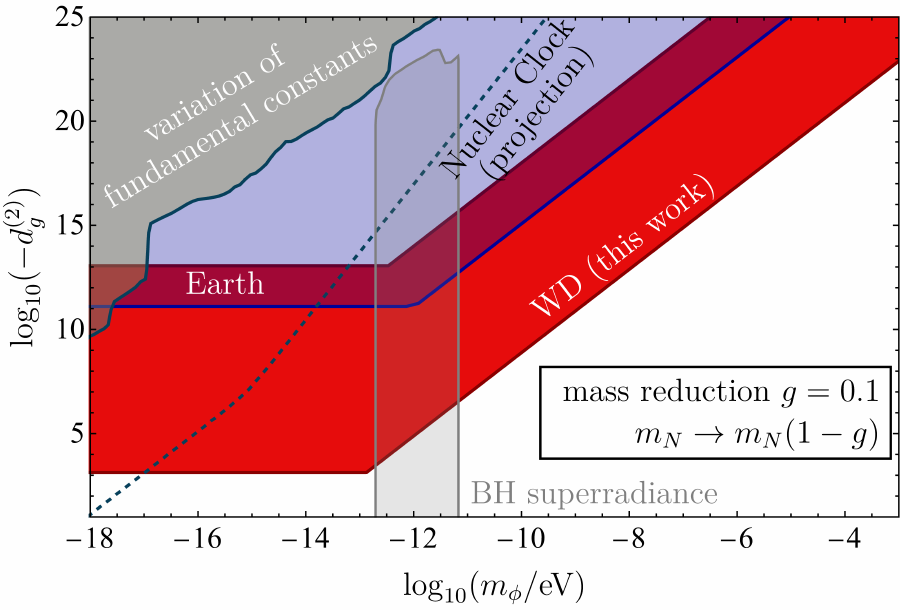}
    \end{subfigure}
    
    \begin{subfigure}{0.49\linewidth}
        \centering
        \includegraphics[width=\linewidth]{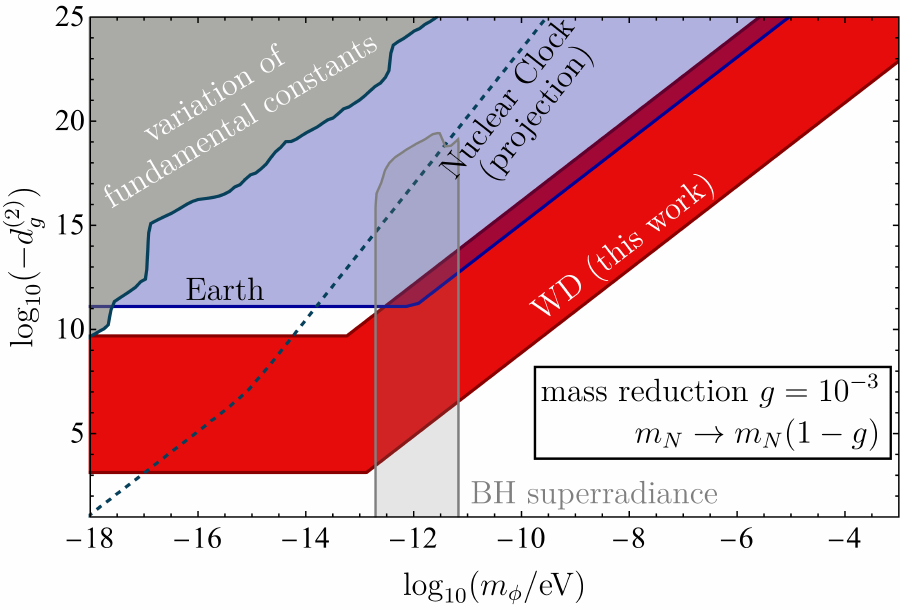}
    \end{subfigure}
    \begin{subfigure}{0.49\linewidth}
        \centering
        \includegraphics[width=\linewidth]{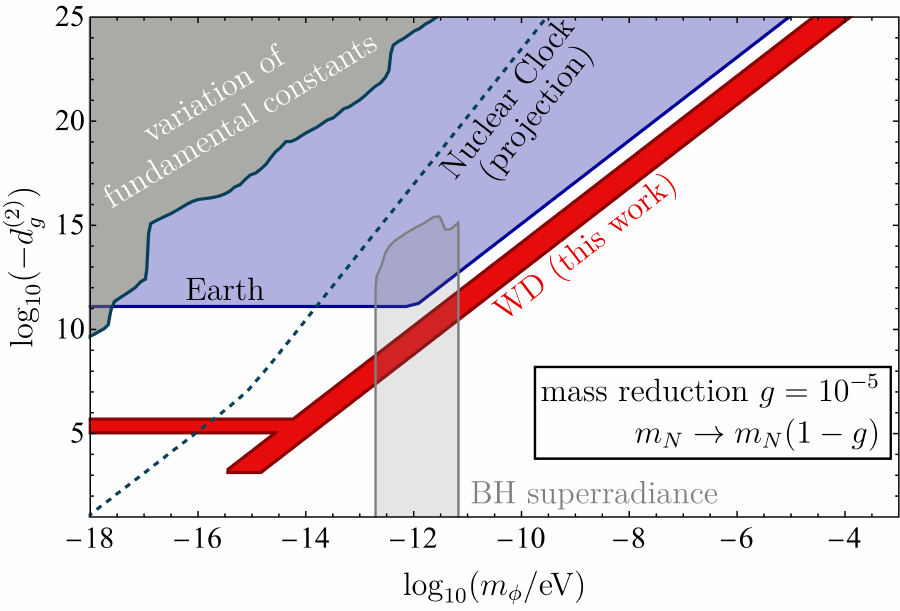}
    \end{subfigure}

    \caption{
    Constraint on quadratic scalar gluon coupling from sourcing in white dwarfs, combining limits from negligible gradient solutions and negligible potential estimates.
    Different panels show different values of the maximal mass reduction $g$.
    As discussed in \Sec{sec:SunEarthBounds}, if the scalar field is sourced in the Earth, it is also excluded, this is marked in blue.
    In the gray shaded region, black hole superradiance leads to constraints \cite{Witte:2024drg,Hoof:2024quk} (assuming $\mathrm{d}^4f/\mathrm{d}\theta^4 \sim \mathcal{O}(1)$).
    If the scalar constitutes a significant portion of dark matter constraints from atomic clocks and other experiments looking for a time variation of fundamental constants \cite{Fuchs:2024edo,Hees:2016gop,Campbell:2020fvq,Kobayashi:2022vsf,Zhang:2022ewz,Sherrill:2023zah,Banerjee:2023bjc,Oswald:2021vtc}, apply.
    A fully operational nuclear clock is expected to achieve the sensitivity marked by the blue dashed line \cite{Fuchs:2024edo}.
    }
    \label{fig:exclusionBoundNucDmG}
\end{figure}
In \Fig{fig:exclusionBoundNucDmG} we show our new constraint in comparison with atomic clock and interferometry bounds \cite{Fuchs:2024edo,Hees:2016gop,Campbell:2020fvq,Kobayashi:2022vsf,Zhang:2022ewz,Sherrill:2023zah,Banerjee:2023bjc,Oswald:2021vtc} and projected nuclear clock sensitivity \cite{Fuchs:2024edo} for different values of the maximal mass reduction $g$. 
Note that in addition to the negligible gradient limit, similar to the case of an electron-coupling, finite gradient effects can stabilize the mass gap at low masses and couplings.
To quantify this effect, we again use the analytic estimates derived in \App{app:finiteGradient}.
Assuming $\theta_\ast = \pi$ leads to the horizontal component of the white dwarf exclusion in the top right and both bottom panels of \Fig{fig:exclusionBoundNucDmG}, the full $\theta_\ast$ dependence is shown in \App{app:finiteGradient}.

For the white dwarf to be sensitive to scalar sourcing, we need a minimal amount of change in the nucleon mass $\Delta m_N / m_N = g \gtrsim \num{7e-6}$.
At larger values of $g$, the constraints from sourcing in white dwarfs are much stronger than the results from precision experiments on Earth.
At lower $g$ the white dwarf constraints become weaker and disappear at $g \lesssim \num{7e-6}$.
In this region of parameter space, the clock experiments, which are only sensitive to the small field excursion of dark matter $\phi_\mathrm{DM} \sim \sqrt{\rho_\mathrm{DM}}/m_\phi$ can start to dominate again.

We briefly comment on the case of an unbounded nucleon mass, \ie, the case where sourcing would bring the nucleon mass arbitrarily close to zero.
The mass of the nucleons is mostly due to gluons or the confinement of QCD.
Consequently, making nucleons massless (or light compared to electrons) in white dwarfs requires pushing the confinement scale $\Lambda_\mathrm{QCD}$ well below any typical white dwarf scale.
Therefore, physics in white dwarfs would be as complicated as the interior of a neutron star once the scalar field is sourced.
All statements concerning the metastable branch would still hold, but the phenomenology of the stable branch would be very different. 
This interesting scenario is beyond the scope of this work.

\section{Constraints from other astrophysical objects}
\label{sec:SunEarthBounds}

Beyond the \phiDwarf\space\MR relationship, there exist further astrophysical signatures to probe the sourcing of scalar fields, like the \phiDwarf\space and solar spectrum, stellar cooling, or exotic effects on Earth. 

The electromagnetic spectrum of \phiDwarfs~provides an additional probe.
In particular, in the NGS case, where the scalar field only transitions to $\phi=0$ at the edge of the star, the electron (nucleon) mass in the \phiDwarf~atmosphere will be lighter and can shift spectral lines, which can be in tension with observations.
This is particularly relevant if the radius gap happens to be at larger radii than where white dwarfs are typically observed, potentially extending the lower bound in $c_{m_e}$ ($c_{m_N}$), see 
\Fig{fig:NGSBoundaryAndMRBoundElectrons} (\Fig{fig:NGSBoundaryAndMRBoundNucleons}).
Similar limits also apply in the FOPT case if large field values leak out of the star due to gradient effects.
Furthermore, depending on the density of the NGS $n^\ast$ and the gradient scale $\lambda_\phi$, the typical low-density atmosphere responsible for the absorption lines might not be present at all, further changing the spectrum. 
Understanding the sensitivity of measured white dwarf spectra on electron- and nucleon-mass changes requires additional astrophysical input and is beyond the scope of this work.
See also \cite{Bai:2025yxm} for superradiant axion-induced spectral changes for stars around black holes.

Neutron stars are another excellent laboratory for light scalar fields \cite{Hook:2017psm,Balkin:2020dsr}.
Due to their high central density, they can, in principle, probe parts of the parameter space where sourcing does not occur in white dwarfs, see \cite{Balkin:2023xtr}.
Further anomalously fast cooling can also be used to constrain scalar-induced new ground states in neutron stars, see
\cite{Gomez-Banon:2024oux}.
Additional observables of scalar sourcing in neutron stars are discussed \eg in \cite{Kumamoto:2024wjd,Zhang:2021mks,Hook:2017psm,Draft2025}.

The Sun and Earth are complementary probes of scalar sourcing; see \cite{Hook:2017psm}.
They are less dense than white dwarfs and therefore probe lower values of the mass-to-coupling ratio $c_{m_e}$ (or $c_{m_N}$). However, for small values of the mass reduction $g$, they become competitive due to the availability of precision observables.

If the scalar field is sourced in the Sun but not on Earth, 
the spectral lines of the Sun would be shifted due to the change in fermion masses.
Spectral lines are produced in the outermost layers of the Sun, and thus are particularly affected if the scalar field leads to an NGS.
In the case of a phase transition, where the scalar transition happens in the interior of the Sun, the spectrum is only affected for a small range of couplings, where the gradient scale is large enough that the scalar profile significantly leaks out of the star and thereby changes fermion masses in the atmosphere.

In the case of a purely quadratic coupling to electrons, any choice of parameters that leads to sourcing in the Sun also leads to an NGS. Electrons in this case become (nearly) massless, which results in an extreme change of solar spectra, which is clearly excluded.
If $m_e(\theta)$ is bounded, the minimal value of the mass-reduction-parameter $g$ that can still be excluded directly depends on the accuracy of solar spectrum measurements.
For $g \gtrsim \mathcal{O}(10^{-3})$, any sourcing in the Sun will be an NGS and can be probed by the spectrum.
For $g$ smaller, the exclusion becomes $g$ dependent and tracks the NGS boundary \Fig{fig:NGSBoundaryAndMRBoundElectrons}.
For parameters leading to a FOPT, a bound exists for large enough leakage due to gradient effects, corresponding to couplings of order $\dme\sim 10^9$.
The sensitivity to nucleon couplings is reduced by a factor of $m_e/m_N$, which is the reason why we do not show the Sun bound in Fig.~\ref{fig:exclusionBoundNucDmG}.
A detailed study of the solar spectrum in regards of sourcing is left for future work.

Although the Earth is less dense, the terrestrial environment can still yield competitive limits - especially at low $g$.
We again consider the case of an NGS (or any scenario with a significantly long-range gradient tail).
The induced reduction in fermion masses contributes an additional binding energy.
For sufficiently small test bodies, the field gradient prevents appreciable sourcing inside the object. In this regime, the associated binding energy is released as the object approaches Earth and must be supplied as it departs. 
The figure of merit is the (scalar induced) binding energy per nucleon, defined as
\begin{equation}
    \frac{E_\mathrm{bind}^\phi}{N} \sim \left\{
    \begin{matrix}
        g m_N,& \text{coupling to nucleons}\\
        g m_e, & \text{coupling to electrons}
    \end{matrix}
    \right.,
\end{equation}
to be compared with the gravitational binding energy on the surface of Earth
\begin{equation}
    \frac{E_\mathrm{bind}^\mathrm{grav}}{N} \sim \frac{G M_\mathrm{Earth} m_N}{R_\mathrm{Earth}}.
\end{equation}
For gradient scales of the order of meters and larger, $\lambda_\phi \gtrsim \text{m}$, a naive conservative bound excludes the parameter space where these two are comparable, for example, from rocket launches.
This yields a bound from sourcing in Earth for
\begin{align*}
    g \gtrsim \num{e-6}& \qquad \text{for} \qquad \num{e15} \lesssim \dme \lesssim \num{e26} \qquad \text{(coupling to electrons)}\\
    g \gtrsim \num{e-9}&\qquad \text{for} \qquad \num{e11} \lesssim \dmn \lesssim \num{e23} \qquad \text{(coupling to nucleons)}.
\end{align*}
For shorter gradient scales, \ie larger couplings $\dme$ or $\dmn$, the scalar field is sourced inside the body of the rocket, and rocket launches cease to be a clean probe.
Other terrestrial tests, \eg objects falling onto Earth or behavior of matter inside vacuum chambers,\footnote{For example, objects of the same material but different sizes or shapes could accelerate at different rates, as sourcing happens differently within them. Also, the behavior close to the boundary of the vacuum chamber can be significantly modified.} should remain sensitive.
Even lower values of the mass-reduction $g$ can be tested with precision experiments, \eg EP-tests and fifth-force searches can see some effects due to scalar sourcing within their test masses.
This is only relevant at very low values of $g$, and the specific exclusion due to scalar sourcing depends on $\dme$, $\dmn,$ and $m_\phi$, as well as on the geometry and setup of the experiment.
A detailed study of these interesting bounds is beyond the scope of our work.

\section{Conclusion}\label{sec:Conclusion}

In this paper, we studied how quadratically coupled scalar fields can turn white dwarfs into \phiDwarfs, which affects the white dwarf mass-radius relation.
As first discussed in \cite{Balkin:2022qer,Balkin:2023xtr}, if the sign of the coupling is such that the fermion mass gets reduced, sourcing can lead to a new ground state of matter.
We showed that this occurs for a wide range of theories, namely, scalar fields coupling to electrons or to nucleons.
Relative to laboratory searches, which additionally require the scalar to constitute the dark matter, the stellar bounds derived here are typically much stronger over large regions of parameter space.

The existence of such a new ground state leads to a rich, testable phenomenology for white dwarfs, in particular, observable in their mass-radius relation.
At comparable high scalar mass to coupling ratios, the new ground state induced splitting of the equation of state into a stable and a metastable branch leads to a \emph{forbidden gap} in the ($M\,$-$\,R$) curve, as first observed in \cite{Balkin:2022qer} for the case of light QCD axions, see Fig.~\ref{fig:MRScalarElectron} (left).

In this work, we found two additional changes in the \emph{shape} of ($M\,$-$\,R$) curve, depending on the scalar field fermion coupling: 
\begin{itemize}
    \item If electrons are considerably lighter in the new ground state, they quickly become ultrarelativistic, leading to white dwarf masses close to the Chandrasekhar limit for a large range of radii, see Fig.~\ref{fig:MRScalarElectron} (right).
\item If the nucleon mass is changed by the scalar field by an $\mathcal{O}(1)$ amount, the maximal white dwarf mass changes significantly, see Fig.~\ref{fig:NGSBoundaryAndMRBoundNucleons} (right).
\end{itemize}

Confronting these predictions with observational data on the white dwarf mass-radius relationship, which became available in recent years, allowed us to place constraints.
In particular, using the radii of the global white dwarf population, we excluded all scalar fields that lead to an \emph{forbidden gap} at radii where white dwarfs have been observed.
To bound \emph{shape} distortions, we focused on the two nearest white dwarfs with precisely measured properties - Sirius B and Procyon B -  and excluded parameter regions whose ($M\,$-$\,R$) curves fail to fit both stars simultaneously within their uncertainties.

Our ($M\,$-$\,R$) calculations incorporate stellar composition variations and finite temperature effects. Temperature is essential to account for observed low-mass white dwarfs in the absence of a scalar. 

Extending the bounds to higher scalar masses follows two paths. 
If the new ground state persists to high values of coupling to mass ratio, continuing the exclusion to higher mass requires higher central densities.
While in principle, observation of more massive white dwarfs would lead to a slight improvement of the bounds presented here, a significant increase would require studying denser objects, namely, neutron stars.
Given the substantial theoretical uncertainty in neutron-star equations of state, we leave this discussion for future work.

If, in contrast, the new ground state ends at mass-to-coupling ratios that still allow for sourcing in white dwarfs, a first-order phase transition region also becomes accessible.
This is, for example, the case for a quadratic coupling to electrons.
If there is a scalar-induced phase transition in white dwarfs, their structure does change as well, but in general, the changes will be less prominent.
Consequently, a more precise astrophysical modeling and understanding of the underlying thermal structure, as well as the data on the mass-radius relation, is necessary, which will be discussed in an upcoming publication \cite{Bartnick:2025lbv}.

We conclude by stressing the complementarity of stellar and precision laboratory experiments.
If the coupling does not lead to a significant mass reduction, laboratory experiments provide the best sensitivity, assuming the scalar constitutes a significant fraction of the dark matter. 
Elsewhere in parameter space, stellar constraints from \emph{shape} distortions or \emph{forbidden gaps} offer unique coverage, probing regions that are otherwise inaccessible.

\acknowledgments

We thank Reuven Balkin for discussions during early stages of the work and Detlev Koester and Rolf-Peter Kudritzki for helpful discussions, patience, and collaboration on related work.
KB and AW are supported by the Collaborative Research Center SFB1258 and by the Excellence Cluster ORIGINS, which is funded by the Deutsche Forschungsgemeinschaft (DFG) under Germany's Excellence Strategy – EXC-2094 – 390783311.
KS is supported by a research grant from Mr. and Mrs. George Zbeda and by the Minerva Foundation.
SS is supported by the Swiss National Science Foundation under contract 200020-213104 and acknowledges the hospitality of the CERN theory group. This research was additionally supported by the Munich Institute for Astro-, Particle and BioPhysics (MIAPbP) funded by the DFG.
KB would like to thank the Particle Theory group at Cornell University for their hospitality during the final stages of this work.

\appendix

\section{Details about finite temperature}\label{App:finiteT}
In this appendix, we will discuss our treatment of finite temperature effects and, in particular, the derivation of equation \eqref{eq:EOSfiniteT} in some more detail.

We follow chapter 37 of \cite{1990sse..book.....K}, key results that lead to the finite temperature EOS will be summarized below.
In this approximation, the white dwarf is split into two regions: an interior core and an outer envelope.
In the interior core of the white dwarf, due to electron degeneracy, the thermal conductivity is so high that the temperature $T_0$ is nearly constant.
In the envelope, outside the central core, this approximation no longer holds.
The matter is still very dense and thus opaque. 
In particular, the photon mean free path $l_p$ is much smaller than any other length-scale, and transport of radiation energy can be described as diffusion. 
The diffuse flux $j$ of some density $n$ is thus given as
\begin{equation}
    \mathbf{j} = - D \bm{\nabla} n,
\end{equation}
with diffusion coefficient $D$.
This means that the flux of radiation $\mathbf{F}$ is proportional to the gradient of the radiation energy density $U$, with a diffusion coefficient $D = \frac{1}{3} l_p$, since photons move at the speed of light.
Due to spherical symmetry, both the gradient and the flux are purely radial, and we find
\begin{equation}
    F_r = - \frac{1}{3} l_p \frac{\partial}{\partial r} U = - \frac{16 \sigma}{3} \frac{T^3}{\kappa_\omega \varepsilon} \dv{T}{r}. \label{eq:radiationDiffusion}
\end{equation}
For the second equality, we defined $l_p = \frac{1}{\kappa_\omega \varepsilon}$ with frequency averaged opacity $\kappa_\omega$ and used $U = 4 \sigma T^4$ where $\sigma = \frac{\pi^2}{60}$ is the Stefan-Boltzmann constant.
The flux is related to the Luminosity $l = 4 \pi r^2 F_r$, the total radiated power at radius $r$. Combining \Eq{eq:radiationDiffusion} with the equation of hydrostatic equilibrium \Eq{eq:TOV_p} and neglecting corrections from General Relativity, we obtain the equation of radiative transport
\begin{equation}
     \pdv{T}{p} = \frac{3}{64 \pi \sigma G} \frac{\kappa_\omega l}{T^3 m}. \label{eq:envelopeStructure}
\end{equation}
To obtain a stellar solution in full generality, this equation, in combination with \Eq{eq:TOV} and an energy balance equation, has to be solved as a coupled system.
If the envelope is thin, the situation simplifies.
In this case, the mass is dominantly in the core, so $m = \text{const.}$ in the envelope, and it is given by $m=M$, the total white dwarf mass.
Similarly, there are no sources or sinks of energy in the envelope, so we can also approximate $l = L$, the total white dwarf luminosity.
Lastly, for the opacity, we assume a power law, in particular Kramers law
\begin{equation}
    \kappa_\omega(p,T) = \kappa_0 \cdot p \cdot T^{-\frac{9}{2}},\label{eq:kramerslaw}
\end{equation}
for more details, see \cite{Kramers:1923, Cox_Giuli_StellarStructure}.
Now, it is easy to integrate \Eq{eq:envelopeStructure}, which yields
\begin{equation}
    T^{-\frac{17}{2}}= \underbrace{\frac{17}{4} \frac{3}{64 \pi} \frac{\kappa_0}{\sigma G} \frac{L}{M}}_{=B} \left(p^{2}+C\right). \label{eq:integratedEnvelopeStructure}
\end{equation}
where $C$ is an integration constant, which we set to zero \cite{1990sse..book.....K}. 
The pressure in the envelope can be approximated as an ideal gas of electrons and fully ionized nuclei (since temperature dominates over degeneracy), and thus is given by
\begin{equation}
    p = p_e + p_\mathrm{ion} = \left(n_e + n_\mathrm{ion}\right) T = n \frac{Z+1}{Z} T = \frac{\varepsilon}{Y m_n}\frac{Z+1}{Z} T,\label{eq:pTEnvelope}
\end{equation}
where again $Z$ is the average number of protons per ion. 

To find the transition point from the envelope to the core and to fix envelope quantities in terms of core properties, we require thermodynamic quantities to be continuous.
In particular, we define the envelope-core-boundary at the density $n_b$ and pressure $p_b$ where the electron degeneracy pressure is equal to the ideal gas pressure at the central temperature $T_0$. 
After this point, electron degeneracy dominates over temperature effects, and we can safely continue with the 0-temperature EOS in the interior of the white dwarf.\footnote{
Using a finite but constant temperature EOS in the interior tends to shift the transition point to the envelope and thus, in principle, can have a noticeable effect on the mass-radius curves at high central temperatures. 
This effect is largely degenerate with a shift in central temperature.
Moreover, the same behavior is observed for the actually measurable effective temperatures: Using a finite temperature interior EOS yields higher effective temperatures at a given central temperature. However, this effect is compensated by the same shift in central temperature as for the mass-radius curves, leaving the relationship between observables largely unaffected.
}
Assuming non-relativistic electrons at the boundary between core and envelope (the error of this approximation is at most a few percent for the temperatures we consider and thus has no significant influence on mass and radius), we find (also compare \Eq{eq:EOSfinteTMatchingEnergy})
\begin{equation}
    n_b = \frac{\varepsilon_b}{Y m_N} = \frac{5^{3/2}}{3 \pi^2} \left(\frac{Z+1}{Z}\right)^{3/2} \left(m_e T_0\right)^{3/2}
\end{equation}
Secondly, we also require the temperature profile to be continuous.
In particular, we need to find $T=T_0$ for $p=p_b$ in \Eq{eq:integratedEnvelopeStructure}, which fixes the prefactor $B$. 
Equivalently, the central temperature fixes the luminosity $L$ for a given $\kappa_0$.
\begin{equation}
    B = \left(\frac{Z}{Z+1}\right)^5 \frac{\left(3 \pi^2\right)^2}{\left(5 m_e\right)^3} T_0^{7/2}.
\end{equation}
Using this together with \Eq{eq:integratedEnvelopeStructure} in \Eq{eq:pTEnvelope}, we find the EOS in the envelope, so for $\varepsilon<\varepsilon_b$,

\begin{equation}
 p(\varepsilon,T_0)=  \frac{\varepsilon}{ Y_e m_N} \left( \frac{3 \pi^2}{ 5^{3/2} }\left(\frac{Z + 1}{Z}\right)^{7/4} \frac{ \varepsilon}{ Y_e m_N }  \frac{T_0^{7/4}}{{m_e}^{3/2}}\right)^{4/13}.
\end{equation}

Scalar sourcing can also be included in this EOS.
In particular, the thermal corrections to the energy density scale as $n T$ and are thus independent of the scalar mass.
Consequently, we can solve for $\theta(n)$ first and use this to (numerically) find $n_b$ and $B$ from the continuity of $p$ and $T$ as described above, while taking sourcing into account.

\section{Analytic estimates for the radius gap} \label{App:RadiusGap}
As described in the main text, in the negligible gradient limit, particularly if the scalar field leads to an NGS, a \emph{forbidden gap} emerges in the \phiDwarf~\MR curves.
As outlined above, the gap position can be found by numerically solving the scalar EOM and the TOV equations.
As first presented in \cite{Balkin:2022qer} for the case of lighter QCD axions coupling to nucleons, the forbidden radius range for any given scalar mass and coupling (as long as it leads to an NGS) can be found analytically up to a prefactor which is scalar mass and coupling independent.
In this appendix, we perform a similar analysis for the scalar with quadratic coupling to electrons discussed in \Sec{sec:QuadCouplingElectrons} and, in particular, derive \Eq{eq:RgapAnalyticEstimates}.
The generalization to the other couplings discussed above is straightforward and, in particular, closely follows the axion case discussed in \cite{Balkin:2022qer}.

The critical density, which sets the end of the metastable branch, is given by
\begin{equation}
    n_c = m_e^3 c_{m_e} \label{eq:nCAnalytic}
\end{equation}
as discussed in the main text.
Here we use a non-relativistic limit, which is a good approximation for values of $c_{m_e}$ leading to an NGS.
The NGS-density $n^\ast$, controlling the stable branch, is defined at vanishing total pressure $p(n^\ast) = p_e -V = 0$, \ie
\begin{equation}
    p_e\left(n^\ast,m_e\left(\theta\left(n^\ast\right)\right)\right) = V\left(\theta\left(n^\ast\right)\right).
\end{equation}
In the unbounded mass cases, $\theta(n^\ast) < 1$ typically, and $\theta(n^\ast)$ has to also be solved for self-consistently, see also \cite{Balkin:2023xtr}. 
We pick a recursive approach.
In the NGS, the electrons are nearly massless, \ie, we can use an ultrarelativistic approximation for the electron pressure, which yields
\begin{equation}
    n^\ast = m_e^3 \left(\frac{1}{3 \pi^2}\right)^{1/4} \left(4 c_{m_e} \theta(n^\ast)^2\right)^{3/4}. \label{eq:rhoNGSRecursive}
\end{equation}
As we approach the asymptotic value $\theta \to \theta_\infty = 1$, the scalar density approaches
\begin{equation*}
    n_s \to n_{s,\infty} = \abs{\frac{\partial V(\theta)/\partial\theta}{\partial m(\theta) /\partial\theta}}_{\theta = \theta_\infty} = m_e^3 c_{m_e}.
\end{equation*}
Using the ultrarelativistic approximation we have $n_s = \frac{1}{2} \left(\frac{3}{\pi}\right)^{2/3} m_e(\theta) n^{2/3}$, which gives an expression for $\theta(n)$.
Solving \Eq{eq:rhoNGSRecursive} recursively to second order, we find
\begin{equation}
    n^\ast \approx m_e^3 \frac{2}{3^{1/4}} \sqrt{\frac{2}{\pi }} \left(c_{m_e} \left(1-\pi \sqrt{\frac{c_{m_e}}{3}}\right)\right)^{3/4}, 
    \label{eq:rhoNGSAnalytic}
\end{equation}
which we checked to agree well with numeric results.

Given the analytic expressions for $n_c$ and $n^\ast$, \Eqs{eq:nCAnalytic}{eq:rhoNGSAnalytic}, we can now estimate the position of the gap.
In particular, just two points of the \MR curve are needed: The minimal radius on the metastable branch and the maximal radius on the stable branch.
Both cases can be understood as a pressure balance between the outward pressure of electron gas and scalar field and the inward pressure of gravity,
\begin{equation}
    p = p_e - V = p_\mathrm{grav}.
\end{equation}
Neglecting, $\mathcal{O}(1)$ prefactors, we have
\begin{equation}
	p_\mathrm{grav} \sim G n^2 R^2 m_\mathrm{N}^2 \label{eq:pGrav}
\end{equation}
On the metastable branch, the outward pressure is given by the degeneracy pressure of the electrons.
The minimal radius is achieved for maximal central density, \ie for $n = n_c$.
In the non-relativistic limit where $p = p_e \sim n^{5/3}/m_e$, we find
\begin{equation}
	R_\mathrm{min}^\mathrm{meta} \approx 1.124 \cdot \frac{1}{\sqrt{G} m_n m_e} \left(\frac{m_e^3} {n_c}\right)^{1/6},
\end{equation}
Here, the prefactor was set by comparison with numeric solutions and is independent of $n_c$.

On the stable branch, the pressure counteracting gravity is the sum of matter pressure $p_e$ and the negative scalar potential $-V$.
For $n \sim n^\ast$ one has $p_e - V \ll p_e \sim V$. As the density increases, this near cancellation quickly gets lifted, leading to stars at approximately constant density with a cubic mass-radius relation, $M \propto R^3$.
Once the cancellation is spoiled, $p_e > V$, we can neglect the potential and one finds the typical behavior of ($M\,$-$\,R$) curves, where increasing the mass reduces the radius. 
Consequently, the maximal radius is reached at the transition density $n_\mathrm{trans}$ between these two behaviors, which happens if
\begin{equation}
    (p_e - V)(n_\mathrm{trans}) \sim p_e ( n^\ast) \sim (n^\ast)^{4/3}. \label{eq:pAtRMaxStable}
\end{equation}
The last relation assumes an ultra-relativistic limit, which is valid for the nearly massless electrons in the NGS.
Including the potential at most gives an $\mathcal{O}(1)$ modification.
Equating \eqref{eq:pAtRMaxStable} with \eqref{eq:pGrav} and and fixing the overall, $n^\ast$ independent, prefactor with a numeric calculation, one finds
\begin{equation}
	R_\mathrm{max}^\mathrm{stable} = 0.291 \frac{1}{\sqrt{G}m_N m_e} \left(\frac{m_e^3}{n^\ast}\right)^{1/3}.
\end{equation}
Here, $n^\ast$ is given by \Eq{eq:rhoNGSAnalytic}.

\section{Gradient effects} \label{app:finiteGradient}
Going beyond the negligible gradient limit, the scalar-field gradient energy can stabilize white dwarf solutions and lead to a \emph{forbidden gap} at typical white dwarf radii, even if the potential contribution is negligibly small, as first discussed in \cite{Balkin:2022qer}.
In principle, a full numerical solution of the coupled scalar field -- TOV equations would be needed to precisely include this effect.
Here, we instead focus on some simple analytic estimates, which allow us to again estimate the position of the \emph{forbidden gap} and obtain $\mathcal{O}(1)$ estimates for the exclusion limits on the scalar field due to gradient effects.

Following \cite{Balkin:2022qer}, as in \App{App:RadiusGap}, we model the white dwarf by a simple pressure balance. 
Here, it has to be modified to also include the gradient pressure $\Delta p_\mathrm{grad}$ that the scalar wall exerts inwards and reads
\begin{equation}
    p = p_\mathrm{grav} + \Delta p_\mathrm{grad}. \label{eq:pressureBalanceGradient}
\end{equation}
The radial derivative, $\mathrm{d} \Delta p_\mathrm{grad} / \mathrm{d}r$ is, by definition, given by the scalar gradient contributions to the TOV equation for $p'$ \cite{Balkin:2023xtr}.
Neglecting general relativistic effect and integrating over the scalar wall, we have,
\begin{equation}
    \Delta p_\mathrm{grad} = - \int \dd{r} \tilde{\theta}' \left(\pdv{V}{\tilde{\theta}}+n_s \pdv{m_*}{\tilde{\theta}}\right) = - \int \dd{r} \tilde{\theta}' f^2 \left(\tilde{\theta}'' + \frac{2}{r} \tilde{\theta}'\right),
\end{equation}
for a scalar $\phi = \tilde{\theta} f$ which couples to some fermion with scalar dependent mass $m_*$.
For the second equality, we use the scalar equation of motion.
The term containing $\tilde{\theta}''$ is a total derivative and drops when integrating over the whole wall, since outside of the wall the field is roughly constant.
Estimating $\tilde{\theta}' \sim 1/\lambda_\phi$ for thin walls ($\lambda_\phi\ll R$) and $\tilde{\theta}' \sim 1/R$ in the thick wall limit, we find
\begin{equation}
    \Delta p_\mathrm{grad} = \left\{ \begin{matrix}
        \frac{f^2}{R \lambda_\phi} & \lambda_\phi \ll R\\
        \frac{f^2}{R^2} & \lambda_\phi \sim R
    \end{matrix}
    \right..
\end{equation}
Here $\lambda_\phi \sim \frac{f}{\sqrt{\delta m n_R}}$, with density $n_R$ at $r=R$ and mass reduction $\delta m$.
To obtain estimates for the two main scenarios discussed (coupling to electrons and coupling to nucleons), we define
\begin{equation}
    \delta m \equiv g m_{N/e} \qquad \text{and} \qquad f \equiv M_p \frac{\theta_\ast}{\pi} \sqrt{\frac{g}{\dq{m_{N/e}}}},
\end{equation}
where $\theta = \theta_\ast$ minimizes the fermion mass, defined as in the main text, while $\tilde{\theta}$ is normalized such that the minimal mass, \ie the expectation value of the field in the sourced case, is always achieved at $\tilde{\theta} = \pi$, independent of $\theta_\ast$.
The normalization to $\pi$ is arbitrary, but convenient when matching to the axion case discussed in \cite{Balkin:2022qer}.
We focus only on the bound case ($g<1$), since for the unbound case, even if the potential is negligibly small, the \emph{shape} of the \MR curves already yields an exclusion without considering gradient effects.

Using these expressions in \Eq{eq:pressureBalanceGradient}, we can again estimate the maximal radius on the stable branch.
Working in the limit where the potential is negligibly small, the maximum radius is now parametrically reached when gravity starts to dominate over the gradient contribution to the pressure.
This happens at a density $n_\mathrm{eq}$ defined by $p_\mathrm{grav}(n_\mathrm{eq}) = \Delta p_\mathrm{grad}(n_\mathrm{eq})$.
The corresponding radius can be found from the full pressure balance again, using $p \sim p_e = p_\mathrm{grav}+ \Delta p_\mathrm{grad} \sim p_\mathrm{grav}$ at density $n=n_\mathrm{eq}$.
In the thin wall limit, if the electrons stay non-relativistic, we find
\begin{equation}
    R_\mathrm{max}^\mathrm{stable,thin,NR} = \alpha_1 \left\{
    \begin{matrix}
         (8\pi)^{1/12} \frac{1}{\sqrt{G} m_e^{3/4} m_N^{5/4} (1-g)^{7/6}} \left(\frac{\pi}{\theta_\ast}\right)^{1/6} \frac{\left(\dmn\right)^{1/12}}{g^{1/6}} & \text{nucleon coupling}\\
        (8 \pi)^{1/12} \frac{1}{\sqrt{G}} \left(\frac{1}{m_N}\right)^{7/6} \frac{\left(\dme\right)^{1/12}}{(1-g)^{3/4} m_e^{5/6} g^{1/6}} \left(\frac{\pi}{\theta_\ast}\right)^{1/6} & \text{electron coupling}
    \end{matrix}
    \right.. \label{eq:RmaxStableGradienThinNR}
\end{equation}
Similarly, for ultrarelativistic electrons in the thin wall limit, we have
\begin{equation}
    R_\mathrm{max}^\mathrm{stable,thin,UR} = \alpha_2 \left\{
    \begin{matrix}
         (8\pi)^{1/3} \frac{1}{\sqrt{G} m_N^2 (1-g)^{5/3}} \left(\frac{\pi}{\theta_\ast}\right)^{2/3} \frac{\left(\dmn\right)^{1/3}}{g^{2/3}} & \text{nucleon coupling}\\
          (8 \pi)^{1/3} \frac{1}{\sqrt{G}} \left(\frac{1}{m_N}\right)^{5/3} \left(\frac{\pi}{\theta_\ast}\right)^{2/3} \left(\frac{\dme}{g^2 m_e}\right)^{1/3} & \text{electron coupling}
    \end{matrix}
    \right..
\end{equation}
Lastly, for ultrarelativistic electrons in the thick wall limit, one finds
\begin{equation}
     R_\mathrm{max}^\mathrm{stable,thick,UR} = \alpha_3 \left\{
        \begin{matrix}
            \sqrt{8\pi} \frac{1}{\sqrt{G} m_N^2 (1-g)^2} \frac{\pi}{\theta_\ast} \sqrt{\frac{\dmn}{g}} & \text{nucleon coupling}\\
            \sqrt{8\pi} \frac{1}{\sqrt{G}} \left(\frac{1}{m_N}\right)^2 \frac{\pi}{\theta_\ast} \sqrt{\frac{\dme}{g}} & \text{electron coupling}
        \end{matrix}
     \right.
     .
\end{equation}
Here, the $\alpha_i$ again have to be fixed by comparison to a numerical calculation.

Along the metastable branch, the scalar field remains at $\phi=0$ and consequently there is also no gradient pressure.
Instead, metastability is exactly achieved due to the scalar gradient energy preventing sourcing.
Thus, the maximal density along the metastable branch (which then corresponds to the minimal radius reached), is given by $\lambda_\phi \sim R$, which gives $n(R)$. 
From the pressure balance equation, one quickly finds
\begin{equation}
   R_\mathrm{min}^\mathrm{meta} = \beta\left\{ \begin{matrix}
       (8\pi)^{1/4} \frac{1}{\sqrt{G} m_N^{5/4} m_e^{3/4}} \sqrt{\frac{\pi}{\theta_\ast}} \left(\dmn\right)^{1/4} & \text{nucleon coupling}\\
       (8\pi)^{1/4} \frac{1}{\sqrt{G}} \left(\frac{1}{m_N}\right)^{3/2} \frac{1}{m_e^{1/2}} \sqrt{\frac{\pi}{\theta_\ast}} \left(\dme\right)^{1/4} & \text{electron coupling}
   \end{matrix}
    \right..
\end{equation}
As in the previous sections, along the metastable branch, the electrons stay non-relativistic to a good approximation, and $\beta$ has to be matched from a numerical calculation.

Using these results, we can now turn to the derivation of exclusions in the gradient-dominated limit.
We start with the case of a scalar field coupled to nucleons, which in particular contains the case of a lighter-than-usual QCD axion, which was numerically solved in \cite{Balkin:2022qer}.
Since we explicitly parametrize the dependence on the mass reduction $g$, $\alpha_i$, and $\beta$ are universal and can simply be fixed from the axion case.

\begin{figure}[h]
    \centering
    \begin{subfigure}{0.49\linewidth}
        \centering
        \includegraphics[width=\linewidth]{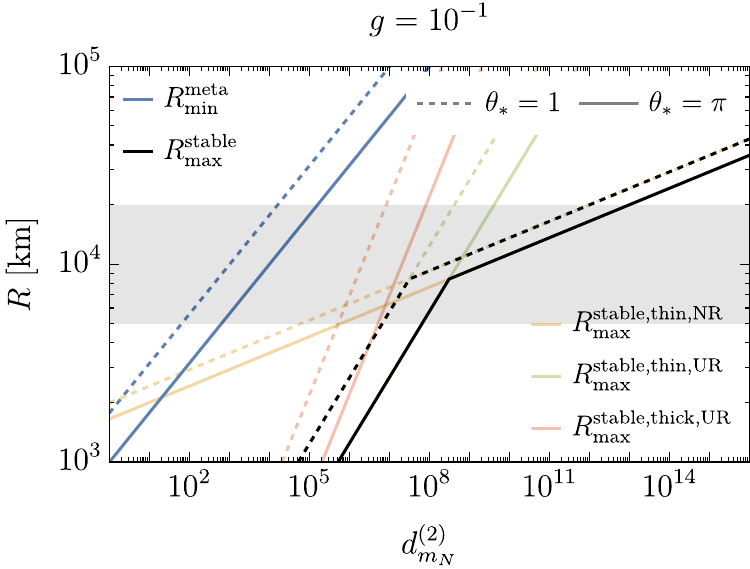}
    \end{subfigure}
    \begin{subfigure}{0.49\linewidth}
        \centering
        \includegraphics[width=\linewidth]{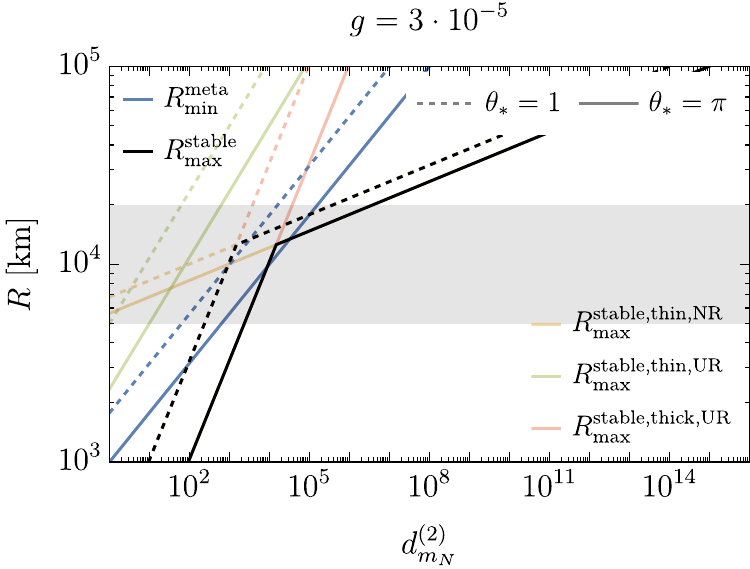}
    \end{subfigure}
    \caption{
    Analytic estimates for the position of the \emph{forbidden gap} in the gradient-dominated limit as a function of $\dmn$.
    The gap spans between the minimal radius of the metastable branch $R_\mathrm{min}^\mathrm{meta}$ (blue) and the maximal radius of the stable branch $R_\mathrm{max}^\mathrm{stable}$ (black).
    The realized value for $R_\mathrm{max}^\mathrm{stable}$ follows the lowest of the three estimates derived above (faint colors).
    Solid and dashed lines show the effect of a variation of $\theta_\ast$.
    The gray band shows the range of white dwarf radii considered for deriving exclusion limits.}
    \label{fig:RGapGradienDmn}
\end{figure}

Consequently, we obtain the gradient-induced radius gap as a function of $\dmn$ for different values of $g$, as shown in \Fig{fig:RGapGradienDmn}.

The maximal stable radius at a given $\dmn$ is obtained by taking the minimum of the three estimates:
At large $\dmn$ (large radii and thus low densities) the wall is thin and the electrons are non-relativistic and we follow $R_\mathrm{min}^\mathrm{stable,thin,NR}$. 
As $\dmn$ decreases, so does the radius, and the density increases.
Consequently, electrons become more relativistic, and we have to switch to $R_\mathrm{min}^\mathrm{stable,thin,UR}$. This happens roughly when the two corresponding radius estimates are equal.
Similarly, one transitions from the thin to the thick wall limit as the radius decreases further.
For low $g$, the transition to the thick wall limit can happen before the thin-wall, ultrarelativistic solution becomes relevant (see \eg \Fig{fig:RGapGradienDmn}, right).

In contrast to the negligible gradient limit, in the gradient-dominated case, an explicit dependence on the maximum field excursion $\theta_\ast$ remains.\footnote{In principle, when working purely in the gradient-dominated limit, the $\theta_\ast$ dependence could be absorbed into a redefinition of the coupling $\dmn$. In terms of this new coupling, the bounds from the negligible gradient limit would pick up a $\theta_\ast$ dependence instead.}
While we show the variation with $\theta_\ast$ in \Fig{fig:RGapGradienDmn}, for concreteness, we fix to the axion-inspired value $\theta_\ast = \pi$ for the limits in the main text.
Additionally, the exact \MR~curves and thus the maximal/minimal radius reached will also depend on the shape of the potential and coupling (\ie the function $F$ defined in the main text), but we expect these effects to be small compared to the variation with $\theta_\ast$.

From these results, we can set an excluded interval in $\dmn$ based on the gradient-induced \emph{forbidden gap}.
For large $\dmn$, the exclusion ends if the gap is at radii larger than where white dwarfs are typically observed.
In contrast to the main text, here we conservatively set the cut-off at $R=\SI{20000}{\km}$.
This accounts for the thermal increase of the white dwarf radii, which is not included in the analytic estimates and starts to become sizable at roughly this radius.

\begin{figure}[h]
    \centering
    \includegraphics[width=0.7\linewidth]{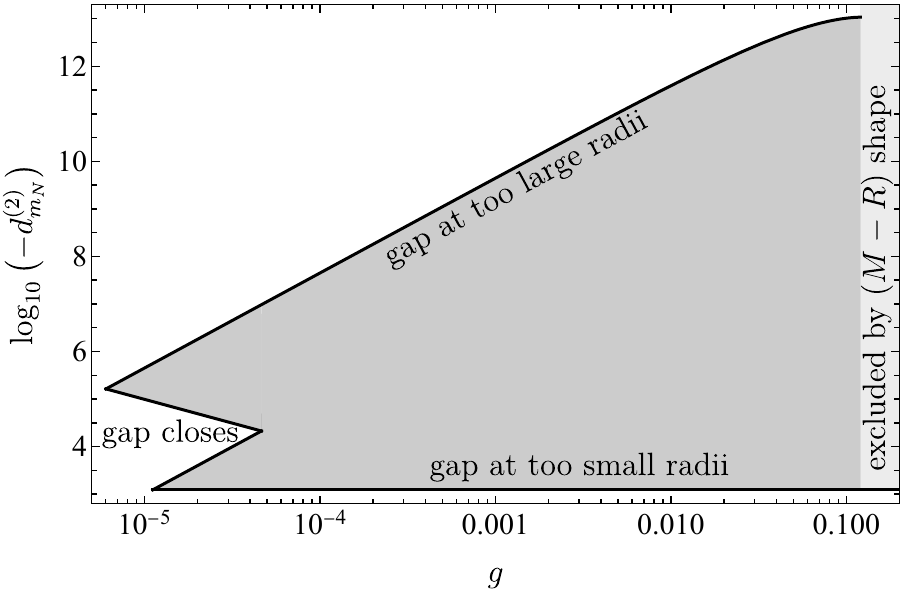}
    \caption{Exclusion limits on $\dmn$ in the gradient dominated limit as a function of $g$ with $\theta_\ast = \pi$ fixed for concreteness.
    For large $\dmn$, the gradient-induced \emph{forbidden gap} is at too large radii to be in tension with observation. For low $\dmn$, gradient effects stabilize the metastable branch enough to explain all observed white dwarfs. 
    At small $g$, the gap can close for intermediate values of $g$, leading to the more complicated features. For even lower $g$, there is no gap remaining at typical radii where white dwarfs are observed, preventing any exclusion.
    For large $g\gtrsim0.1$, the change in the \emph{shape} of the $($M\,$-$\,R$)$ curves is drastic enough to exclude all couplings and masses, which lead to an NGS in white dwarfs, making the gradient-analysis redundant.}
    \label{fig:exclNucGradG}
\end{figure}

For $g\gtrsim \num{5e-5}$ the gap remains open continuously as $\dmn$ decreases (see \eg \Fig{fig:RGapGradienDmn}, left).
Consequently, the exclusion only ends at small $\dmn$ where the metastable branch explains all observed white dwarfs, \ie when $R_\mathrm{min}^\mathrm{meta} \lesssim \SI{5000}{\km}$.
In practice, we assume the same lowest excluded $\dmn$ as in the negligible gradient limit, obtained by comparing the typical gradient scale with white dwarf radii (see \Eq{eq:GradientScaleEstimate}), which describes the same physics and turns out to be slightly more conservative.
For smaller values of $g$ as $\dmn$ is decreased, the gap can close, but then will reopen, once the ultra-relativistic thick wall limit becomes valid for the stable branch (\Fig{fig:RGapGradienDmn}, right).
In practice, this means that the excluded region splits into two intervals in $\dmn$.
At intermediate $\dmn$, where the forbidden gap closes, no exclusion can be placed.
There might be slight variations in a full numerical calculation, but if the region in $\dmn$ where the gap is closed is sufficiently large, we expect this prediction to be robust.
Consequently, we can exclude $\dmn$ as a function of $g$ as shown in \Fig{fig:exclNucGradG} and derive the gradient-induced bounds included in figure \ref{fig:exclusionBoundNucDmG}.

Let us quickly comment on the unusual shape of the exclusion in the bottom right panel of \Fig{fig:exclusionBoundNucDmG}.
There, $g=10^{-5}$, which means that the gap closes at intermediate values of $\dmn$ and in principle two intervals in $\dmn$ might be probed.
Only the upper interval in $\dmn$ is excluded; when the gap reopens for low $\dmn$, the metastable branch can already explain all observed white dwarfs.
Meanwhile, for larger scalar masses, \ie in the potential-dominated limit, the potential keeps the gap open until gradient effects prevent sourcing entirely, so till lower $\dmn$.
The steep step between the two cases is an artifact of just working in those two limits.
In a full numeric calculation, including potential and gradient effects simultaneously, the discontinuity would be replaced by a smooth transition.

For a coupling to electrons, the analysis proceeds in much the same way.
In particular, we take the same $\alpha_i$ and $\beta$ as matched above from the axion case.
Besides the known model dependence due to $\theta_\ast$ and the unknown but presumed small dependence on the exact shape of the potential, an additional uncertainty is introduced here: We use the numerical $O(1)$ coefficients that were found for nucleon couplings directly for a coupling to electrons, and assume a potential change to be small.

\begin{figure}[h]
    \centering
    \begin{subfigure}{0.49\linewidth}
        \centering
        \includegraphics[width=\linewidth]{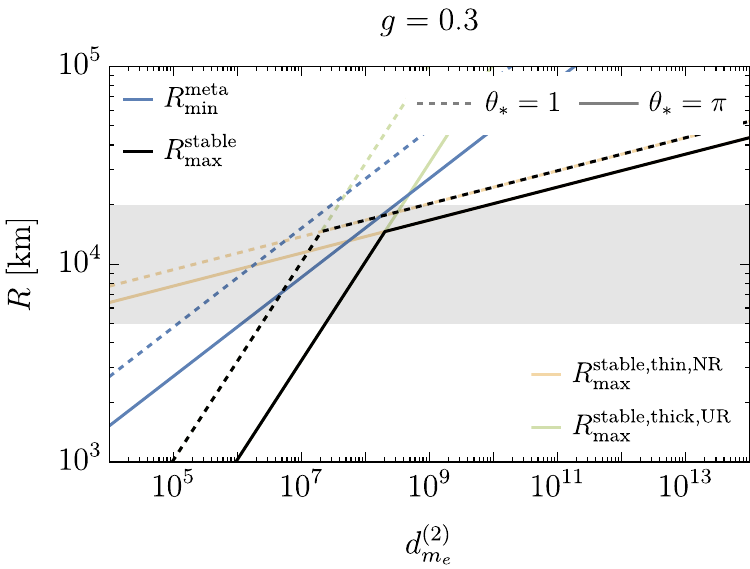}
    \end{subfigure}
    \begin{subfigure}{0.49\linewidth}
        \centering
        \includegraphics[width=\linewidth]{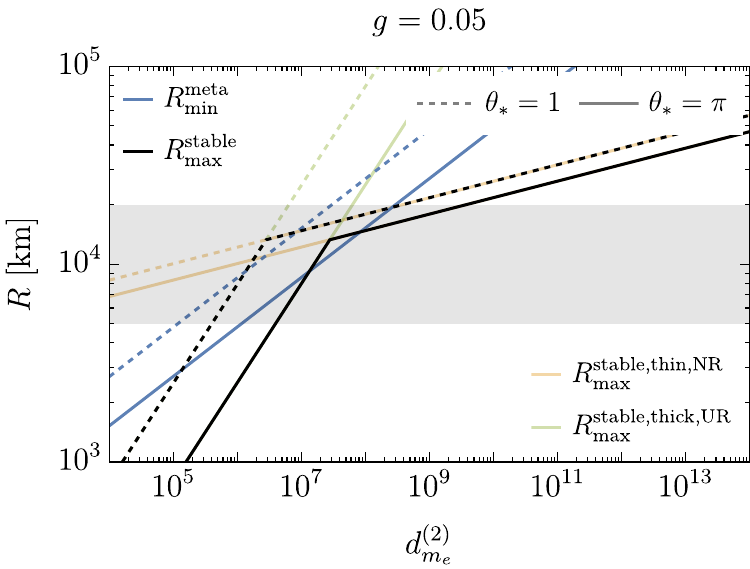}
    \end{subfigure}
    \caption{Analytic estimates for the position of the \emph{forbidden gap} in the gradient-dominated limit as a function of $\dme$: The gap is spanned between $R_\mathrm{min}^\mathrm{meta}$ (blue) and $R_\mathrm{max}^\mathrm{stable}$ (black). The latter is obtained from the minimum of the two relevant analytic estimates (orange and green); $R_\mathrm{max}^\mathrm{stable,thin,UR}$ turns out to be irrelevant.
    Effects of a variation of $\theta_\ast$ (solid vs dotted) and the range of white dwarf radii (gray) are shown as well.}
    
    \label{fig:RGapGradienDme}
\end{figure}

\begin{figure}[h]
    \centering
    \includegraphics[width=0.7\linewidth]{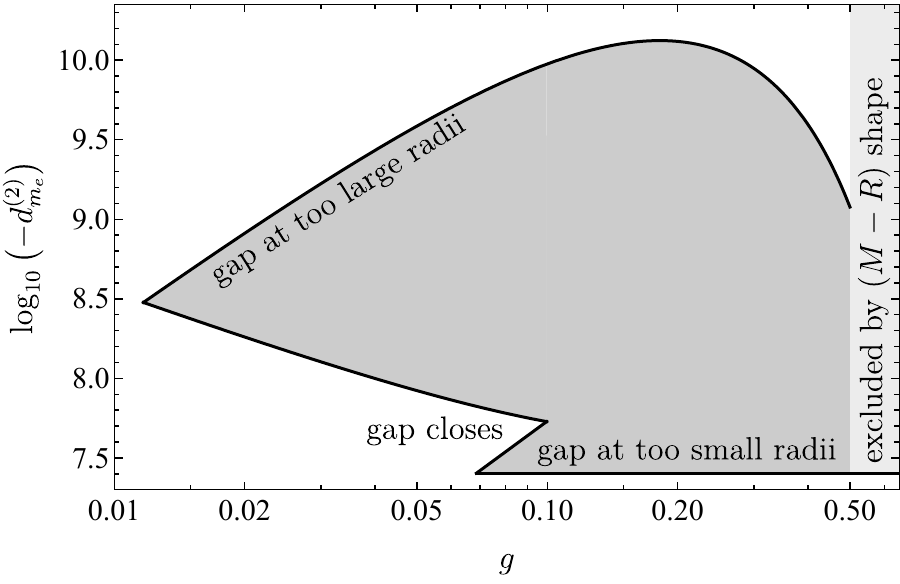}
    \caption{Exclusion limits on $\dme$ in the gradient dominated limit as a function of $g$ with $\theta_\ast = \pi$. The exclusion ends if the \emph{forbidden gap} is at too large radii (large $\dme$), if gradient effects stabilize the metastable branch (small $\dme$) or if the gap only remains open for radii where no white dwarfs are observed (small $g$). At large $g$, there is a $\dme$ independent exclusion due to the \emph{shape} of the \MR\space curve.
    Additionally, at large $g$, the effect that electrons become lighter due to sourcing starts to affect the pressure ($1-g$ term in \Eq{eq:RmaxStableGradienThinNR}), leading to larger maximal radii and thus a weaker gradient limit.
    }
    \label{fig:exclElGradG}
\end{figure}

We again find $R_\mathrm{max}^\mathrm{stable}$ and $R_\mathrm{min}^\mathrm{meta}$ as a function of $\dme$ and $g$ (\Fig{fig:RGapGradienDme}).
It turns out that for all $g$-values of interest, $R_\mathrm{max}^\mathrm{stable,thin,UR}$ is irrelevant and thus will be omitted in the plots.
The gap analysis proceeds exactly as described for the nucleonic case above and leads to the exclusions in \Fig{fig:exclElGradG} and the gradient bands in \Fig{fig:exclusionBoundElectron}, where we again fix $\theta_\ast = \pi$.

Similarly to the case of a nucleonic coupling, we predict the gap to close for intermediate values of $\dme$ where the stable branch transitions from the thin to the thick wall limit, this leads to the features at low $g$ in \Fig{fig:exclElGradG}.
Again, for even smaller $g$, only the gap at larger $\dme$/radii remains at radii where white dwarfs are observed, while the gap at small $\dme$/radii quickly moves to regions where no white dwarfs are seen.
This explains the different $\dme$ probed in the potential versus gradient dominated limit in the bottom panels of \Fig{fig:exclElGradG}, precisely as for the nucleonic case above.
Again, the sharp discontinuity is an artifact of just working in either the potential or the gradient-dominated limit and will be cured by a full numeric calculation.

\section{Combined Couplings} \label{app:combinedCouplingsMR}
As described in \Sec{sec:MixedQuadraticCouplings}, in large parts of the parameter space, it is sufficient to focus on a quadratic coupling of the scalar field to either just electrons or just nucleons.
In this appendix, we show some results for two benchmark points where both couplings are active.
In particular, this includes the case of a dilatonic coupling, a scalar that couples to the standard model energy-momentum tensor.
We denote the coupling as
\begin{equation}
    \mathcal{L} \supset - g \frac{m_\phi^2 M_p^2}{|\dq{m}|} F\left(\theta\right) - m_e \left(1  + \kappa_e g F(\theta)\right) \bar{\psi}_e \psi_e - m_N \left(1 + \kappa_N g F(\theta)\right) \bar{\psi}_N \psi_N
    \label{eq:dilatonCoupling}
\end{equation}
with $\theta = \sqrt{\frac{|\dq{m}|}{g}} \frac{\phi}{M_p}$, $\kappa_e = \kappa_N = \mathrm{sign}\left(\dq{m}\right)$ and $F$ as in \Eq{eq:boundElectronCoupling}.
Again, we focus on the case $\dq{m}<0$, which leads to sourcing.

Compared to a model where the scalar field just couples to nucleons, introducing an electron coupling leads to smaller critical densities.
The change in critical density is however given by $n_c^\mathrm{both}/n_c^\mathrm{nucleons} \approx (1-m_e/(Ym_N)) \approx 1/4000$, so in practice negligible.
Similarly, when the scalar becomes sourced, there is an additional reduction of the energy per particle due to the electrons becoming lighter as well. Again, it is suppressed by the same factor $m_e/(Ym_N)$ and thus negligible.
Consequently, the effect of the additional coupling on the NGS-phase transition boundary is negligible, and the boundary shown in \Fig{fig:NGSBoundaryAndMRBoundNucleons} (left) is still valid to a good approximation.

The same still holds if the absolute value of the coupling between nucleons and electrons is the same, but the sign is different, \ie, where the electrons become heavier by a factor of $(1+g)$ when the scalar becomes sourced.
This corresponds to $\kappa_N = -1$ and $\kappa_e = + 1$ in \Eq{eq:dilatonCoupling}.

\begin{figure}[h]
    \centering
    \begin{subfigure}{0.49\linewidth}
        \centering
        \includegraphics[width=\linewidth]{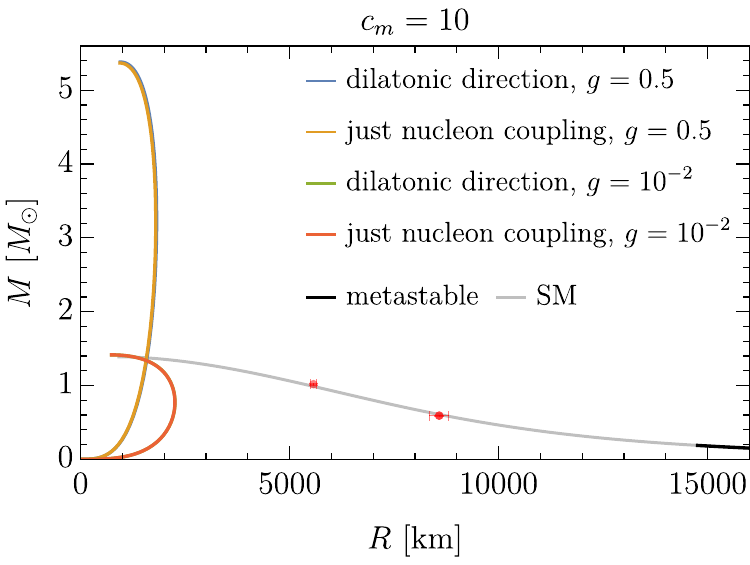}
    \end{subfigure}
    \begin{subfigure}{0.49\linewidth}
        \centering
        \includegraphics[width=\linewidth]{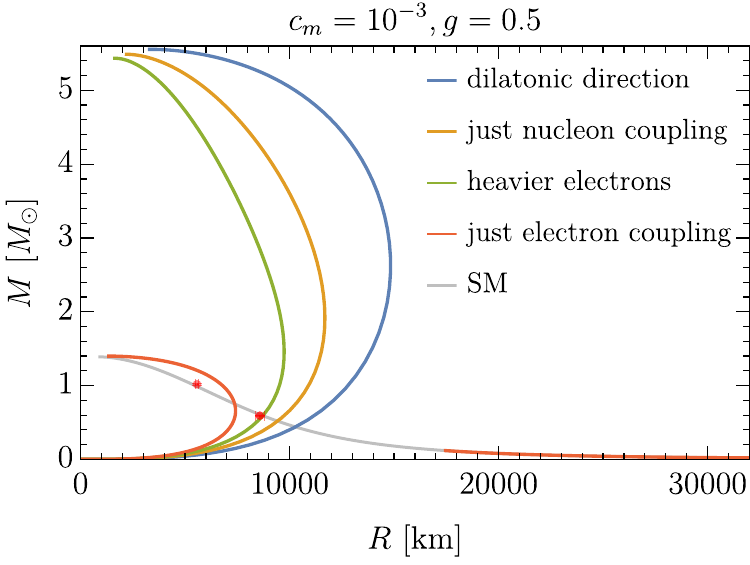}
    \end{subfigure}
    \begin{subfigure}{0.49\linewidth}
        \centering
        \includegraphics[width=\linewidth]{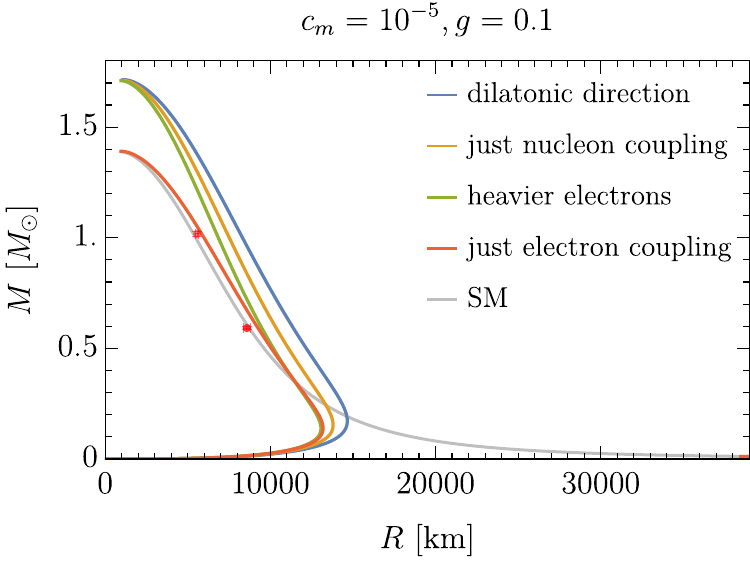}
    \end{subfigure}
    \begin{subfigure}{0.49\linewidth}
        \centering
        \includegraphics[width=\linewidth]{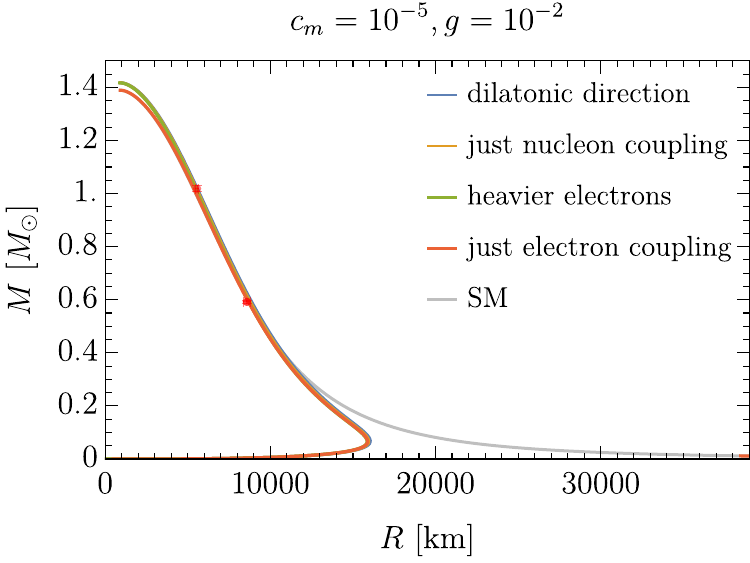}
    \end{subfigure}
    \caption{($M\,$-$\,R$) curves comparing the effects of different SM-scalar couplings at different values of $c$ and $g$. Well-known white dwarfs Sirius B and Procyon B are shown in red. In the top left panel, the metastable branch (black) is the same for all couplings considered. In the other panels, the shown one (red) only applies to the coupling to electrons; if the scalar also (or exclusively) couples to nucleons, the metastable branch ends at far larger radii.}
    \label{fig:dilatonCouplingMR}
\end{figure}

To analyze the effect of the model on white dwarfs, we are again focusing on the ($M\,$-$\,R$) relation, some results are shown in figure \Fig{fig:dilatonCouplingMR}.
If $c_{m_e}$ is large (top left), as long as an NGS is present, there is barely a visible difference between just coupling to nucleons and the dilatonic coupling.
The difference becomes more pronounced for smaller $c_{m_e}$ and large $g$ (top right).
In contrast to the large $c_{m_e}$ case, electrons are still mostly non-relativistic at the NGS density, consequently making them considerably lighter has the effect of leading to blown-up ($M\,$-$\,R$) curves, as also described in \Sec{sec:QuadCouplingElectrons}.
Consequently, in principle, the gap becomes smaller (larger) when electrons become lighter (heavier) due to scalar sourcing.
Nonetheless, this is not enough to significantly close the gap. 
Moreover, the blown-up ($M\,$-$\,R$) curves are also in tension earlier with the observation of Sirius B and Procyon B.
Lastly, reducing the mass reduction $g$ (bottom row) reduces this effect:
If the change in the electron mass is small, they also do not become relativistic much earlier.
In particular, let us also point out the case of very small $g \lesssim 10^{-2}$ (bottom right).
In this limit, the nucleon and electron mass reduction effect on the ($M\,$-$\,R$) curves is negligible.
The dominant effect now lies in the addition of the vacuum potential to the EOS, so all models, no matter if coupling to just nucleons, just electrons or both, lead to nearly the same stable ($M\,$-$\,R$)-branch.\footnote{Note, that the critical density is of course still very much influence by the species coupled to, since the same relative change in the nucleon mass is linked to a much higher change in energy density compared to a change in the electron mass.}
This can be understood as follows: 
The modifications of the energy density (pressure) due to sourcing can be split into two parts, an increased (decreased) matter energy-density $\delta \varepsilon_m$ (pressure $\delta p_m$) due to the changed mass and the addition (subtraction) of the vacuum potential.
To leading NR-order $\delta \varepsilon_m$ and $\delta p_m$ are both suppressed with the mass change $\delta m/m \ll 1$ relative to $\varepsilon$ and $p$ respectively and thus can be neglected. 
On the one hand, the vacuum potential is, by definition of $n_c$, exactly canceled by the change in energy density at the critical density $V = \delta \varepsilon_m(n_c)$.
Since $\delta \varepsilon_m$ increases with density but remains small compared to $\varepsilon$, $V \lesssim \delta \varepsilon_m \ll \varepsilon_m$ can also be neglected.
On the other hand, at the NGS density, $V$ is given by the total matter pressure $V = p_m$, \ie., in general, $V\sim p_m$, which cannot be neglected.
Consequently, if the mass reduction is small, but an NGS is present, the change of the EOS in the NGS phase can be approximated as
\begin{equation*}
    \varepsilon \to \varepsilon \qquad \qquad p \to p - V,
\end{equation*}
which is independent of the specific coupling between the scalar field and fermions.
Let us finally note that in the NR-limit, we have $\varepsilon \sim m_N m_e^3 (n/m_e^3)$, while $p \sim m_e^4 (n/m_e^3)^{5/3}$, \ie, for $n\ll m_N^3$, $p \ll \varepsilon$, so both conditions on $V$ can be fulfilled, leading to an NGS.

Summarizing the above, while sometimes distinguishable in ($M\,$-$\,R$) curves, the difference between a coupling to just nucleons and a dilatonic coupling does not change our exclusion limits due to white dwarf observation.
Consequently, the limit shown in \Fig{fig:exclusionBoundNucDmG} remains valid for the dilatonic case.

\bibliographystyle{JHEP}
\addcontentsline{toc}{section}{Bibliography}
\bibliography{bibliography}

\end{document}